# Latin American Giant Observatory: Contributions to the 34th International Cosmic Ray Conference (ICRC 2015)

## LAGO Collaboration

W. Alvarez[13], C. Alvarez[9], C. Araujo[18], O. Areso[7], H. Arnaldi[2], H. Asorey[2], M. Audelo[5], H. Barros[20], X. Bertou[2], M. Bonnett[18], R. Calderon[14], M. Calderon[4], A. Campos-Fauth[21], A. Carramiñana[8], E. Carrasco[8], E. Carrera[19], D. Cazar[19], E. Cifuentes[13], D. Cogollo[22], R. Conde[1], J. Cotzomi[1], S. Dasso[7,24,25], A. De Castro[14], J. De La Torre[18], R. De León[13], A. Estupiñan[14], A. Galindo[8], L. Garcia[13], M. Gómez Berisso[2], M. González[2], W. Guevara[3], A.M. Gulisano[6,7], H. Hernández[11], A. Jaimes[14], J. López[10], C. Mantilla[4], R. Martín[10], A. Martínez-Mendez[14], O. Martínez[1], E. Martins[22], J.J. Masías-Meza[25], R. Mayo-García[23], T. Melo[22], J. Mendoza[19], P. Miranda[15], E. Montes[23], E. Morales[18], I. Morales[13], E. Moreno[1], C. Murrugarra[20], C. Nina[15], L.A. Núñez[14], A. Núñez-Castiñeyra[11], L. Otiniano[3], J. Peña-Rodríguez[14], J. Perenguez[12], H. Pérez[13], Y. Perez[11], G. Perez[17], S. Pinilla-Velandia[14], E. Ponce[1], R. Quishpe[19], F. Quispe[3], M. Ramelli[7], K. Reyes[12], H. Rivera[15], J. Rodriguez[18], M. Rodríguez-Pascual[23], M. Romero[18], A.J. Rubio-Montero[23], H. Salazar[1], J. Salinas[15], C. Sarmiento-Cano[14], I. Sidelnik[2], M. Sofo Haro[2], M. Suárez-Durán[14], M. Subieta[15], J. Tello[20], R. Ticona[15], I. Torres[8], L. Torres-Niño[14], J. Truyenque[3], M. Valencia-Otero[14], S. Vargas[4], N. Vásquez[4], L. Villasenor[16], M. Zamalloa[18], L. Zavala[5]

[1] Benemérita Universidad Autónoma de Puebla, México.
[2] Centro Atómico Bariloche, Instituto Balseiro, Argentina.
[3] Comisión Nacional de Investigación y Desarrollo Aeroespacial, Perú.
[4] Escuela Politécnica Nacional, Ecuador.
[5] Escuela Superior Politécnica de Chimborazo, Ecuador.
[6] Instituto Antártico Argentina, Dirección Nacional del Antártico, Argentina.
[7] Instituto de Astronomía y Física del Espacio, IAFE, Argentina.
[8] Instituto Nacional de Astrofísica, Óptica y Electrónica, México.
[9] Universidad Autónoma de Chiapas, México.
[10] Universidad Central de Venezuela, Venezuela.
[11] Universidad de Los Andes, Venezuela.
[12] Universidad de Nariño, Colombia.
[13] Universidad de San Carlos, Guatemala.

[14] Universidad Industrial de Santander, Colombia.
[15] Universidad Mayor de San Andrés, Bolivia.
[16] Universidad Michoacana de San Nicolás de Hidalgo, México.
[17] Universidad Politécnica de Pachuca, México.
[18] Universidad San Antonio Abad del Cusco, Perú.
[19] Universidad San Francisco de Quito, Ecuador.
[20] Universidad Simón Bolivar. Venezuela.
[21] Universidade Estadual de Campinas, Brazil.
[22] Universidade Federal de Campina Grande, Brazil.
[23] Centro de Investigaciones Energéticas Medioambientales y Tecnológicas, Spain.
[24] Departamento de Ciencias de la Atmósfera y los Océanos, Facultad de Ciencias Exactas y Naturales, Universidad de Buenos Aires, Argentina.
[25] Departamento de Física, Facultad de Ciencias Exactas y Naturales, Universidad de Buenos Aires, Argentina.
[26] Instituto Geofísico del Perú, Perú.

# Acknowledgments

The operation and installation of the Latin American Giant Observatory would not have been possible without the continuous technical and financial support of the following institutions, agencies and organizations, to whom we are most grateful:

Comisión Nacional de Energía Atómica (CNEA), Agencia Nacional de Promoción Científica y Tecnológica (ANPCyT), Consejo Nacional de Investigaciones Científicas y Técnicas (CONICET), Fondo para la Investigación Científica y Tecnológica (FonCyT), Ministerio de Ciencia, Tecnología e Innovación Productiva (MinCyT), Universidad Nacional de Cuyo (UnCuyo), Universidad Nacional de Buenos Aires (UBA), Centro Atómico Bariloche (CAB), Instituto Balseiro (IB), Instituto Antártico Argentina (IAA), Instituto de Astronomía y Física del Espacio, (IAFE), Argentina; Ministerio de Defensa de Argentina, Grant PIDDEF 2014-2017 (8); Universidad Mayor de San Andrés (UMSA), Bolivia; Universidade Estadual de Campinas, Universidade Federal de Campina Grande (UFCG), Brasil; Vicerrectoría de Investigación y Extensión de la Universidad Industrial de Santander, Departamento Administrativo de Ciencia, Tecnología e Innovación (COLCIENCIAS), Universidad de Nariño, Colombia; Universidad San Francisco de Quito (USFQ), Escuela Politécnica Nacional (EPN), Escuela Superior Politécnica de Chimborazo, Fundación CEDIA, Red CEDIA, Grant: CEPRAIX-2015-4-Astropartículas, Ecuador; Consejo Nacional de Ciencia y Tecnología (CONACYT), Grants 194547, 105666, Red de Física de Altas Energías (FAE), Benemérita Universidad Autónoma de Puebla (BUAP), Instituto Nacional de Astrofísica, Óptica y Electrónica (INAOE), Universidad Autónoma de Chiapas (UNACH) Universidad Michoacana de San Nicolás de Hidalgo (UMSNH), Universidad Politécnica de Pachuca (UPP), Mexico; Universidad de San Carlos, Guatemala; Comisión Nacional de Investigación y Desarrollo Aeroespacial (CONIDA), Instituto Geofísico del Perú (IGP), Consejo Nacional de Ciencia, Tecnología e Innovación Tecnológica (CONCYTEC), Fondo para la Innovación, Ciencia y Tecnología (FINCyT), Universidad San Antonio Abad del Cusco, Perú; Centro de Investigaciones Energéticas, Medioambientales y Tecnológicas (CIEMAT), Spain; Consejo de Desarrollo Científico y Humanístico, Tecnológico y de las Artes (CDCHTA) la Universidad de los Andes, Consejo de Desarrollo Científico y Humanístico, Tecnológico y de las Artes (CDCHTA) de la Universidad Central de Venezuela, Fondo Nacional de Ciencia Tecnología e Innovación (FONACIT), Universidad Central de Venezuela (UCV), Universidad de Los Andes (ULA), Universidad Simón Bolívar, Teleférico de Mérida, grants CDCH-UCV PG-03-8808-2013/1, CDCHTA-ULA C-1949-15-05-EM, CDCHT-ULA C-1598-08-05-A, Venezuela; Pierre Auger Observatory, Centro Latinoamericano de Física (CLAF), Fondo Regional para la Innovación Digital en América Latina y el Caribe (FRIDA), International Centre for Theoretical Physics (ICTP).

# Contributions

# 1

# Latin American Giant Observatory Project

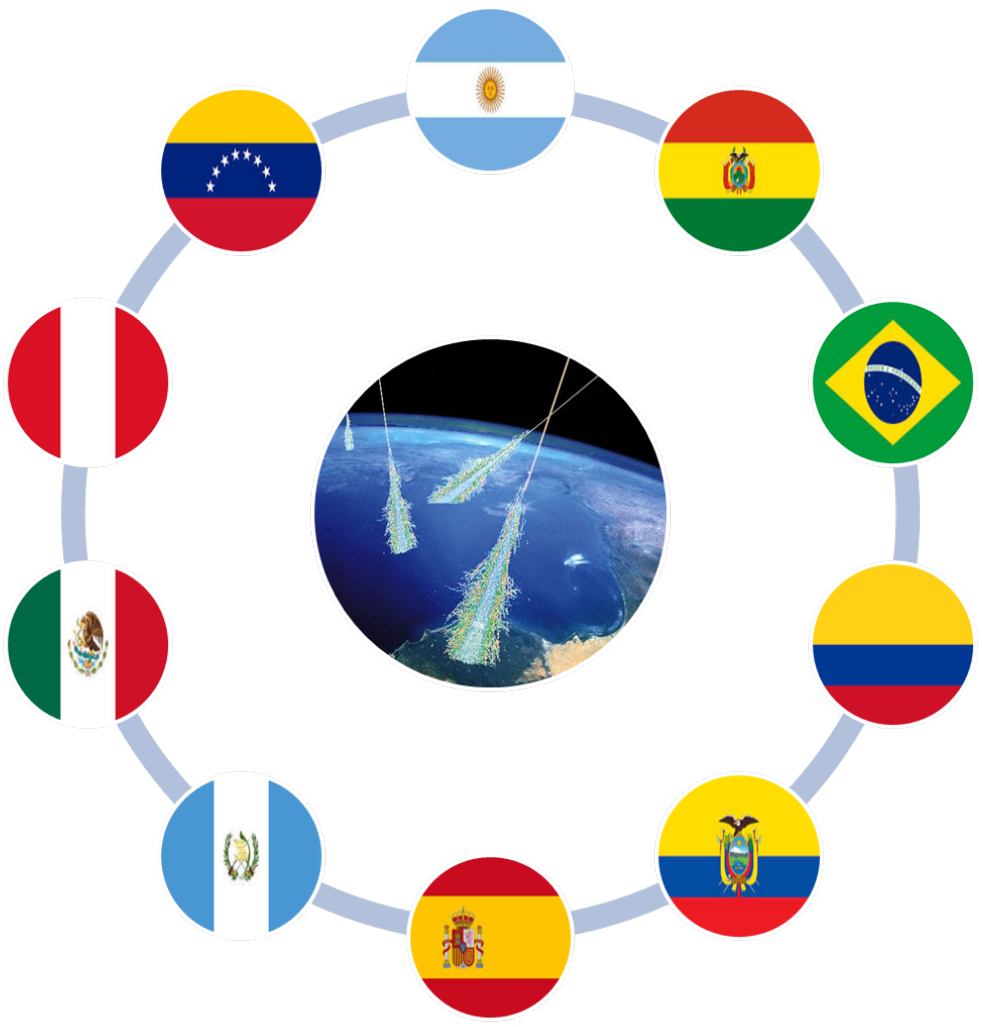
[1]

# LAGO: the Latin American Giant Observatory

**H. Asorey**[1,2,3] **and S. Dasso**[*4,5] **, for the LAGO Collaboration**[6]

[1] *Laboratorio Detección de Partículas y Radiación, Instituto Balseiro y Centro Atómico Bariloche, S.C. de Bariloche, Argentina.*
[2] *Sede Andina, Universidad Nacional de Río Negro, S.C. de Bariloche, Argentina.*
[3] *Escuela de Física, Universidad Industrial de Santander, Bucaramanga, Colombia.*
[4] *Instituto de Astronomía y Física del Espacio (UBA-CONICET), Buenos Aires, Argentina*
[5] *Departamento de Ciencias de la Atmósfera y los Océanos and Departamento de Física, Facultad de Ciencias Exactas y Naturales, Universidad de Buenos Aires (UBA), Buenos Aires, Argentina*
[6] *lagoproject.org, see the full list of members and institutions at lagoproject.org/collab.html*
*e-mail:* asoreyh@cab.cnea.gov.ar

The Latin American Giant Observatory (LAGO) is an extended cosmic ray observatory composed by a network of water-Cherenkov detectors spanning over different sites located at significantly different altitudes (from sea level up to more than 5000 m a.s.l.) and latitudes across Latin America, covering a huge range of geomagnetic rigidity cut-offs and atmospheric absorption/reaction levels. This detection network is designed to measure the temporal evolution of the radiation flux at ground level with extreme detail. The LAGO project is mainly oriented to perform basic research in three branches: high energy phenomena, space weather and atmospheric radiation at ground level. LAGO is built and operated by the LAGO Collaboration, a non-centralized collaborative union of more than 30 institutions from ten countries.

In this work, we will describe several scientific and academic programs that are conducted within the LAGO framework, its present status and future perspectives.

*Speaker.

## 1. Introduction

Astroparticle physics is nowadays one of the scientific fields that evidence large interdisciplinary contributions. This is not only possible but even needed, given the large array of topics this discipline covers: from high energy astrophysics (e.g. acceleration mechanisms) to computational sciences (e.g., data lineage and provenance). Cosmology, radiation-matter interactions, magnetohydrodynmics, Earth sciences, atmospheric science, detector physics, chemistry, biology and medical physics, have to be turned up to correctly interpret our results and to asses the extent of our predictions.

A large, enthusiastic and very active community has been emerging around this discipline as it can be observed at this conference. Several space borne and ground based cosmic rays observatories have been built or are being designed, comprising hundreds or even thousands of scientists working in large international collaborations. In the particular case of Latin America, the successful installation and commissioning of the Pierre Auger Observatory [1] in Malargüe, Argentina, generated an outstanding opportunity to develop Astroparticle physics and High Energy Physics in this region. Nowadays this is one of the main academic objectives of the Latin American Giant Observatory (LAGO), formerly known as Large Aperture Gamma Ray Bursts Observatory, a project conceived in 2006 [2] to detect the high energy component of Gamma Ray Bursts (GRBs) using low cost water Cherenkov detectors (WCDs) at high altitude sites across the Andes.

From this initial milestone, LAGO project has been evolved toward an extended astroparticle observatory at regional scale, currently operating WCDs and other particle detectors at nine countries in LA. Currently, LAGO has three main scientific objectives: to study high energy gamma events at high altitude sites, to understand space weather phenomena and monitoring it at continental scale and decipher the impact (direct and indirect) of the cosmic radiation on atmospheric phenomena. These objectives are complemented by two main academic goals: to train students in astroparticle and high energy physics techniques, and foremost, to support the development of astroparticle physics in LA. This paper is organized as follows: in the next section, the project, its organization, detector concepts and main objectives are described. Then, in section 3, some of our programs are described. Finally, in section 4, the future perspectives are presented.

## 2. The Latin American Giant Observatory

The Latin American Giant Observatory was originally designed to search for the high energy component of GRBs (with typical energy of primaries $E_p \gtrsim 20\,\mathrm{GeV}$) by installing $10\,\mathrm{m}^2 - 20\,\mathrm{m}^2$ of sensitive area at very high altitude sites across the Andean ranges [3]. Later in 2013, a new data analysis procedure was developed, allowing the possibility to study, at the same time, Space Weather phenomena by observing the Solar modulation of Galactic Cosmic Rays (GCR), and taking the advantage of several WCDs installed at low altitude sites [4].

At present, the LAGO Project is being built as an extended astroparticle observatory at a regional scale, operated by a cooperative and non-centralized collaboration of 26 institutions from 9 Latin American countries (Argentina, Bolivia, Brazil, Colombia, Ecuador, Guatemala, Mexico, Perú and Venezuela) and the recent incorporation of institutions from Spain. The geographic distribution of our sites, involving different countries with different cultures, idiosyncrasies and procedures, requires a highly dynamic and extended organizational schema. The "Coordination Committee" is the decision making entity of the project and consists of a principal investigator and one democratically elected representative for each participating country. All the collaboration members are organized in four working groups: WG1-Physics, WG2-Detectors, WG3-Data and WG0-Management.

The LAGO detection network, shown in figure 1 on the left, consist of single or small arrays of astroparticle detectors installed in different sites across the Andean region [5]. Currently, ten WCDs are in operation and we expect to have another eleven detectors starting their operation and calibration in the 2016-2017 biennium. As it is clearly visible in the map, our detection network spans from the south of Mexico, with a small array installed at Sierra Negra ( 4550 m a.s.l.), to the Patagonia, with three WCDs installed at Bariloche (865 m a.s.l.). Moreover, it was recently funded the installation of two WCDs at the Marambio Base (Arg., 200 m a.s.l.) located in the Antarctic Peninsula [6], mainly oriented for Space Weather studies and monitoring [7]. In the same figure, on the right, the geomagnetic rigidity cut-off (in GV) of each site as a function of its geographic latitude (connected with the map on the left) is shown. In this sense, the distribution of the network, comprises quite similar geographical longitudes but large differences in geographical latitudes and altitudes. Then, by combining simultaneous measurements at different rigidity cutoffs and atmospheric absorption reactions places we are able to produce near-real-time information at different energy ranges of, for example, disturbances induced by interplanetary transients and long term space weather phenomena.

Each LAGO water Cherenkov detector consist of a plastic tank containing 1 $m^3$ to 40 $m^3$ of purified water, where one [8] to four [9] on-top large photomultiplier tubes (PMT) collect the Cherenkov light produced by ultra-relativistic particles moving through the water volume. An internal coating of an ultra-violet light highly reflective and diffusive fabric is used. The WCD signals are shaped and digitized by a custom made 40 MHz electronic board controlled by a Digilent Nexys2 FPGA [10, 9].

The need to measure very different physical magnitudes in remote places with almost no infrastructure requires autonomous and reliable detectors and components. Moreover, we based our design in commercial off-the-shelf devices, which have to be available in all the countries where LAGO operates. For this, a different approach was introduced by changing our station concept and to consider the WCD just as an additional sensor that accounts for cosmic radiation. In the same way, the main component is the electronic system that operates the complete station. Other sensors are integrated in an Arduino controlled common sensor module to measure atmospheric pressure, air temperature, solar radiance, cloudiness, and even $CO_2$ and other greenhouse and smog gasses concentrations such as CO, $NO_x$, $CH_4$ and $H_2O$. A commercial global navigation satellite system module is included for timing synchronization and to allow the calculation of ionospheric total electron content (TEC) calculations in some sites [11]. All these devices, including the WCD electronics boards, are powered by solar panels and are controlled by a single board computer that includes a GPRS module to transfer telemetry and on board pre-analyzed data by using mobile phone networking data services. This new concept produces data that extends the typical objectives of a cosmic ray observatory, allowing several scientific, academic and even citizen science communities to take advantage of our regional scale detection network [13, 14, 15].

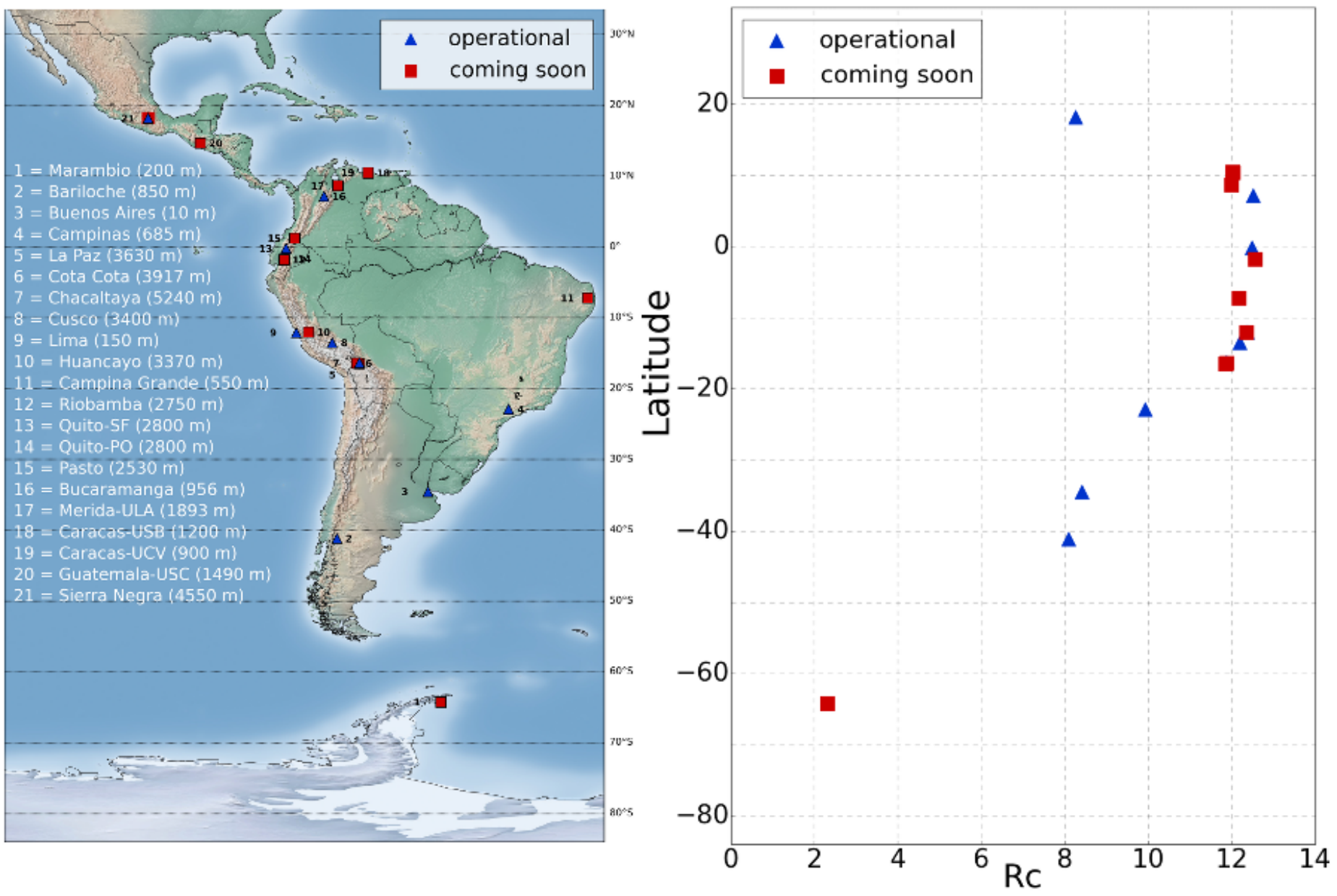

**Figure 1:** Geographical distribution and altitudes of the operational LAGO water Cherenkov detectors (blue triangles), and those that are being deployed and will start its operation during the 2016-2017 biennium (red squares). At the right panel, the vertical rigidity cut-off of each LAGO site is shown.

## 3. The LAGO Programs

Scientific and academic objectives are organized in different programs and are carried out by the corresponding working groups. LAGO programs cover several aspects of the project, from the installation [6, 8], calibration and operation [16, 9] of the detectors to the search for pathways to transfer data from remote sites [13]. Complete simulation chains involving all the related aspects (from CR propagation to detector response) [16, 7], data analysis techniques specially designed for the very different energy and temporal scales of the studying phenomena [7, 17, 9], data preservation [13], and the design of new experiments to be conducted in graduate and undergraduate lectures in the participating Universities [15, 14], are just some examples of the range of the objectives of the LAGO project. In this section some of these programs will be described and some of the scientific results presented at this conference will be highlighted.

High altitude sites ($h > 4500$ m) are designed and operated mainly for the search of high energy components of GRB. The main purpose to climb to this high altitudes mountains is to diminish the atmospheric absorption of extensive air showers (EAS) initiated by low energy cosmic rays. As at this altitudes the muonic component of EAS are not fully developed (see e.g. [4] for the background muon content at the LAGO site of Chacaltaya), it is not possible to use the vertical equivalent muon (VEM) technique to calibrate WCD (see e.g. [1, 7]). In this case, instead of looking for the characteristic muon hump in the histogram of time-integrated pulse signals (charge histograms), the VEM is obtained by looking for a slight change in the charge histogram slope. However, to increase electromagnetic-muon separation at single pulse level, a method based on the total charge and pulse rise-time analysis is also implemented, as can be observed in the left panel of figure 2. The increase in separation performance can be used to improve the search for possible GRB candidates, as gamma initiated showers evidence lower muon fractions at detector level when compared with hadronic primaries.

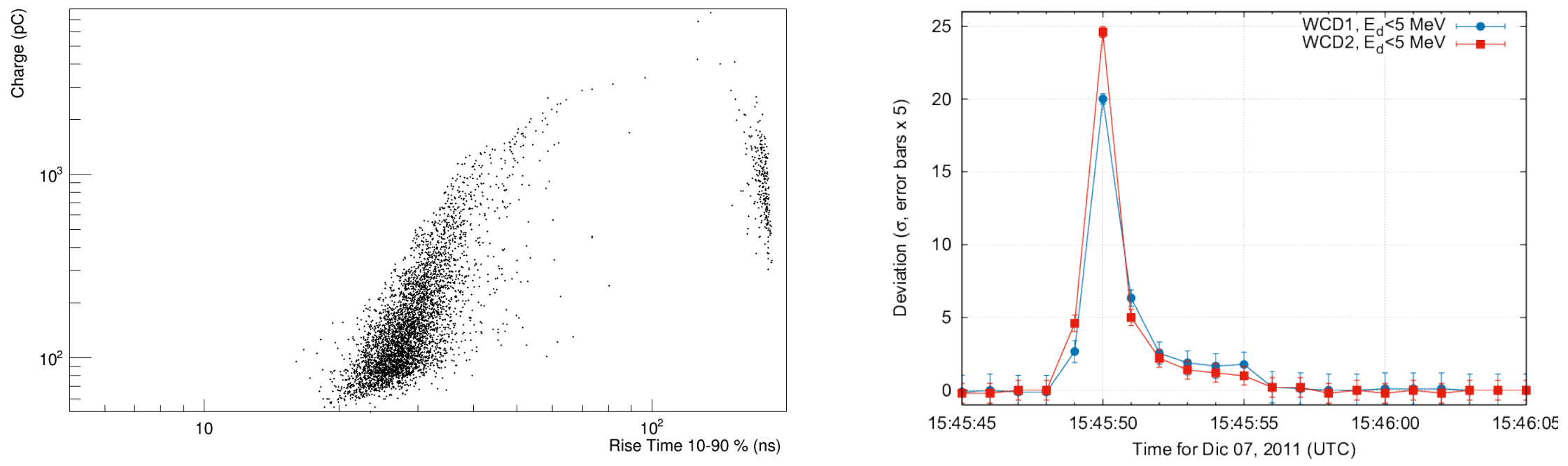

**Figure 2:** Left: component separation by using pulse shape analysis of individual signals: here, the total integrated charge as a function of the 10% → 90% rise-time shows a clear separation of low and short signals (mainly electromagnetic) from typical muon signals (from [9]). Right: Signal alert for a potential event candidate registered on Wed Dec 07 15:47:02.378 UTC 2011 detected by LAGO at Chacaltaya, Bolivia, 5250 m a.s.l. (from [17]).

All the LAGO programs exploit the single particle technique (SPT) [18] by looking for significant excesses in the counting of background signals at different sites. In particular, the LAGO facilities at Chacaltaya mountain consist of three LAGO WCDs with a total detection area of $10\,\mathrm{m}^2$. In the period 2010-2012 the Chacaltaya station has collected data during $\sim 17000$ hours of detection for WCD1 and WCD2, and $\sim 15700$ hours for WCD3. Recent studies, based on CORSIKA simulations, show that the angular aperture of the Chacaltaya site can be extended up to a zenith angle of $25^{\circ}$ in the energy range of interest for GRB and other Gamma originated signals at Chacaltaya. Combining these results with the uptime of each WCD at Chacaltaya in this period, the total exposure of this site accumulated during this period was $2.7 \times 10^8\,\mathrm{m}^2\,\mathrm{s\,sr}$. Considering the typical variations on the flux at the time scales of interest, we base the first steps of our analysis in the moving window average (MWA) method. Using this method, we look for $\geq |3\sigma|$ instantaneous deviations in the central 5 ms bin on a moving average window of 2 minutes (24,000 time bins) of duration. If such a deviation is observed, the data is tagged as a potential transient candidate. Then, a strict criterion based on searching for simultaneous deviations in different energy sub-channels and detectors is imposed (see [17]). Following this procedure, a new potential candidate was found, started on Wed Dec 07 15:45:49.675±0.005 UTC 2011 and a duration of 5.5 s, as can be seen in the right panel of figure 2. At this time, the equatorial coordinates of Chacaltaya zenith were RA/Dec (J2000) $16^{\mathrm{h}}17^{\mathrm{m}}31.3^{\mathrm{s}}/-16^{\circ}21'00''$, with an acceptance aperture of $\theta \lesssim 25^{\circ}$. At this time, Fermi satellite was outside of our acceptance cone in Chacaltaya.

While this technique has the lack of directional reconstruction, it is however possible to use atmospheric absorption as a selection tool for periodic signals like gamma point sources. Data stacking or summation in two different time systems, solar and sidereal, were made over the data. The idea behind this process is based on the random, poissonian, nature of the majority of the radiation measured by our detectors. When a non random and periodic signal exists, sufficient data can be summed in a scale in which this signal happened at the same time every day or month or week, thus increasing the signal-to-noise ratio. In particular, sidereal and solar time epoch stacking were performed, observing clear indications, both in phase and in amplitude, of solar modulation of the flux of galactic cosmic rays. These observations support the fact that, by using an adapted analysis technique to the characteristics of our small detectors, it is possible to observe space weather phenomena at different time scales from ground level in the LAGO network of WCD.

Simultaneous measurements of galactic cosmic rays flux modulation at different locations on Earth using the same type of detectors, can provide important information, e.g., about certain properties of the global structure of the magnetic clouds reaching the terrestrial environment during interplanetary Coronal Mass Ejection (iCME) pointing towards Earth (e.g. [12]). By using WCDs, we are able to determine the flux of secondary particles at different bands of deposited energy within the detector volume. These bands are dominated by different components of the CR reaching the Earth atmosphere (primaries). This is what we called the multi-spectral analysis technique (MSAT), and constitutes the basis of our Space Weather oriented data analysis [7]. Then, by combining all the data measured at different locations of our detection network, the LAGO project will provide very detailed and simultaneous information of the temporal evolution and of the small and large scales characteristics of the disturbances produced by different transient and long term space weather phenomena.

A complex and complete chain of simulations support this program [7, 16, 8]. For every LAGO site, the directional rigidity cutoff $R_c$ is determined for secular and altered geomagnetic field conditions, such as those produced during intense geomagnetic storms. After that, we determine the expected number of primaries by integrating the measured flux of all the hadronic cosmic rays with $1\,\mathrm{GeV} < E_p < 1\,\mathrm{PeV}$. A set of CORSIKA simulations is then used to determine the expected number of secondary particles at ground level. Only those secondaries originated on primaries with rigidities above local cut-offs are included. As an example, the effect of geomagnetic field corrections for the LAGO site of Bucaramanga, Colombia, is shown in the left panel of figure 3. The expected flux of secondaries without corrections is compared with the corresponding flux after considering geomagnetic effects, as a function of the secondary momentum. The peak observed at $\simeq 0.5\,\mathrm{GeV/c}$ corresponds to the neutron-dominated region. For this particular component, the decrease at ground level in geomagnetic secular conditions represents a diminution of $-36.6\%$. This could be an indication of the sensitivity of secondary neutron flux as a proxy of the changing conditions in the near-earth space environment. Finally, expected secondary particles are injected into a Geant4 based model of the LAGO WCD to obtain the signal flux at detector level [16].

The LAGO implementation of MSAT take advantage of the WCDs capability to measure deposited energy of secondary particles entering the detector. Charge pulse histogram has different features that can be correlated with electromagnetic secondaries flux, muon dominated flux and multiple-particles flux. Thus, with the LAGO MSAT we are able to determine the evolution of the flux observed at those three bands. As the transition points between these different regimes are characterized by changes in the histogram slopes, a fully automated algorithm searches for all these features in 1-hour calibration histograms. From this, we determine the integrated total flux at each band, with typical integration windows between 1 minute and 15 minutes. As an example, in the right panel of figure 3 the measurement of the Forbush decrease occurred on March 8th, 2012, is shown for the electromagnetic dominated band. These measurements evidence WCDs capabilities to extend present studies of space weather phenomena using low cost detectors from ground level.

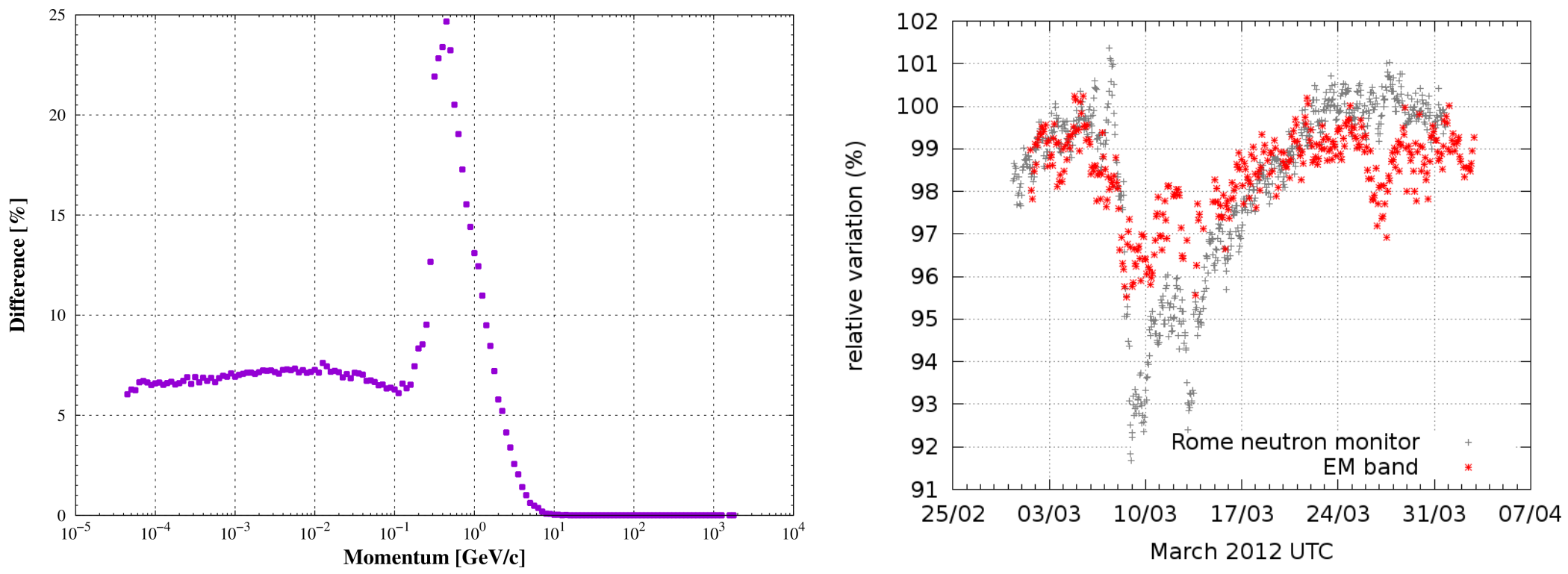

**Figure 3:** Left: The effect of geomagnetic corrections in the flux of secondary particles is shown as a function of their momentum, for the simulations performed to characterize the LAGO site of Bucaramanga, Colombia. Right: Multi-spectral analysis of the Forbush Decrease of March 8th, 2012, measured in a single 1.8 $m^2$ WCD in Bariloche, Argentina (red stars), compared with neutron flux measured by the Rome neutron monitor (gray pluses).

## 4. Conclusions

In this paper we give a basic sketch of the Latin American Giant Observatory, its main scientific capabilities and implications for the development of astroparticle physics at Latin America. We also described the key programs of LAGO, oriented to search for GRBs, high energy phenomena, and space weather, using low cost and reliable detectors at ground level. A complete chain of simulations and different data analysis techniques have been developed and presented, exploiting the WCDs capabilities to measure at different energy regimes. Several examples summarize here the LAGO capabilities to study these very different astrophysical phenomena, and, as new detectors are starting to operate (such as the new LAGO site in Antarctica) a large and unprecedented detection network emerges at regional scale.

Besides its scientific importance, the LAGO project is also a seeder for astroparticle physics development in the Andean region. A proof of this is the fact that most of the results presented at this conference have been obtained by several undergraduate and master students working in coordinated collaboration at several Latin American countries.

In summary, LAGO is a promising observatory that is helping to support the development of several interdisciplinary branches associated with Astroparticles in Latin America.

**Acknowledgments** The LAGO Collaboration is very thankful to all the participating institutions and to the Pierre Auger Collaboration for their continuous support.

# The Sites of the Latin American Giant Observatory

**Iván Sidelnik**[*1 †] **for the LAGO Collaboration**[2]

[1] *Laboratorio de Neutrones y Reactores, Centro Atómico Bariloche and Instituto Balseiro-UNCuyo*
*and Consejo Nacional de Investigaciones Científicas y Técnicas,*
*Av. Bustillo 9500, R8402AGP, S.C. de Bariloche, Río Negro, Argentina*
*E-mail:* sidelnik@cnea.gov.ar
[2] *lagoproject.org, see the full list of members and institutions at lagoproject.org/collab.html*
*e-mail:* asoreyh@cab.cnea.gov.ar

The Latin American Giant Observatory (LAGO) is an extended cosmic ray observatory, which consists on a wide network of water Cherenkov detectors located in nine different countries. The geographic distribution of the LAGO sites, with different altitudes and geomagnetic rigidity cut offs, combined with the new electronic system for control, atmospheric sensing and data acquisition on board of each detector, allow the realization of multiple and variated astrophysics, space physics and atmospheric physics studies at regional scale. In this work it will be described the LAGO sites and the capabilities of the LAGO detection network spanned across Latin America.

[*] Speaker.
[†] Presenter: Alberto Carramiñana

## 1. Introduction

LAGO (Latin American Giant Observatory) is a collaboration that comes from the association of Latin American astroparticle researchers [1]. It started in 2006 and it was originally designed to survey the high-energy component of Gamma Ray Bursts (GRBs) [2]. It is a network of ground-based water Cherenkov detectors (WCDs), located at different altitudes, between sea level and mount Chacaltaya at more than 5200 m a.s.l., with also a large range of geomagnetic rigidity cut-offs (RC). This collaboration was motivated by the experience of the Pierre Auger Observatory, and one of its primary objectives is to install WCDs along different sites in Latin America. At the moment LAGO is composed by ten countries forming a non-centralized and collaborative network of 25 institutions. The countries are: Argentina, Bolivia, Brazil, Colombia, Ecuador, Guatemala, México, Perú and Venezuela and the recent incorporation of institutions from Spain. 86 people integrate the LAGO community. It results very important to notice that this is a collaboration to search for space phenomena and understand its physics, working also as an integration network of science. LAGO operates as a connection for a lot of universities developing educational project to help physics students to deal with the construction of detectors, data acquisition and analysis, along to its physics interpretation.

Although the original plan comprised the detection of the high energy component of GRBs [3, 4], recently it has been shown that WCDs can also be used to study the Solar Modulation (SM) of Galactic Cosmic Rays (GCR) and other transient phenomena, by measuring the variations of the flux of secondary particles at ground level [5]. Measuring the flux modulation of the GCR at different locations on Earth that span over a wide range of rigidity cut offs and using the same type of detectors, can provide important information about the global structure of the magnetic cloud reaching the space environment surrounding the Earth. Astroparticle studies in the context of GRBs [6, 7, 8], space weather and background radiation at ground level [9] are the main scientific objectives of the Latin American Giant Observatory [1].

This works describes the new detector concept as an instrument that measures the physical quantities that are necessary to perform the analysis of low energy astroparticles at ground level. This is shown in section 2. In section 3 there is an overview of the different characteristics of the sites of LAGO in the Latin American region. Finally in section 4 we give some final remarks.

## 2. The detector

The WCDs of LAGO are composed by a water container with a reflective material inside, and a photomultiplier tube (PMT) to convert the Cherenkov radiation produced by the interaction of the high energy particles with water in to an electric signal. The response of each WCD, like other particle detectors, are short electric pulses ($\sim 200\,$ns) that are recorded by high speed (240 MHz) data acquisition systems. A typical WCD of one of the LAGO sites is shown in the right side of Figure 1.

This respond to the concept of a detector working as an isolated station that transmit data to a centralized place. The new idea that is showed in section 2.1 develop a concept of a multi use station.

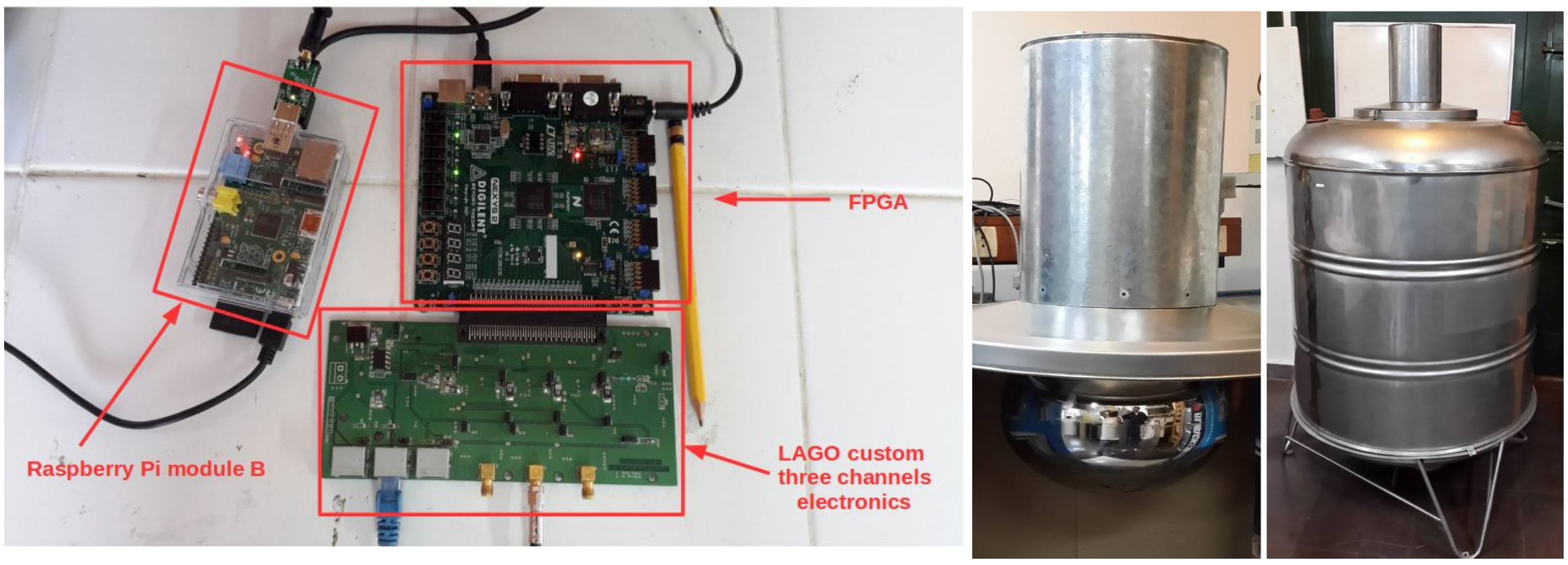

**Figure 1:** LAGO electronics: It can be seen the raspberry pi model B, the board with the Nexys2 FPGA and the custom three channels digitazer board.

Regarding the WCD, the PMT is connected to a electronic board that have a voltage divider, a high voltage supply, and an amplifier of the output pulse. The high voltage supply can be controlled typically from 0 V to 2000 V with a signal from 0 V to 2.5 V. The output pulse shape of the PMT depends on the geometry of the detector, for a tank of 1000 l the FWHM[1] of the pulse is around 150 ns. The data acquisition and high voltage parameters of the experiment can be settled by a PC program and the acquire pulses are transfered in real time to the PC or SBC via a USB interface. This system has the capability of acquire three or four independent channels and provide the control signal for the three or four HV supply.

The data acquisition system includes a custom electronic board that integrate the front-end electronics: three ADC, anti-aliasing filters, baseline control control circuit, high voltage control signal, and the necessary service circuits. This board is connected to a NEXYS-2 board via a high speed FX2-Hirose connector. The NEXYS2 have a FPGA Spartan3E that is connected to the PC with a USB interface. The USB interface between the PC and the FPGA is managed with a Cypress chip included in the board. In the FPGA there is an implementation of a module of command, that provided connectivity with the sensors the GPS and with a pre-processor that filter the input pulses. The pulses that have an amplitude over a trigger level are transferred to the PC.

Recent developments of Single Board Computers (SBC) as the Raspberry Pi and Arduino, allow us to made an integral and autonomous station, with low power consumption that has the capability of perform analysis, encoding, and data transfer. The Raspberry Pi board that controls the station is a model B, showed in Figure 1, left. Also, it is shown the whole LAGO electronic that can be seen with the different components. The NEXYS-2 FPGA board is showed with the digitalized board custom made with three or four available channels, depending the number of detectors available.

Since the LAGO data analysis needs to take into account the influence of atmospheric effects, such as pressure or air temperature, each LAGO WCD is equipped with several environmental sensors. These measurements represent an opportunity to provide environmental information to other communities, like for example, ecologists studying the high altitude environments to correlate

[1] Full Width at Half Maximum

for possible climate change and global warming effects.

### 2.1 New detector concept

The need to measure different groups of physical magnitudes in remote and difficult to access places, or with almost no infrastructure, requires the implementation of autonomous and reliable devices. This kind of apparatus might be able to handle the access to different peripheral sensors that should be controlled independently and may need to manage different time stamps for the incoming data.

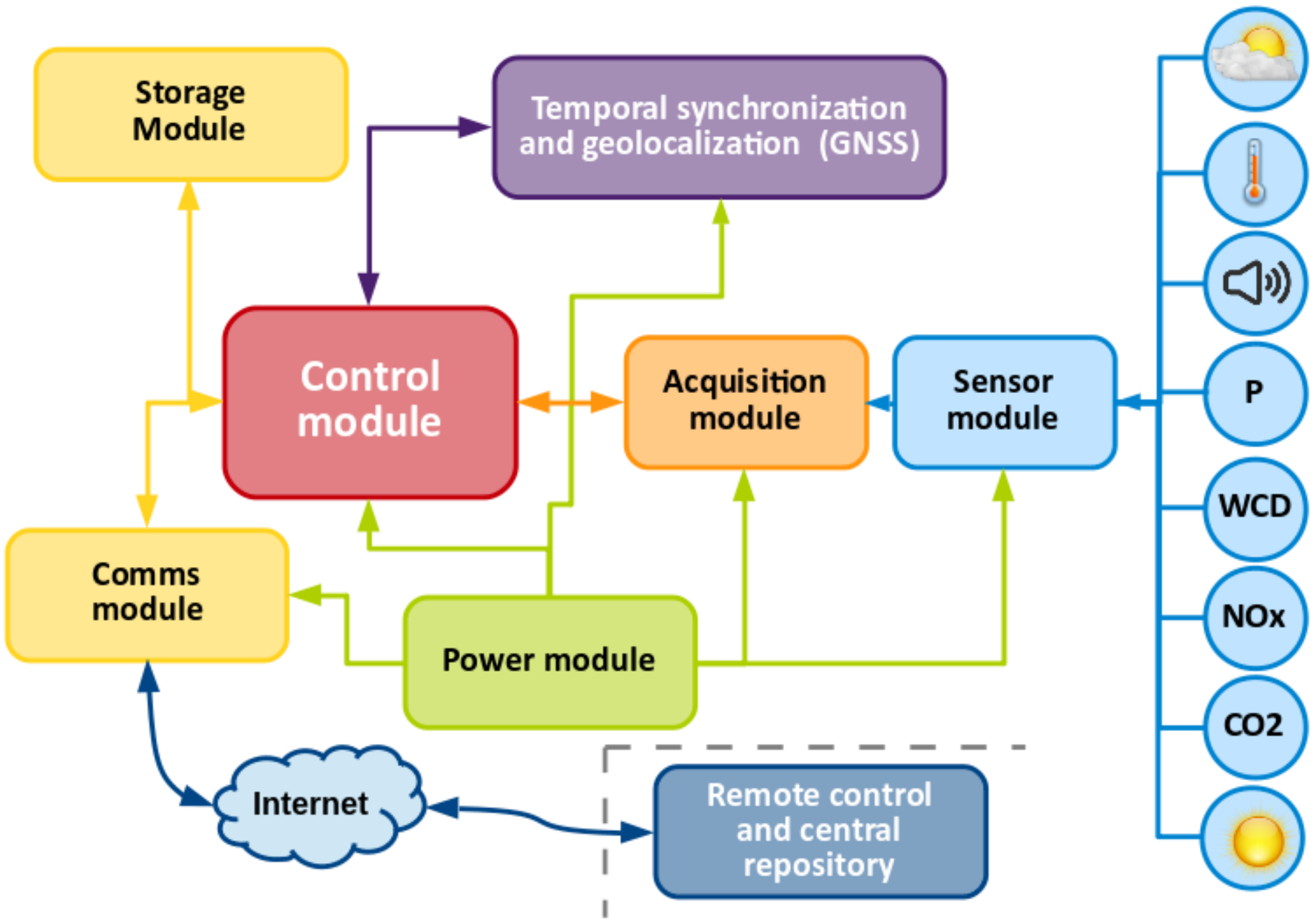

**Figure 2:** Block diagram of the autonomous station for the new LAGO detector concept. The constituents of the station are highlighted: the data acquired by the different sensors that are in the peripheral are sent to the sensor module and from there to the acquisition module for its digitalization or channeling trough the $I^2C$ bus. The digitalized data are resented to the control module for temporal identification, storage and transmission to a remote repository.

In Figure 2 it is shown a diagram of the autonomous and remote measurement station. This new concept for the LAGO detector shows how the instrument that measures the particles (the WCD) can be treated as an peripheral element. This autonomous measurement is built from the integration of different independent modules that adds different functionalities to the station.

The modules that composes a LAGO station are listed below:

**Sensor module:** built to support the different sensors. These sensors are integrated in to a board Arduino-Shield of customary design. We use commercial, low cost and low consumption sensors.

**Acquisition and digitization module:** data acquisition and digitization is performed in a board based in the Arduino are microcontrollerATmega328. The data bus allows the reading and digitization of up to six analogous sensors, fourteen digital sensors and a communication bus based on the $I^2C$ standard.

**Control, encoding, storage and transport module:** This module is supported by and SBC Raspberry Pi B/B+ and recently Pi 2 All station operations are controlled within this module. The digitized data by the acquisition module are conformed, classified and stored following a priori established parameters. The flexibility that gives the operative system of the SBC, GNU/Linux based, gives the chance to perform some on-line local analysis of the data. In case of need, it also gives the possibility of remotely access, through a communication module. Data is storage using SSD solid state devices of 1 TB, that guarantees low power consumption and with that the capability of operation at remote places.

**Power supply module:** Low power consumption has been a mandatory design parameter to take in to account for the assemble of an autonomous station. With a maximum consumption of 10 W this detector can be operate in remote places using solar panels and batteries.

**Geolocalization module:** This module allows to incorporate *Global Navigation Satellite System* (GNSS) devices compatible with the NAVSTAR-GPS/GLONASS/GALILEO system. This gives the chance to determine the position of the station and temporal synchronization of the signal with a resolution of $\lesssim 20$ ns [10]. The selected devices will allow also to perform total electron content (TEC) analysis [11].

**Communication module:** The design of this module depends on the communication network available at the place of operation of the detector. Currently there are different sub modules available that can stablish links at different velocities through the standards: IEEE 802.3 (ethernet), IEEE 802.11 (WiFi), IEEE 802.16 (WiMax), IEEE 802.15.1 (Bluetooth), GPRS (2.5G and 3G), 4G-LTE.

## 3. The sites of the Latin American Giant Observatory

LAGO spans over many different countries in Latin America. In this section we describe the sites according to the location in each of the countries that have or will have a LAGO detector. In Fig.3, left, we show a map of the Latin American region from Mexico to the Antarctic. In this map the blue triangles shows the sites that have at least one operational detector, and in red squares those who are to be installed in the 2015-2016 period.

The plot in Figure 3, right, shows the rigidity cut offs, in GV, for the latitudes connected with the map on the left [9, 12]. It can be appreciated the spanning around similar geographical longitudes but different values of rigidities that can be characterized by the LAGO network. Then, by combining all the data measured at several locations of our detection network and analyzed by using the multi-spectral technique[9], the LAGO project will provide detailed and simultaneous information at different energy ranges of the disturbances produced by transients and long term space weather phenomena.

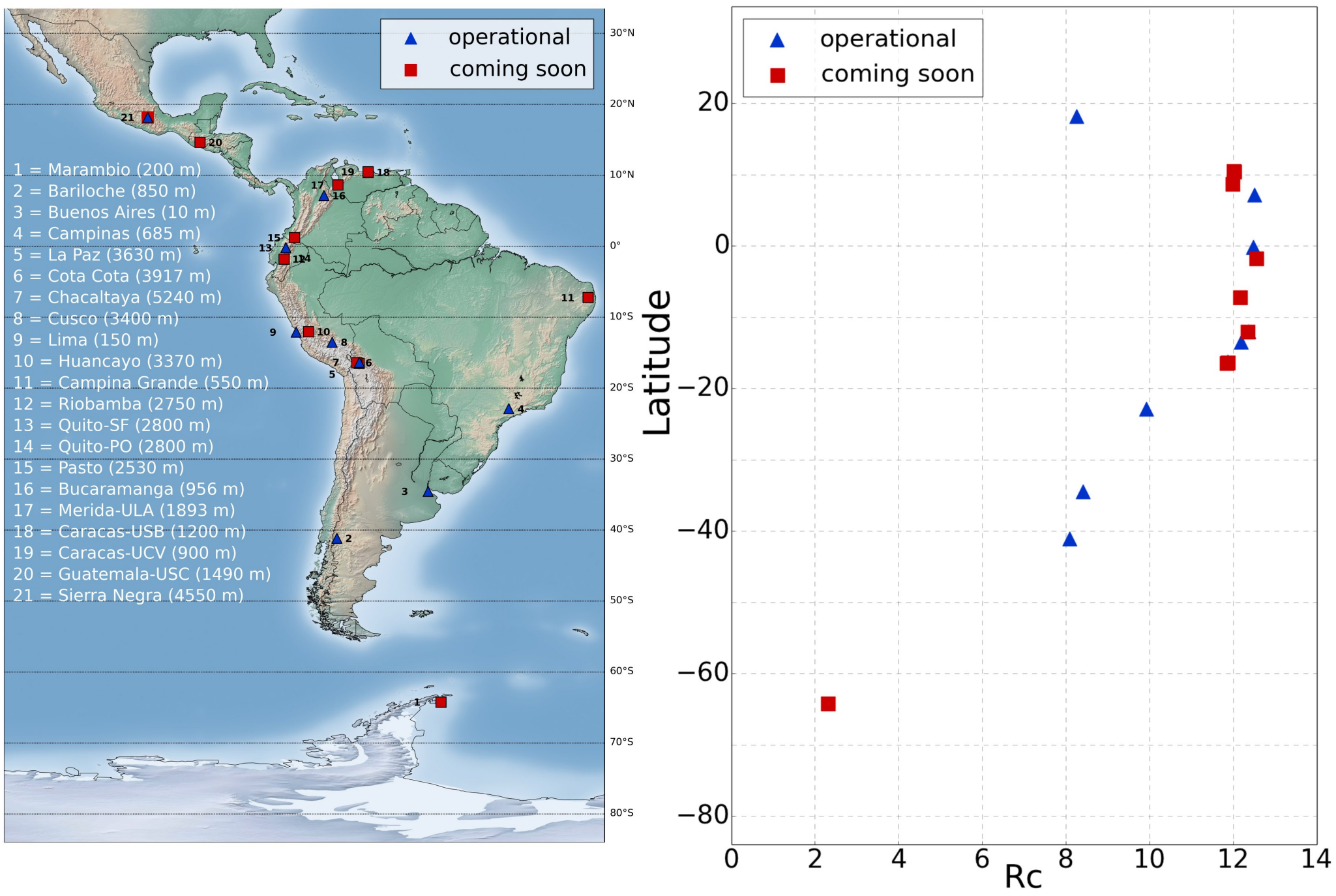

**Figure 3:** Left: map showing all LAGO sites in the Latin American region. The markers shows if the detector is operation, blue triangles, or if it is going to start measurements in the period 2015-2016, red squares. Right: rigidity cut off for the different LAGO sites, it can be seen in concordance with the map on the left. The different rigidities cut off shows the important of having a detection network like LAGO[9].

A compilation of the locations, number of detectors, altitudes, rigidity cut off, latitude and longitude, is made in table 1. It is worth mention some highlights of this table related to the LAGO science capabilities.

## 4. Conclusion and final remarks

In this paper we have shown a new concept developed by the LAGO Collaboration of an integral detector that includes the WCD working with an environmental station that has the capability of provides data for others communities.

It is also a highlight of the LAGO locations the very important difference in altitude sites across the Andes mountain range. This allow us to perform analyses of background radiation, space weather and low energy astroparticles.

The ultra long baseline "array" of WCD that goes from México to Antarctica bring different rigidity cut offs but with similar geographical longitudes

| Country | Site | Number of WCD | Altitude [m a.s.l.] | Rigidity cut off [GV] | Latitude [deg] | Longitude [deg] | Observations |
|---|---|---|---|---|---|---|---|
| Argentina | Bariloche | 3 | 850 | 8.10 | 41.15 S | 71.30 W | |
| | Buenos Aires | 1 | 10 | 8.41 | 34.54 S | 58.44 W | |
| | Marambio[13] | 1 | 200 | 2.31 | 64.24 S | 56.62 W | RC lowest value (south) |
| Bolivia | Chacaltaya | 3 | 5240 | 11.87 | 16.35 S | 68.13 W | Highest place |
| | Cota Cota | 1 | 3917 | 11.88 | 16.41 S | 68.5 W | |
| Brazil | Campinas | 1 | 685 | 9.93 | 22.9 S | 47.06 W | |
| | Campina Grande | 1 | 550 | 12.17 | 34.54 S | 58.44 W | Eastern place |
| Colombia | Bucaramanga | 1 | 956 | 12.52 | 7.14 S | 73.12 W | |
| Ecuador [14] | Riobamba | 1 | 2750 | 12.55 | 1.81 S | 78.74 W | RC highest value |
| | Quito-SF | 1 | 2800 | 12.49 | 0.2 S | 78.5 W | |
| | Quito-PO | 1 | 2800 | 12.49 | 0.2 S | 78.5 W | |
| Guatemala | Guatemala | 1 | 1490 | 10.32 | 14.63 N | 90.59 W | |
| México [7] | Sierra Negra | 4 | 4550 | 8.26 | 18.16 N | 97.95 W | RC lowest value (north) and western |
| Perú [8] | Lima | 2 | 150 | 12.37 | 12.10 S | 77.02 W | Large difference in altitude at same RC |
| | Cusco | 1 | 3400 | 12.19 | 13.52 S | 71.96 W | |
| | Huancayo | 1 | 3370 | 12.34 | 12.04 S | 75.30 W | |
| Venezuela | Caracas-UCV | 1 | 900 | 12.01 | 10.49 N | 66.89 W | |
| | Caracas-USB | 1 | 1200 | 12.02 | 10.41 N | 66.88 W | |
| | Mérida-ULA | 1 | 1893 | 11.99 | 8.63 N | 71.15 W | |

**Table 1:** Table of the different locations of the LAGO sites.

The LAGO Project is being built as an extended astroparticle Observatory at a regional scale, operated by a collaborative and non-centralized network of institutions of nine Latin American countries and Spain.

### Acknowledgements

The LAGO Collaboration is very thankful to the Pierre Auger Collaboration for its continuous support. Iván Sidelnik also wish to thank CONICET, CNEA and Instituto Balseiro.

# 2

# Data Science

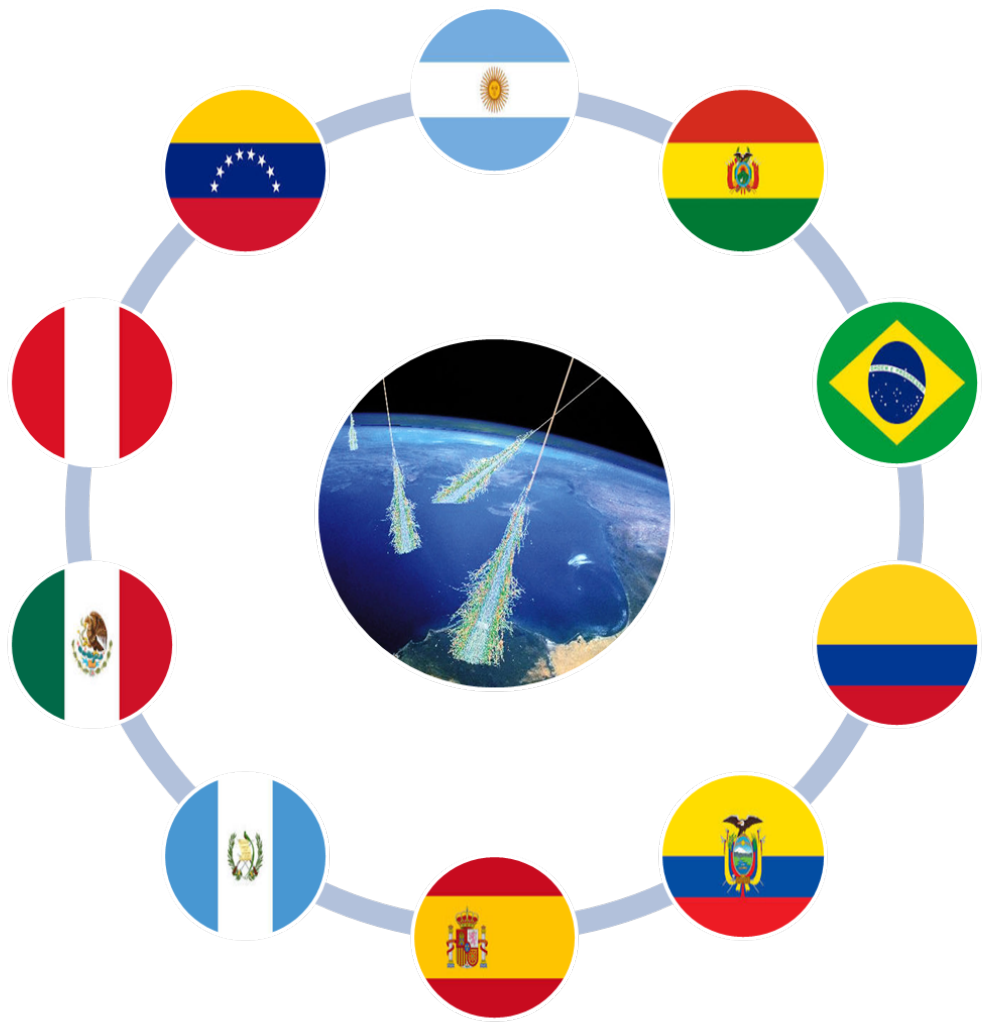

# Data Accessibility, Reproducibility and Trustworthiness with LAGO Data Repository

**H. Asorey[1,2,3], D. Cazar-Ramírez*[4], R. Mayo-García[5], L.A. Núñez[3,6], M. Rodríguez-Pascual[5], L.A. Torres-Niño[7,8], for the LAGO Collaboration[9]**

[1] *Laboratorio Detección de Partículas y Radiación, Instituto Balseiro y Centro Atómico Bariloche, S.C. de Bariloche, Argentina.*
[2] *Sede Andina, Universidad Nacional de Río Negro, S.C. de Bariloche, Argentina.*
[3] *Escuela de Física, Universidad Industrial de Santander, Bucaramanga, Colombia.*
[4] *Colegio de Ciencias e Ingenierias, Universidad San Francisco de Quito, Cumbayá, Ecuador.*
[5] *Centro de Investigaciones Energéticas, Medioambientales y Tecnológicas, Madrid, Spain.*
[6] *Centro de Física Fundamental, Dept. de Física, Universidad de Los Andes, Mérida Venezuela*
[7] *Escuela de Ingeniería en Sistemas, Universidad Industrial de Santander, Bucaramanga, Colombia*
[8] *Centro de Supercomputación y Cálculo Científico, Universidad Industrial de Santander, Bucaramanga, Colombia*
[9] *lagoproject.org, see the full list of members and institutions at lagoproject.org/collab.html*
*e-mail:* `lago-pi@lagoproject.org`

Nowadays, one of the most challenging scenarios scientists and scientific communities are facing is the huge amount of data emerging from vast networks of sensors and from computational simulations performed in a diversity of computing architectures and e-infrastructures. In this work we present the strategy of the Latin American Giant Observatory (LAGO) to catalog and preserve a vast amount of data produced by the water-Cherenkov Detector network and the complete LAGO simulation workflow that characterize each site. Metadata, Permanent Identifiers and the facilities from the LAGO Data Repository are described. These initiatives allow researchers to find data and directly use them in a code running by means of a Science Gateway that provides access to different clusters, Grid and Cloud infrastructures worldwide.

*Speaker.

## 1. Introduction

Data-intensive scientific analysis is a completely new way of doing science. How to deal with large datasets is still in evolution and has a long way to go in almost all disciplines -physical, life sciences and humanities- which are becoming increasingly data-driven and data intensive.

An increasing computing power and a deluge of connected sensor devises -now ready accessible for research communities- have driven a dramatic improved in our ability to collect, preserve and share complex multi-dimensional measured and simulated data. This emerging panorama has set new challenges around the complex discovery environments in this bursting Data Centered Science where, two concepts are identified as main key requirements for any contemporary scientific finding: *trustworthiness* and *reproducibility*.

Trustworthiness can be associated with the description and traceability of scientific protocols[1] involved in any research, and is the base of the confidence on the answers results obtained in any scientific inquiry. We trust a particular result from a researcher or a research group because their meticulousness adherent to a detail and rigorous protocol supporting the finding.

Reproducibility of a scientific discovery is a very debatable concept which ranges from strict "replication" -as in re-running a simulation in exact detail- through "reproduction" in the sense of independent reproduction of the essential aspects of the experiment [1, 2].

In this paper we shall describe the ecosystem of data tools and services and how they are implemented to help solving the Data Accessibility, Reproducibility and Trustworthiness (DART) challenge in the LAGO (for Latin American Giant Observatory formerly known as Large Aperture Gamma Ray Observatory) Project.

It is organized as follows: the coming section presents the DART Challenge; next, in section 3 the LAGO Project is introduced; LAGOData e-infrastructure and LAGO computational resources/services are discussed in section 4 and 5, respectively; we close with some remarks in 6.

## 2. The Dart challenge

The DART initiative was launched by CHAIN-REDS[2] (Coordination and Harmonisation of Advanced e-infrastructure for Research and Education Data Sharing): an European Commission co-project focused on promoting and supporting technological and scientific collaboration across different communities in various continents[3, 4]. This initiative provided a set of interrelated tools and services, based on worldwide adopted standards, to provide easy/seamless access datasets, data/documents repositories and the applications that could generate and/or make use of them.

*Trustworthiness* can be associated to data curation, particularly on the quality of the metadata describing the experimental protocol and data provenance[5], while *reproducibility* and *replicability* are closely connected to the accessibility to data sources and the possibility to manipulate/analyze data contained in them.

[1] We are considering as a scientific *protocol* the detailed specifications for carrying out an experiment, either the detector setup & configuration which condition the registered data or simulation environments that lead to synthetic datasets coming from computational applications[1].

[2] http://www.chain-project.eu

CHAIN-REDS approach to data trustworthiness and reproducibility is based on the integration of computational resources and services, with three main cornerstones:

1. adoption of standards for data discoverability, provenance and recoverability: OAI-PMHfor metadata retrieval, Dublin Core as metadata schema, SPARQL for semantic web search and XML as potential standard for the interchange of data;

2. enablement of datasets authorships and user authentication with the corresponding assignments of specific roles on data services, which can be implemented by two strategies: assignment of Persistent Identifiers (PIDs)[6] to name data in a unique and timeless manner, ensuring that future changes on URIs or internal organization of databases will be transparent to the user; and implementation of a federated identity provision, a secure, flexible and portable mechanism to access e-infrastructures worldwide, based on agreements and standards[3].

3. access to a plethora of computing power to analyze the retrieved data or to contrast them to simulations through an intuitive web-interface. Over the last years, Science Gateways[7, 8] have risen as an ideal tool to allow scientists across the world to seamlessly access different ICT-based infrastructures for research activities to support their day-by-day work and do better (and faster) research.

In other words: user identification, execution of distributed applications with a simple web interface, usage of Open Access Document Repositories/Data Repositories and reproducibility of the experiments conform the DART challenge, where the whole research cycle is covered and can be seamlessly performed by non-expert users, hiding complex processes under simple interfaces and minimizing the need for learning new tools.

## 3. The LAGO Collaboration

The LAGO project is a collaboration [9] of more than eighty Ibero American astroparticle researchers, motivated by the experience of the Pierre Auger Observatory [10] in Argentina and devoted to study space weather effects [11] and Gamma Ray Bursts (GRB) signals on ground-based detectors [12]. Long-term modulation and transient events can also be characterized by using the LAGO detection network, as it spans over a big area with different sites at different latitudes, longitudes and geomagnetic rigidity cut-offs.

Today it is a network of 10 ground-based Water Cherenkov Detector (WCDs) [13], located at different altitudes from Mexico through Patagonia, and other detectors are expected to be up and running by 2016 including two WCD in the Antarctica Peninsula[14]. These data are of interest for two different scientific communities:

- Gamma Astronomy Community: The LAGO WCDs installed at high altitude sites are sensitive to detect the effects of GRB. A significant number of LAGO detectors are over 3,000 m.a.s.l. and three of them are above 4,300 m.

- Space Weather: The LAGO energy range covers a plethora of phenomena related to low-energy cosmic rays physics and space weather phenomena. Nowadays, the study of such phenomena is crucial because levels of radiation in the atmosphere and near-Earth space environment may be established.

Since the LAGO data analysis needs to take into account the influence of atmospheric effects -such as pressure or even air temperature- on the flux of particles at the detector level, each LAGO WCD is equipped with several environmental sensors. These measurements represent an opportunity to provide environmental information to other communities, like ecologists studying the high altitude environments to correlate for possible climate change and global warming effects. Additionally, since the data in the LAGO repositories is open and freely accessible, it is used as motivation to train the general public -mainly the secondary school teachers and students- in statistical data analysis and related techniques, and raising the awareness of the general public in global warming and climate change impact in everyday life. This important citizen science initiative is one of the main objectives of the LAGO collaboration and is implemented in the so-called LAGO-CS (Citizen Science) program[15].

## 4. LAGOData e-infrastructure

Typically, each detector generates 150 GB of data per month and the entire collaboration generates 1.5 TB/month. The LAGO dataset not only refers to those measured by WCD detectors but also to data generated by simulation of cosmic rays phenomena in the $\sim 2\,\mathrm{GeV}$ to $\sim 1\,\mathrm{PeV}$ energy range, by using CORSIKA (COsmic Ray SImulations for KAscade)[16], a software for detailed simulation of extensive air showers initiated by high energy cosmic ray particles. The CORSIKA particle flux simulations carried out generate 50GB/site and these synthetic data are also preserved in the data repository. The low energy limit depends on the geomagnetic coordinates of the site, while the high energy limit is determined by the collection area at each site and is limited by statistics as the flux becomes lower and lower at higher energies.

This raw data collected by the LAGO detectors and produced by the simulation tools developed are shared trough LAGOData[17], a platform conceived to promote data curation and sharing among LAGO collaborators, which is part of a more ambitious project, LAGOVirtual[18] a working environment which ensure access to the data recorded in all LAGO Sites and facilitate the analysis of such data.

### 4.1 Data curation through DSpace

Dspace is an open source software that enables sharing of many types of content, it is generally used for institutional repositories, providing basic functionality for saving, storing, and retrieving of digital content. DSpace was adopted for the LagoDATA repository, because it hosts Dublin Core metadata with a straightforward adaptability for non-native metadata schemes. It also supports two important interoperability protocols: OAI-PMH (Open Archive Initiatives Protocol for Metadata Harvesting[3] ) and SWORD (Simple Webservice Offering Repository Deposit[4]). The OAI-PMH protocol at the LAGO repository allows the CHAIN-REDS Knowledge Base search engine to navigate into LAGO curated data.

[3] http://www.openarchives.org/pmh/
[4] http://swordapp.org/

We have overcome one of the most important DSpace limitations: its inability to upload/download multiple records. Dspace offers the possibility to upload the corresponding metadata through a command line option via an *import tool* by using *simple archive format* and including it in a separate way. We have developed a script to ingest data profiting from the above mentioned DSpace capability and found that, the assignment of the PID ID to each dataset become a significant overload to the ingest process and we have solved this difficulty by lowering the number ($\approx 10$) of datasets to be ingested in each group.

### 4.2 LAGOData metadata

The Dublin Core metadata element set is a standard for cross-domain information resource description, it is elaborated and sponsored by DCMI (Dublin Core Metadata Initiative ,the implementation of which makes use of XML. Dublin Core is Resource Description Framework based and comprises fifteen metadata elements: Title, Subject, Description, Source, Language, Relation, Coverage, Creator, Publisher, Contributor, Rights, Date, Type, Format, and Identifier. Despite this functionality is mostly centered on the Dublin Core metadata scheme, the additional non-native metadata can be configured as custom fields which are also stored, searched and displayed as the native ones.

The datasets are classified into two different types: simulated and measured data, with their corresponding associated metadata:

**measured metadata** scheme: **data** corresponds to the version/type of the Digital/Analog electronic board; **site** contains the *name*, *latitud*, *longitude*, *altitude*, *type*, *geometry* of the detector and *orcid-id* of the person in charge of the site operations; **voltage** indicating the voltage of polarization of each one of the independent PMTs controlled by the electronic board; **trigger** with the level of trigger and subtrigger of each signal channel; and **sensor** describing type of sensor and calibration constant for each other physical variables measured on board the detector.

**simulated metadata** scheme: **primary**, corresponding to the CORSIKA `.dbase` output file; **site** with the *latitude*, *longitude* and *altitude* of the site; **libraries** indicating the compiling CORSIKA options and libraries, and the version of the LAGO scripts used for this particular calculation; and **computation**, describing the computational environment given by the standard unix commands: `uname -a`, `lsb_release -a`, `free` and `gcc -v`.

### 4.3 PID and LAGOData

The main interface to register and manage PID services for European Research Communities is EPIC (European Persistent Identifiers Consortium[5]) which is based on the Handle System [6] for the allocation and resolution of persistent identifiers. There are several compatible "flavors" of

[5] http://pidconsortium.eu

[6] The Handle System (http://www.ietf.org/rfc/rfc3650.txt) provides an efficient, extensible, and secure resolution mechanism for unique and persistent identifiers of digital objects. The Handle System includes an open set of protocols (http://hdl.handle.net/4263537/4086), a namespace (http://hdl.handle.net/4263537/4068) and a reference implementation (http://handle.net/download.html) of the protocols.

PIDs. The most common is DOI PID[7], which is the most frequently used for publications while the EPIC PIDs cover a wider range of Digital objects.

The GRNET PID service enables the allocation, management and resolution of PIDs and has been employed to ensure the data persistence and reproducibility of the experiments. It supports the use of part identifiers as they are provided by the Handle system. Part identifiers can compute an unlimited number of handles on the fly, without requiring registering each separately.

### 4.4 SWORD and LAGOData

The SWORD (Simple Web-service Offering Repository Deposit), based on the Atom Publishing Protocol (AtomPub), was first developed to standardize a deposit interface to digital repositories but now, it has been further extended to support the whole deposit lifecycle, i.e. deposit, update, replace, and delete resources [19]. Many interfaces of laboratory equipment allow automatic capture of results in an information system and SWORD permits to upload data directly into a repository, without human intervention, tagging as metadata the conditions in which the data was collected.

A Java SWORD Client has been developed -and is part of the firmware of the WCD electronics- to send each recorded dataset to the DSpace Repository, stamping the basic set of metadada describing the status of the WCD at this moment, i,e. data, site, voltage, TrigLevel, and sensor.

## 5. LAGO advanced computing e-infrastructure

In addition to CORSIKA, LAGO uses intensively other important codes: MAGNETOCOSMIC, GEANT4, ROOT and specific self-designed statistical codes for data analysis, focusing most of the collaboration activities (research & outreach) on a data repository.

Thanks to the cooperation of CHAIN REDS, the advanced computational e-infrastructure for the LAGO project was implemented, creating a fully dedicated Virtual Organization called *lago-project* and integrating it into the European Grid Infrastructure (EGI) activities. The Grid implementation of CORSIKA was deployed in two "flavors" being able to run by using GridWay Metascheduler [20] or through a Catania Science Gateway interface[8]. We have crosschecked the performance of both procedures and found that GridWay approach seams to be more appropriated for massive systematic running while CORSIKA Science Gateway implementations is useful for quick exploratory or specific studies./

In the Science Gateway approach a user can seamlessly run a code on different infrastructures by accessing a unique entry point with an identity provision. He/she only has to upload the input data or use a PID to reference it and click on the run icon. The final result will be retrieved whenever the job will be ended. The underlying infrastructure is absolutely transparent to the user and the system decides on which sites and computing platform the code is performed. More than 21,000 CPU hours have been consumed from Oct 2014 by CORSIKA-GRID jobs executed via the this last interface.

[7]The Digital Object Identifier (DOI) System (http://www.doi.org/) is a service, operated by the International DOI Foundation (IDF), which provides a technical and social infrastructure for the registration and use of persistent interoperable identifiers on digital networks.

## 6. Conclusions

LAGO is an experiment that could handle, with reasonable scale, a distributed community, collaborating across Latin America, building a network of data repositories through the continent, using computational intensive applications and developing an outreach program to promote Data Science as a Citizen Science initiative.

LAGO aims to study cosmic rays in the energy range $\sim 1$ GeV to $\sim 100$ TeV. In this energy range there emerge phenomena related to the physics of low-energy cosmic rays, and also to solar activity and space weather. Nowadays it is crucial to study of these effects because it may establish levels of radiation in the atmosphere and near-Earth space environment. Thus the data repository (and the network of data repositories) will be of interest not only for the LAGO or even the cosmic ray community but useful for the solar physics and space climatology communities. It is then expected that its adoption of the DART Challenge will have important impact, as it will extend the use of LAGO resources (computational and data) for a much wider community.

As has been demonstrated, the implementation of DART applications and methodology has lead to a clear improvement in LAGO robustness, usability and visibility, leading to a wide adoption of the new tools/services by the final research users.

## Acknowledgments

The LAGO Collaboration is very thankful to the Pierre Auger Collaboration for its continuous support. We are also in debt to the EC co-funded project CHAIN-REDS (GA 306819) and to the GRNet Technical staff for the implementation of EPIC PID. One of us (LAN) gratefully acknowledges the financial support from CDCHT-ULA project C-1598-08-05-A and Vicerrectoría Investigación y Extensión Universidad Industrial de Santander. This work was also partially supported by Universidad San Francisco de Quito and CEDIA through CEPRA IX research grants.

# 3

# Space Weather

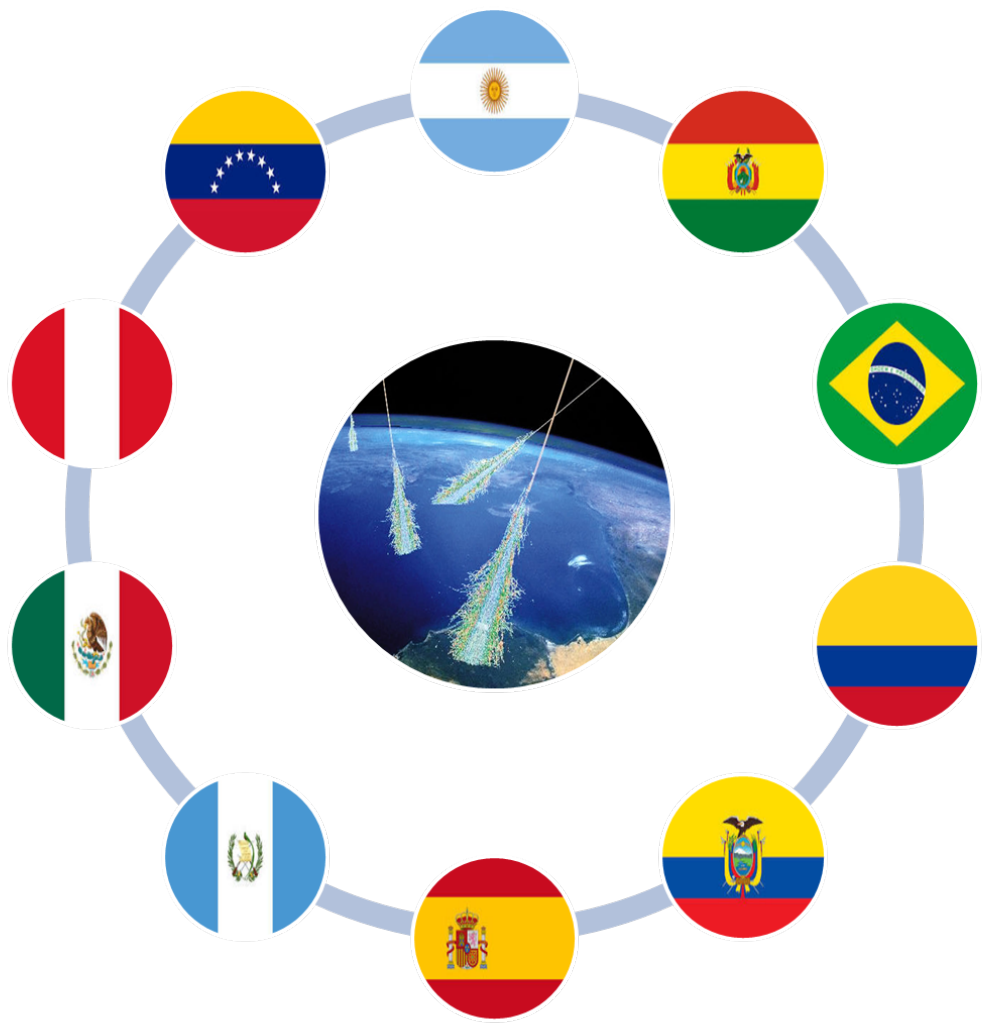

# The LAGO Space Weather Program: Directional Geomagnetic Effects, Background Fluence Calculations and Multi-Spectral Data Analysis

**H. Asorey**[1,2,3]**, S. Dasso**[4,5]**, L.A. Núñez**[3,6]**, Y. Peréz**[6]**, C. Sarmiento-Cano**[∗3]**, M. Suárez-Durán**[†3]**, for The LAGO Collaboration**[7]

[1] *Laboratorio Detección de Partículas y Radiación, Instituto Balseiro y Centro Atómico Bariloche, S.C. de Bariloche, Argentina.*
[2] *Sede Andina, Universidad Nacional de Río Negro, S.C. de Bariloche, Argentina.*
[3] *Escuela de Física, Universidad Industrial de Santander, Bucaramanga, Colombia.*
[4] *Instituto de Astronomía y Física del Espacio (UBA-CONICET), Buenos Aires, Argentina.*
[5] *Dept. de Ciencias de la Atmósfera y los Océanos and Dept. de Física, Universidad de Buenos Aires, Buenos Aires, Argentina.*
[6] *Centro de Física Fundamental, Dept. de Física, Universidad de Los Andes, Mérida Venezuela*
[7] *lagoproject.org, see the full list of members and institutions at lagoproject.org/collab.html*
*e-mail:* `asoreyh@cab.cnea.gov.ar`

The Latin American Giant Observatory (LAGO) is an extended cosmic ray observatory operating in nine Latin American countries. Within the LAGO framework, several scientific and academic programs are being developed and conducted. One of them, the LAGO Space Weather program, aims at producing real time, high resolution and high quality data of the flux of secondary particles at each site of the LAGO detection network, complementing and expanding present measurements of the influence of solar activity from ground level observations. With the LAGO detection network, transient events or long term modulation effects can be measured simultaneously at places having different geomagnetic rigidity cut-offs and atmospheric reaction levels. As the data acquisition system of LAGO registers the individual energy of all secondaries detected, these transients can furthermore be studied in different energy bands. This program is intensively supported by a complex chain of simulations that accounts for geomagnetic and atmospheric effects and the different detectors response. In this work, the complete simulation chain is described, and the first multi-spectral analysis of the Forbush decrease of March of 2012 is presented.

∗Presenter
†Speaker.

## 1. Introduction

Transient solar ejecta and changes in the global structure of the heliospheric and terrestrial magnetic fields modulate the flux of low energy cosmic rays of galactic origin (GCR). Measure of the temporal variations of this flux provide information about the transport of particles in the inner and outer heliosphere.

Long-term modulations in the flux of GCR are associated with the solar cycle, while short-term are produced by interplanetary transient perturbations in the calm solar wind causing phenomena like Forbush Decreases [1]. Observed from ground level, FDs exhibit asymmetrical structures: fast decreases (several hours) of the cosmic ray flux are followed by smooth recoveries (several days). In some particular cases, they can also present more complex structures originated by intricate interplanetary Coronal Mass Ejections (iCME) phenomena such as its interactions with fast streams of plasma or with other iCMEs.

Simultaneous measurements of the GCR flux modulation at different locations on Earth using the same type of detectors, can provide important information about the global structure of the magnetic cloud reaching the near-Earth space environment. Astroparticles studies in the context of GRBs, Space Weather and background radiation at ground level are the main scientific objectives of the Latin American Giant Observatory (LAGO, previously known as Large Aperture GRB Observatory)[2].

The LAGO Project is being built as an extended Astroparticle Observatory at a regional scale, operated by a collaborative and non-centralized network of institutions from nine Latin American countries (Argentina, Bolivia, Brazil, Colombia, Ecuador, Guatemala, Mexico, Peru and Venezuela) and recent incorporation of institutions from Spain.

The LAGO detection network currently consists of single or small arrays of water Cherenkov detectors (WCD) located at different altitude sites and covering a large range of geomagnetic rigidity cut-offs ($R_c$) and atmospherics reaction levels [3]. Current status of the LAGO project and information about the sites characteristics are shown in figure 1, covering from Mexico [4] to Antarctic [5].

This paper is organized as follows: in the coming section we describe the LAGO Space Weather (LAGO-SW) program; then geomagnetic and atmospheric simulations are detailed in section 3, showing a procedure to evaluate how the disturbances in the geomagnetic field (GF) can be linked with variations in the flux of secondary particles at ground level. Then the first steps of the multi-spectral analysis are presented by the analysis of a FD observed on March 8th 2012 in a single WCD located in Bariloche, Argentina; and finally, we close this paper with some concluding remarks and future perspectives in section 5.

## 2. The LAGO Space Weather Program

The LAGO project is organized around scientific and academic objectives [2] that are carried out within specific LAGO programs. The LAGO Space Weather (LAGO-SW) program is oriented toward studying how the variations of the flux of secondary particles at ground level can be linked with the heliospheric and geomagnetic modulation of GCRs, providing precise information of the changing conditions of the near-Earth space environment. These secondary particles are produced during the interaction of a GCR with the atmosphere and are measured at ground level by the LAGO detection network.

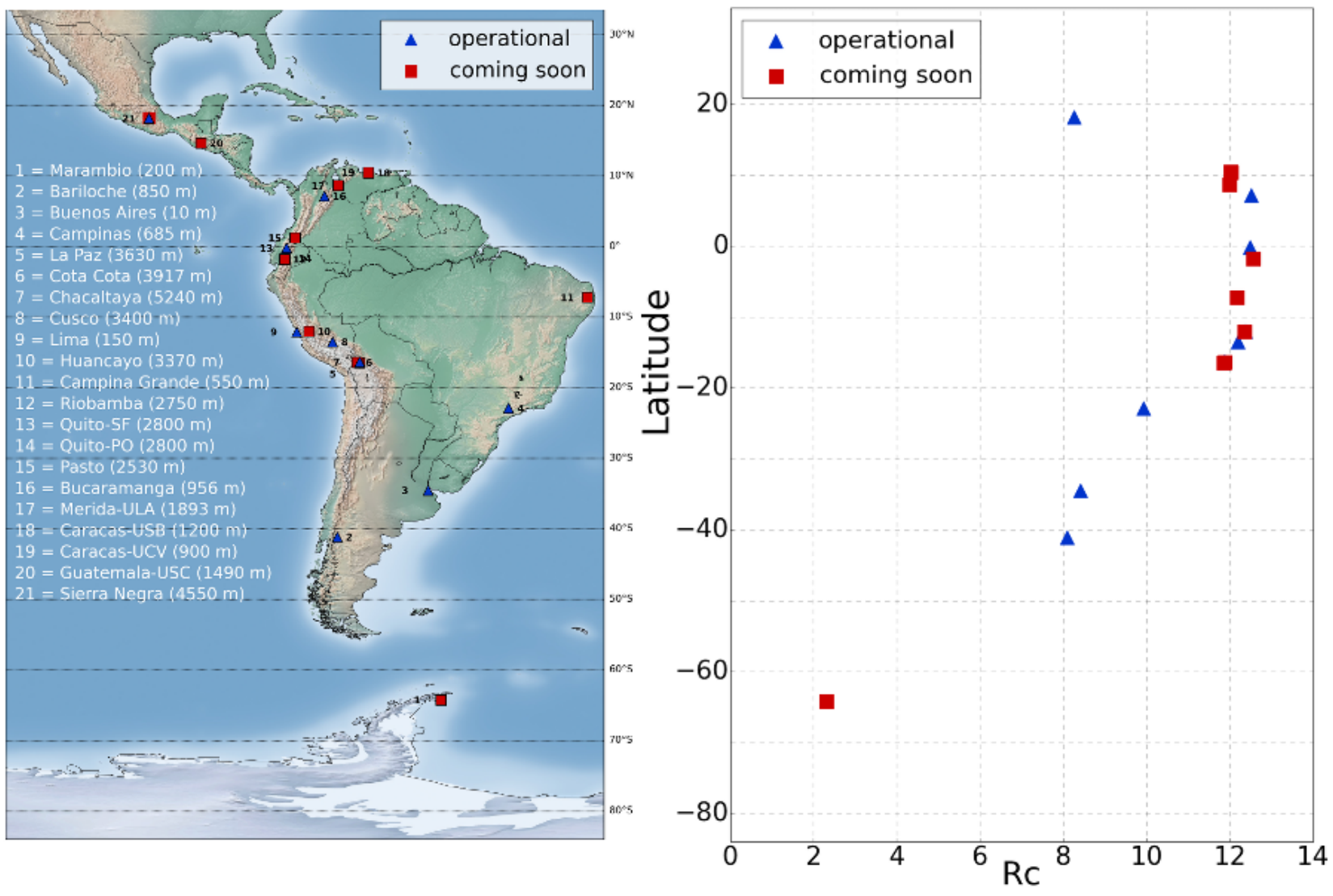

**Figure 1:** Geographical distribution and altitudes of the operational LAGO water Cherenkov detectors (blue triangles), and those that are being deployed and will start its operation during the 2015-2016 biennium (red squares). At the right panel, the vertical rigidity cut-off of each LAGO site are shown.

As will be shown in section 4 we are able to determine the flux of secondary particles at different bands of deposited energy in the detector by using pulse shape discrimination techniques. These bands are dominated by different components of the primary radiation. This is what we called the multi-spectral analysis technique (MSAT). Then, by combining all the data measured at different locations of our detection network, the LAGO project will provide very detailed and simultaneous information of the temporal evolution and of the small and large scales characteristics of the disturbances produced by different transient and long term space weather phenomena.

This complex problem comprises very different processes at different spatial and time scales. So, our approach is based on the following conceptual schema:

$$\begin{array}{ccccc} \text{GCR Flux} & \xrightarrow{\text{Heliosphere}} & \text{Modulated Flux} & \xrightarrow{\text{Magnetosphere}} & \text{Primaries} \rightarrow \cdots \\ \cdots \rightarrow \text{Primaries} & \xrightarrow{\text{Atmosphere}} & \text{Secondary} & \xrightarrow{\text{Detector response}} & \text{Signals.} \end{array}$$

Each step of this chain is evaluated and token into account by direct measurements in our stations [3], or by accessing data from several online servers (e.g. the $D_{st}$ index, see section 3). All our measurements are complemented by a complete chain of simulations, as will be described in the next section.

## 3. Cosmic rays, Extensive Air Showers and the Geomagnetic Field

The LAGO Space Weather program is supported by an intensive and complete chain of simulations accounting for three important factors: the geomagnetic effects, the development of the extensive air showers (EAS) in the atmosphere, and the detector response to the different types of secondary particles at ground level by means of a GEANT4 model [6, 7].

### 3.1 Galactic Cosmic Rays and the Geomagnetic Field

The GF effects on the propagation of charged particles that could contribute to the background radiation at ground level can be characterized by the directional rigidity cut-off $R_c$ at each LAGO site, considered as a function of the geographical position at some altitude (Lat, Lon, Alt) and the arrival direction $(\theta,\phi)$ of the primary at this point. Additionally, a time stamp (TS YYMMDDHHMMSS UTC) should also be included to address transient phenomena during those periods of intense geomagnetic activity, so $R_c \equiv R_c\,(\mathrm{TS},\mathrm{Lat},\mathrm{Lon},\mathrm{Alt},\theta,\phi)$.

This directional $R_c$ can be calculated at each LAGO site by using the Magnetocosmics code [8] by applying the backtracking technique [9] and using the International Geomagnetic Field Reference (IGRF) version 11 [10] for modeling the near-earth GF ($r < 5R_\oplus$) and the Tsyganenkov (TSY01) [11] to describe the outer GF. As it is commonly used, our description of the GF can be tagged by six parameters: solar wind dynamic pressure, $D_{ST}$ index, $B_y$ and $B_z$ components of the geomagnetic field (in GSM system), and the $G_1$ and $G_2$ parameters of TSY01 model.

As the next step in the simulation chain is performed in CORSIKA v 7.3500 [12], that accounts for GF effects during the particle propagation in the atmosphere, $R_c$ is calculated at the edge of the atmosphere, $\sim 110$ km a.s.l. As an example, we will show the calculations performed for the LAGO site in Bucaramanga, Colombia ($+7^\circ 8'$LN, $-73^\circ 0'$ LW) for May of 2005 (initial TS 0505090900). The changing conditions of the GF were obtained from data of the Virtual Radiation Belt Observatory[1]. Results of this calculations are displayed in figures 2: plate 2(a) shows the strong relation between the $R_c$ and the arrival direction of the charged particles hitting the edge of the Earth atmosphere; plate 2(b) displays the predicted changes on $R_c$ during the geomagnetic storm of May 15th 2005, particularly for azimuth angles greater than 150°. To look for the origin of the anomalous behavior observed in the $R_c$ for arrival directions in the range $45^\circ < \theta < 90^\circ$ and $250^\circ < \phi < 300^\circ$, we performed detailed analysis of the individual paths of the particles in this arrival direction range. We found that all these trajectories go through the South Atlantic Anomaly, showing that it could be responsible for this particular behaviour.

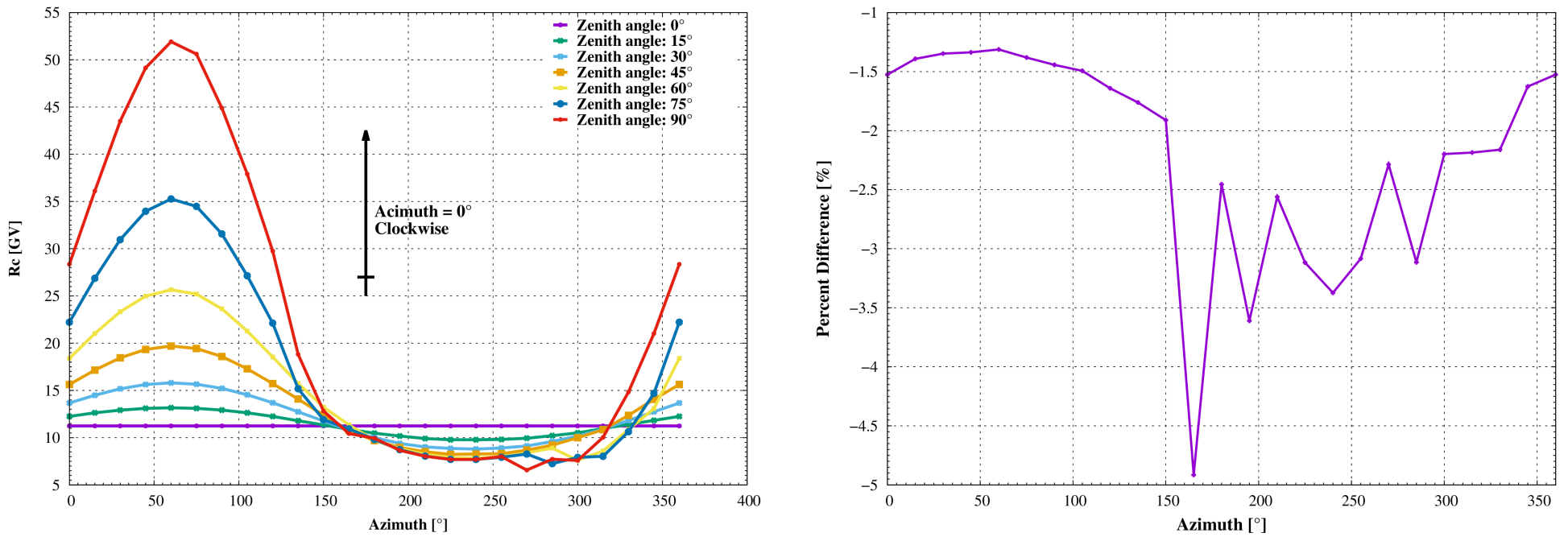

(a) (b)

**Figure 2:** (a) Directional rigidity cut-off ($R_c$) at the atmosphere edge, as a function of the azimuth angle $\phi$ and for different zenith angles $\theta$ for the LAGO site in Bucamaranga, Colombia, during secular conditions of the geomagnetic field. The effect of the SAA is observed for inclined particles in the range $250^\circ < \phi < 300^\circ$. (b) Relative difference for $\theta = 45^\circ$ between the $R_c$ under secular condition (left) and during the geomagnetic storm occurred on May 15th 2005 09:00 UTC.

[1] http://virbo.org

### 3.2 Secondary Particles Flux at Ground Level

The second step of the chain is a complete simulation based in CORSIKA v 7.3500[2] [12] of the secondaries observed at ground level by the EAS produced during the interaction with the atmosphere of the complete flux of primaries. The flux of all nuclei primaries in the range $1 \leq Z \leq 26$ was assumed to be uniform in solid angle and considering a single power law for each nuclei to describe the energy dependence in the range $Z \times R_c \leq E \leq 1\,\mathrm{PeV}$ [13]. The corresponding parameters were obtained from [14]. As the $R_c$ depends on the arrival direction of each primary, the total number of primaries $N$ is strongly angular dependent. To avoid numerical issues during the calculation of $N$, we simulate all the primaries starting from the lowest $R_c$ and then we simply suppress from the final result those secondaries that were originated from primaries that should not be allowed to enter the atmosphere due the geomagnetic field effect.

The calculation performed without GF corrections can be considered as an approximation to the more precise flux of secondaries calculated when the effects of the geomagnetic perturbation have been considered. The effect of the geomagnetic field in the flux of primaries impinging the Earth atmosphere can be observed in figure 3(a), where a comparison of fluxes with/without GF before the edge of the atmosphere is shown. Only primaries that actually produced secondary particles at ground are considered. Low energy nuclei are affected, but it is not a sharp cutoff as the studied effect depends on its $Z$, on arrival directions of the primaries, and also includes the effect of atmospheric absorption of secondary particles during the EAS development.

The GF effect in the secondary flux at ground level can also be obtained by comparing the expected flux without corrections with the corrected corresponding one. In the figure 3(b), the relative difference between these two cases is shown as a function of the particle momentum. The observed differences are large enough to justify the incorporation of geomagnetic corrections when the low energy secondary flux is calculated. A large deviation in the total flux is also observed, which is peaked at $p_{\mathrm{sec}} \simeq 500\,\mathrm{MeV/c}$. At this values, the secondary flux is dominated by neutrons. For this particular component, the decrease at ground level due to GF effect in secular conditions represents a diminution of $-36.6\%$. This could be an indication of the sensitivity of secondary neutron flux as a proxy of the changing conditions in the near-earth space environment.

Based on this preliminary results, different studies are conducted within the LAGO Collaboration to evaluate the sensitivity of WCD via neutron capture in pure water or to add dopants such as gadolinium-boron mixtures.

## 4. Multi-spectral Analysis of Space Weather Phenomena

In this section we shall show one of the actions of the LAGO Space Weather program which proves how a single WCD is able to detect transient solar phenomena affecting the flux of particles at the surface of the Earth. Previous works have shown WCD capabilities to study heliospheric phenomena [13], and here we exploit the possibilities to study the variations on the flux of secondaries at different bands of deposited energy $E_d$.

The charge histogram of the WCD, typically used for detector calibration, is obtained by time integration of the individual pulses measured in the WCD. The resulting histogram, showed in

[2]Options used: QGSJET-II-04; GHEISHA-2002; EGS4; Curve, External Atmosphere, Volumetric Detector

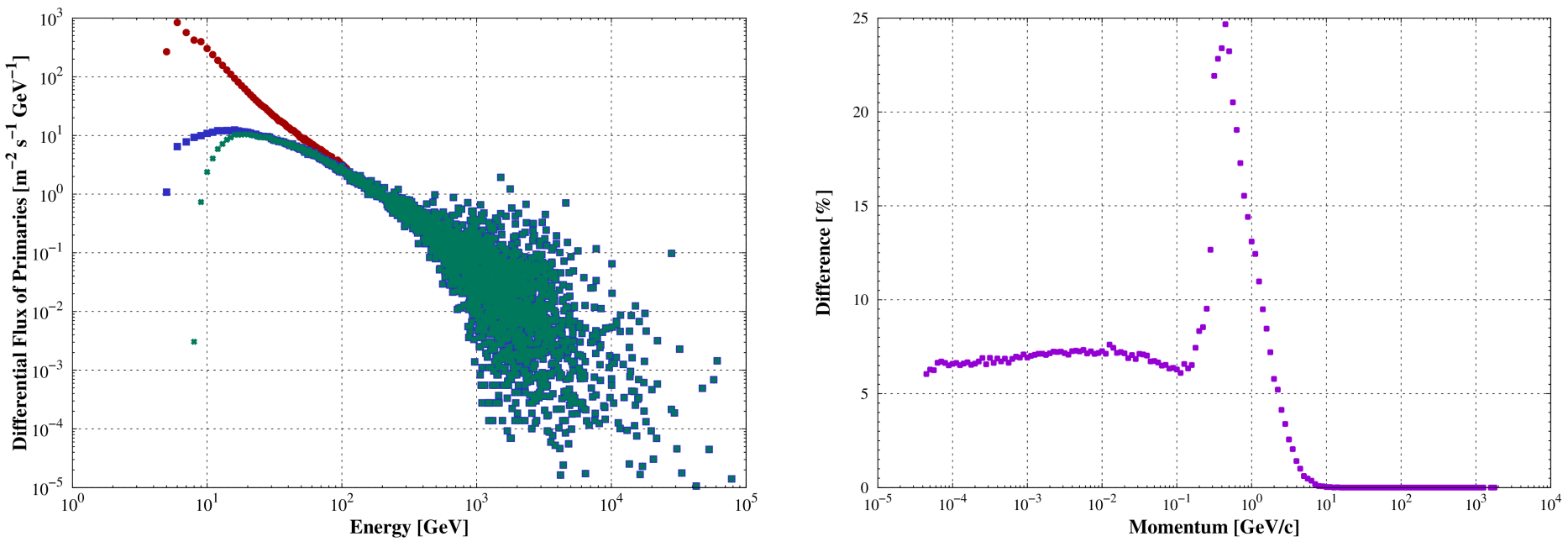

(a) (b)

**Figure 3:** (a) Geomagnetic effects in the flux of low energy GCR, observed when comparing the flux of primaries that actually produced secondary particles at ground when GF corrections for secular conditions are included (green stars) and when they are not considered (blue squares). The total flux of injected primaries is also included (red circles). (b) The GF effect in the secondary particles at ground level as a function of the secondary momentum. In this curve, the relative difference between the flux without GF corrections and the corrected flux is shown. The larger differences are observed at $p_{\rm sec} \simeq 500\,$MeV/c. In this region, the flux of secondaries is dominated by neutrons.

figure 4, is originated by the convolution of the response of the WCD to the different types of EAS particles and the water quality, the inner coating used, the PMT size and the detector geometry. By doing first principle calculations and detailed simulations, it is well established that the small signal region, corresponding to low values of $E_d$, is dominated by the electromagnetic component (EM, $\gamma$s and $e^{\pm}$) of the shower; while high values of $E_d$ correspond to the simultaneous entrance (within the electronic sampling time, 10 ns-40 ns) of multiple particles to the detector volume (the so called mini-shower - MS - regime). Finally, intermediate values of $E_d$, evidenced by a characteristic peak[3] in the histogram called the muon hump, are dominated by single muons through the detector.

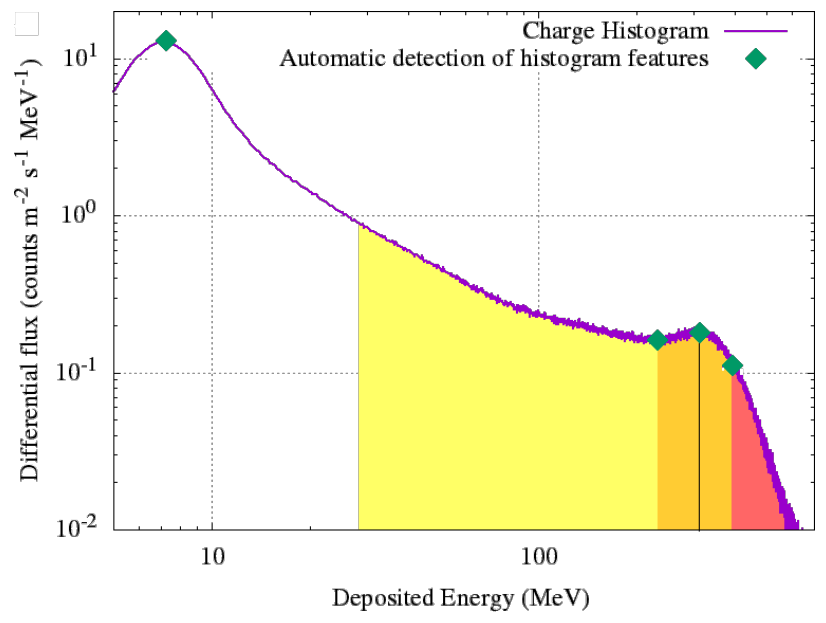

**Figure 4:** Charge histogram of one LAGO WCD in Bariloche, Argentina. A fully automated algorithm look for histogram features (green diamonds) to define integrations bands (shaded regions). Each integration band is dominated by different type of particles: EM particles (yellow), muons (orange) and multiple particles (pink).

The LAGO MSAT determines the evolution of the flux observed at those three bands: EM-band, $\mu$-band and MS-band. As the transition points between these different regimes are characterized by changes in the histogram slopes, a fully automated algorithm searches for all these features in the 1-hour calibration histograms and determines the integrated total flux in those three bands as can be seen in figure 4. Typical integration windows are between 1 minute and 15 minutes long.

[3]The first peak seen at the histogram at very low signals is originated by the detector trigger system and is not considered in these analysis.

As a first attempt, we applied the MSAT to analyze the Forbush event observed on March 8th, 2012, associated with the passage of an iCME and detected by different instruments [15]. The studied data were acquired in a single 1.8 $m^2$ WCD (with 2.5 $m^3$ of pure water and a 8" Hamamatsu R5912 PMT) at the LAGO site of Bariloche, Argentina. After applied flux pressure correction to each band, the FD is clearly visible in our MSAT, with a maximum peak-to-peak decrease of $\sim 5\%$, $\sim 6\%$, and $\sim 4\%$ in the EM, $\mu$ and MS bands respectively. The temporal evolution of the EM and $\mu$ bands are shown in figure 5, where our data is compared with data from the Rome neutron monitor[4]. Solar daily modulation of the flux is also visible on both bands. The particular features observed in our data are currently under careful analysis and will be published soon.

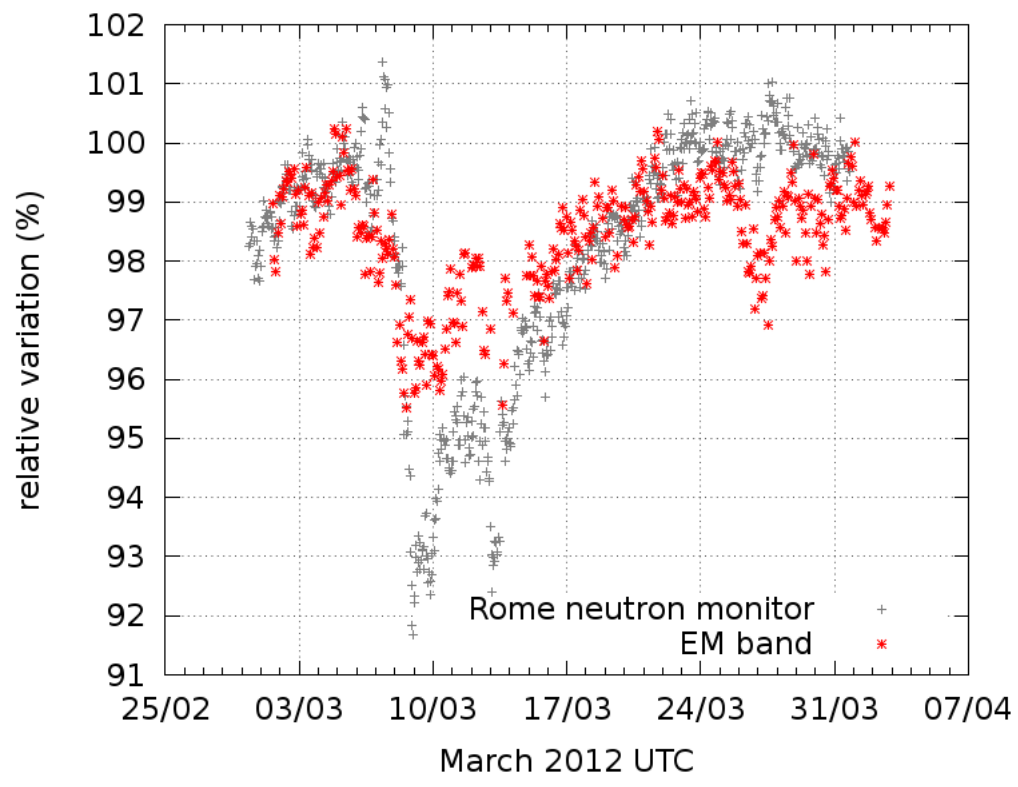

(a)

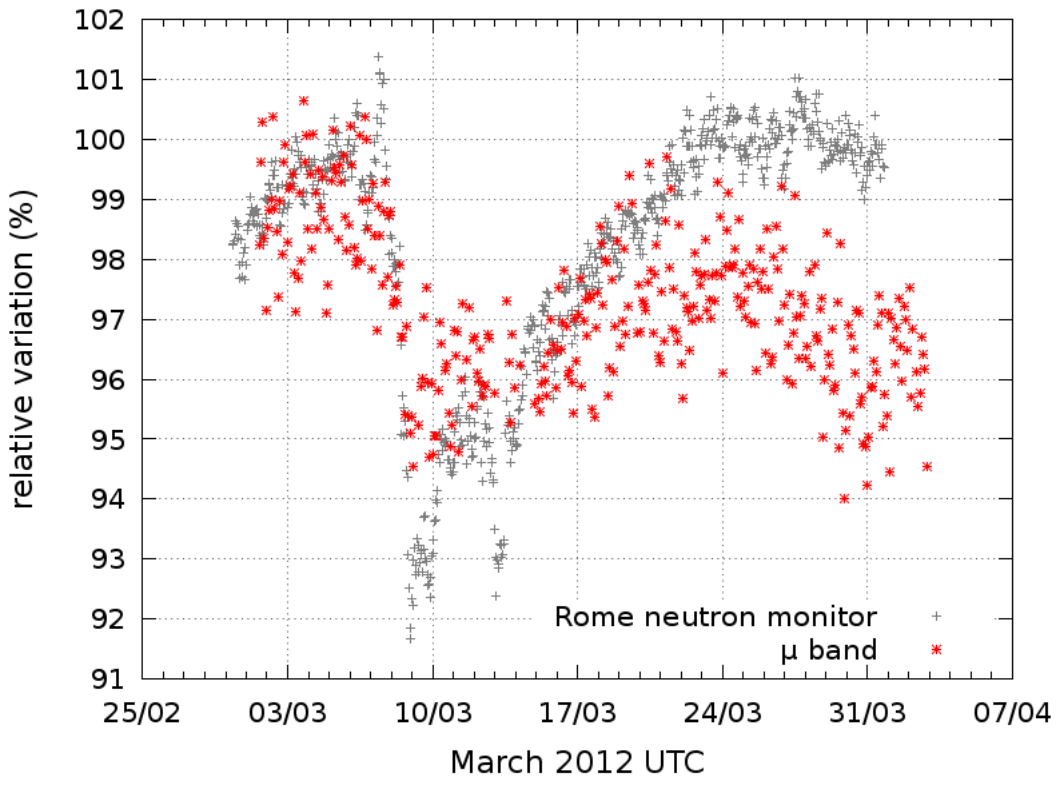

(b)

**Figure 5:** Results of Multi-Spectal Analysis of the Forbush Decrease of March 8th, 2012 measured in a single 1.8 $m^2$ WCD installed in Bariloche, Argentina (red stars), compared with measurements of the Rome neutron monitor (gray pluses), in two bands: EM-band (a), and $\mu$-band (b).

## 5. Summary and Conclusions

In this seminal paper we describe the basis of the LAGO Space Weather program, that accounts for a complete chain of measurement and simulations to study Space Weather phenomena from ground level.

Two important contributions have been presented. First, the simulation steps introduced to obtain a precise calculation of the secondary particle flux at ground level for any current or future LAGO site, that includes now a method to account for geomagnetic effects in the flux of secondaries. We find significant differences on the expected flux of particles at ground level when GF corrections are considered. At this moment, simulations of the background flux expected under different geomagnetic conditions for all LAGO sites are underway.

The second contribution proves the capabilities of single WCD to detect transient space weather phenomena from ground level. The introduction of the MSAT shows also that using even single detector it is possible to determine the flux at bands that are dominated by different types of particles coming from different primaries contribution.

The principal result of this work is, however, to stablish the conceptual basis of one of the main objectives of the LAGO project: the observation of space weather phenomena from ground level. The LAGO detection network, once it is completed and operational, will provide the unique information to delve in the very complex problem of the Sun-Earth connection.

[4]Available at the NEST repository, http://www.nmdb.eu/nest

**Acknowledgments:** The LAGO Collaboration is very thankful to the Pierre Auger Collaboration for its continuous support. LAN and YP gratefully acknowledges the partial financial support from CDCHT-ULA project C-1598-08-05-A. CSC, LAN, MSD and HA are also in debt to Vicerrectoría Investigación y Extensión Universidad Industrial de Santander for its permanent support.

# LAGO Ecuador, Implementing a set of WCD detectors for Space Weather research: first results and further developments

**Stephany Vargas**[*1], **Cristina Mantilla**[1], **Oscar Martínez** [1], **Nicolás Vásquez**[1] **and Dennis Cazar**[2] **for the LAGO Collaboration**[3]

[1]*Escuela Politecnica Nacional, Departamento de Física, Ladron de Guevara E11-258, Quito, Ecuador*

[2]*Colegio de Ciencias e Ingenierías, Diego de Robles y Pampite S/N, Quito, Ecuador*

[3] *lagoproject.org, see the full list of members and institutions at lagoproject.org/collab.html*

*e-mail:* asoreyh@cab.cnea.gov.ar

*E-mail:* stephany.vargas@epn.edu.ec, cristina.mantilla@epn.edu.ec

The Latin American Giant Observatory (LAGO) is an international collaboration focused on the study of the phenomenology of Cosmic Rays (CR) in different energy ranges. Data is retrieved using Water Cherenkov Detectors(WCD) and by the execution of several simulations. Ecuador has been working in the LAGO project for almost 3 years, during this period three detectors had been installed in different universities of the country: one in Riobamba, ESPOCH (2784 m.a.s.l) and two in the city of Quito, USFQ (2430 m.a.s.l) and EPN (2830 m.a.s.l). A complete set of simulations have been performed within the LAGO simulation framework, based on a combination of Magnetocosmics, CORSIKA and GEANT4 codes. In this work, the detector characterization, calibration and the first Ecuadorian data is presented.

*Speaker.

## 1. Background

The Latin American Giant Observatory—LAGO—project is a recent organization created on 2005. There are 10 Latin American countries participating in this project, with more than 80 scientists and students of different disciplines [1]. As objectives, among others, this project studies high energy component of GRBs, transitory events, long term events, space weather and climate. These objectives are accomplished through the detection of the high energy component in GRBs for places located above 4000 meters over the sea level, through the study of solar modulation, the surveillance of the cosmic ray flow as well as other analysis. Academically, the project focuses on the study of astroparticle physics for undergraduate students including analysis and data statistics, detector's physics among other specific topics.

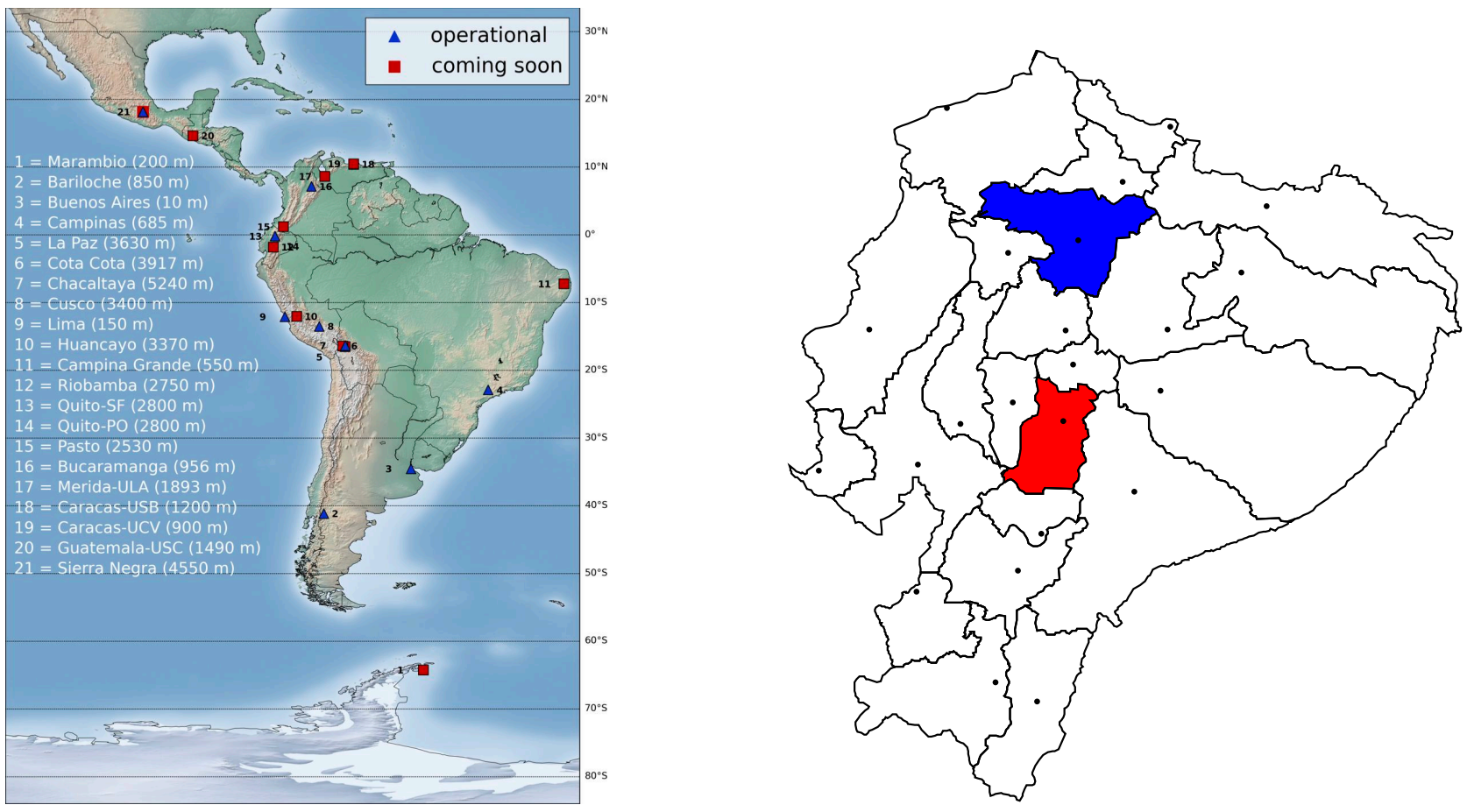

**Figure 1:** *Left:* LAGO locations, where blue triangles show the places associated with functional sites and red squares show the places that will work in 2015 [2]. *Right:* Political map of Ecuador. In red is the Chimborazo province where the "Chimbito" detector is located and in blue is the Pichincha Province where "Panchito" detector and "Politanque" detector are located.

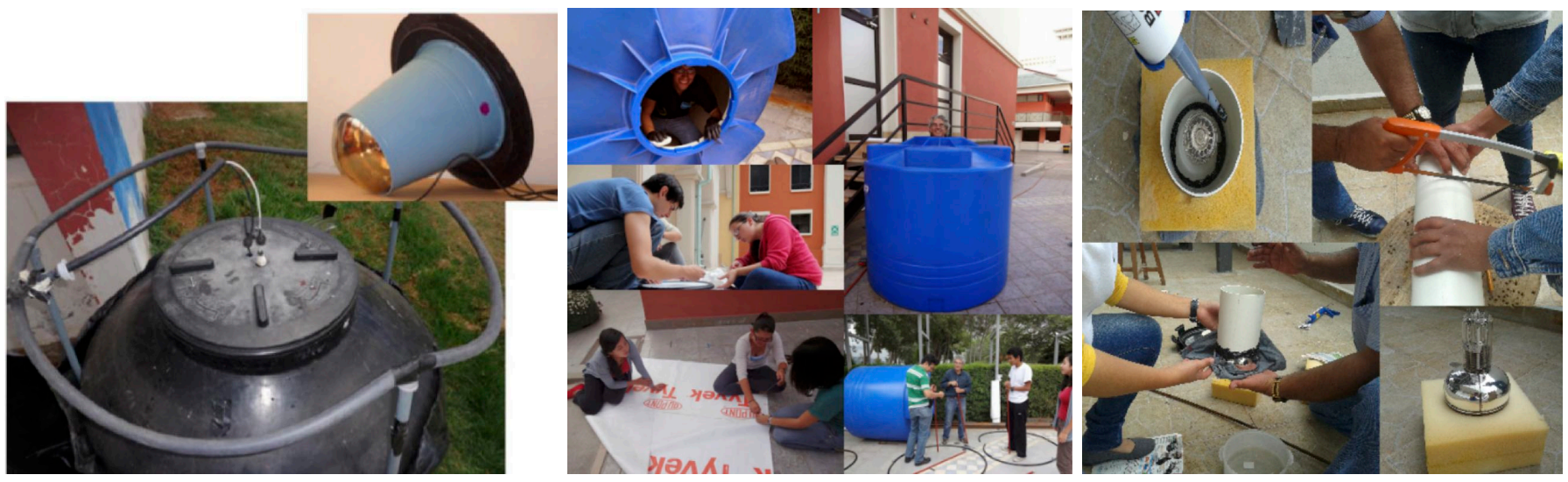

**Figure 2:** *Left:* "Chimbito" detector located in the province of Chimborazo, *Center:* construction of the "Panchito" detector located in the province of Pichincha in the valley of Cumbaya, *Right:* construction of the "Politanque" detector located in the province of Pichincha in Quito.

The LAGO network is unique considering its location (60 and 80 meridians perpendicular to the equatorial plane) and its extension through the southern hemisphere of the globe. The advantage of these characteristics is the possibility to register with great detail transit solar events, such as Coronal Mass Ejections (CME) simultaneously in several sites with very different rigidity cut-offs.

The LAGO project in Ecuador began on 2012 at the Chimborazo Province, Ecuador [3]. The program started with an initial test on August and September at the ESPOCH University Campus with a detector named “Chimbito” [3]. On February 2014 at the USFQ University, located in the Pichincha Province, the construction of a second detector began, its was named “Panchito”. Finally, a third detector, “Politanque”, was implemented at the EPN University in Quito.

## 2. Experimental setup and data

The detector “Politanque” installed at the Escuela Politécnica Nacional (EPN), is the prototype for the Lago-Ecuador collaboration. Is a small-sized detector installed in the facilities of the University, accessible to students and teachers from within the university as well as other scientists within the collaboration. Its aim is to serve as a developing detector set up. “Politanque” has been fully operational for a short period of time.

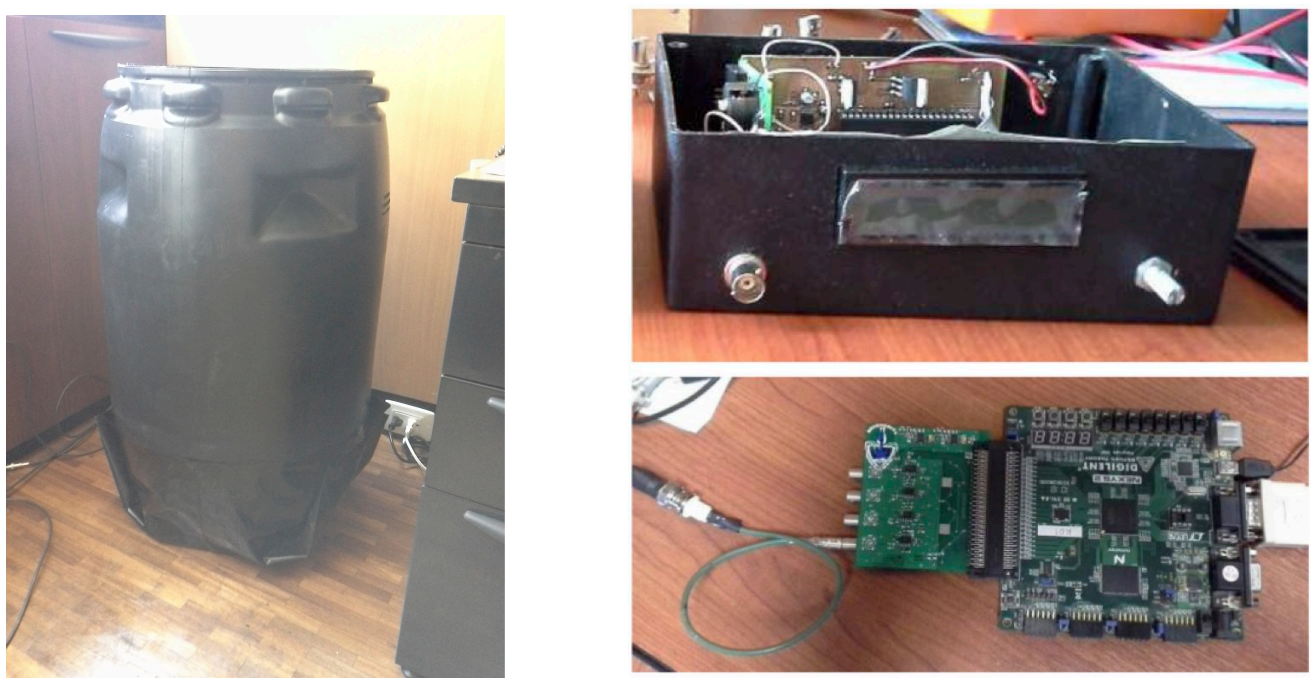

**Figure 3:** *Left:* Polyethylene—WCD, *Rigth up:* high-voltage source, *Right down:* DAQmx and Nexys2 Board, the electronics used for the experiment are from the EAS-BUAP project [4,5].

### 2.1 WCD “Politanque”

The detector consists of a black polyethylene tank that is 0.70 [m] high with a radius 0.23 [m] and with a volume of $0.11\,m^3$. A 5” EMI 9530A photomultiplier is used. The high-voltage source, as well as the hardware and software used for data acquisition are from the EAS-BUAP project [5]. Data acquisition was performed using a voltage input $\sim 1450[V]$ for the PMT. The output signal is collected in a digitizer board–DAQmx (based on Dual ADC AD9216-100MS/s, 10bit resolution, 4-channel input), the information is then processed by a Nexys2 Board programed using VHDL, then it is stored and processed in a PC with Ubuntu 12.04.

### 2.2 Data acquisition

The USB-Serial Port connection from the preconfigured Nexys2 Board to the computer allows us to manipulate, via VHDL programing, the thresholds of the sampled data. The acquired data is in ADC counts in the range of nanoseconds. Each pulse has 16 samples of 10 nanosecond time length.

### 2.3 Data analysis

We present the preliminary results of the data analysis of the Charge Distribution Histogram (CDH) for a sample of 350 000 events, this results are presented in figure 4. We determined the VEM by a Gaussian fit from with we have a value of 456±106 [pC]. In figure 5, we have the zoom of figure 4 from with we get the VEM and its Gaussian fit.

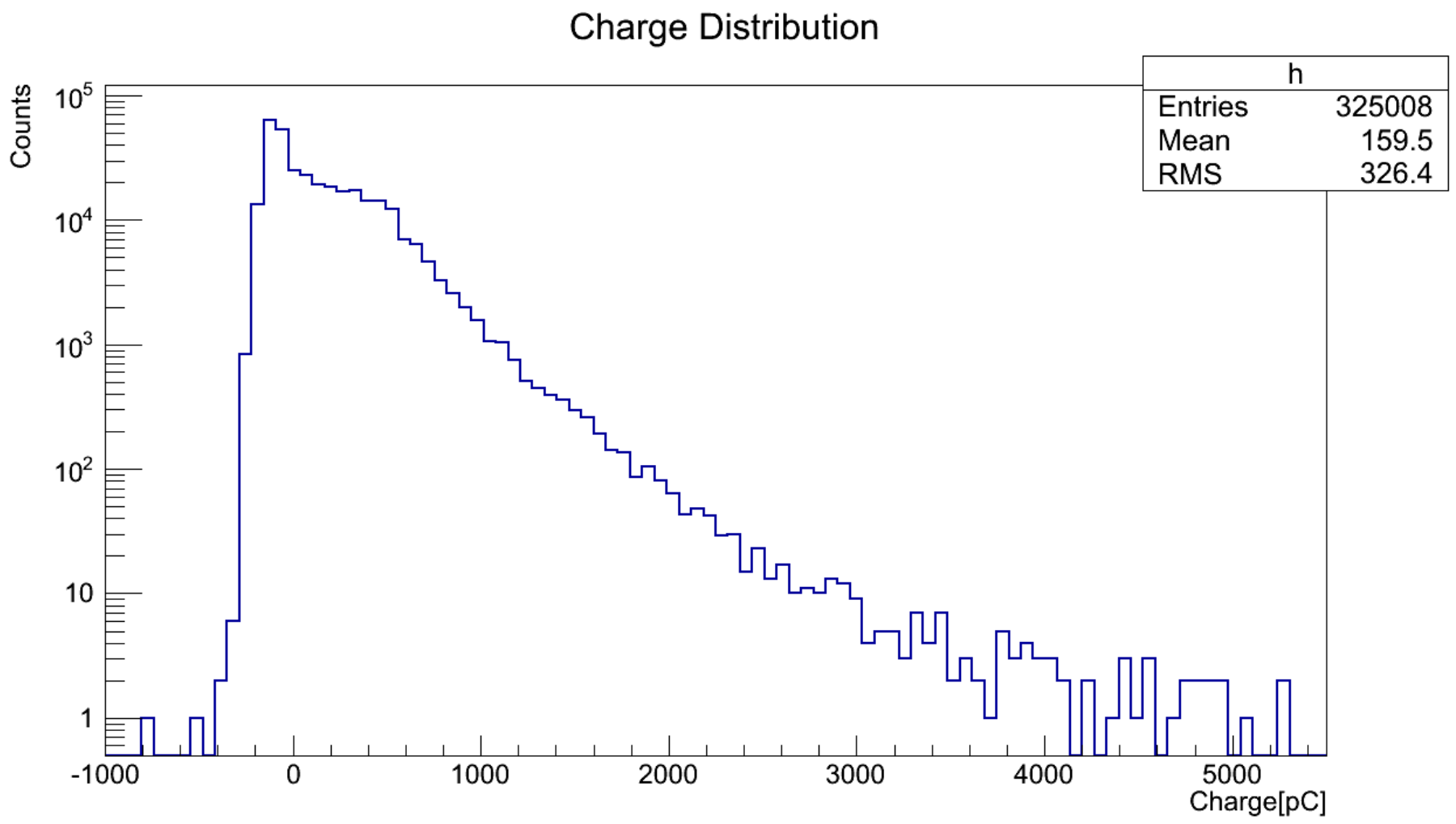

**Figure 4:** Charge Distribution Histogram (CDH) for "Politanque" detector.

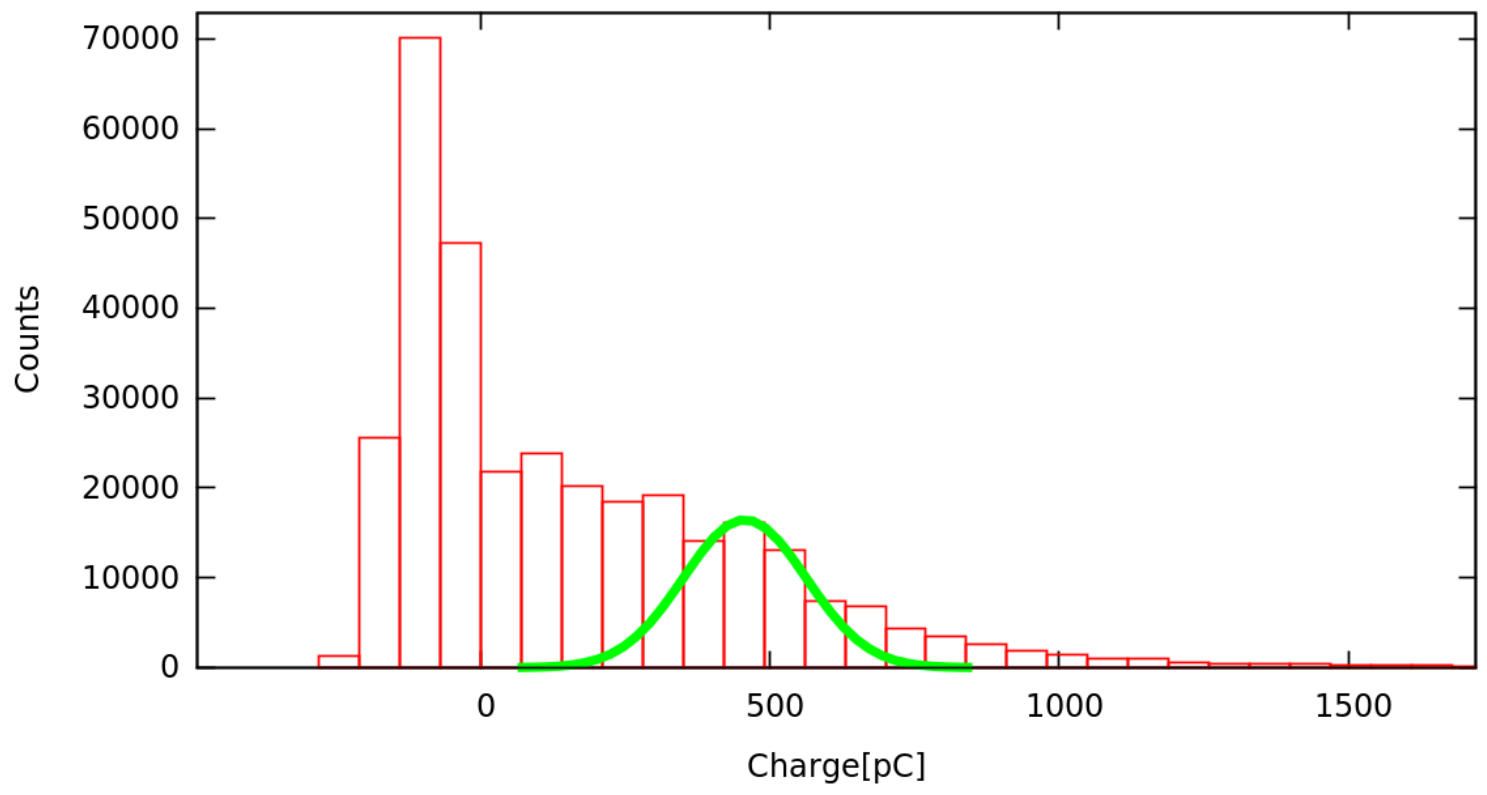

**Figure 5:** VEM (Vertical Equivalent Muon) adjusted by a Gaussian fit. We found a value of 456 [pC] with a statistical error of 106.

## 3. Simulation

The data used in the Collaboration include data registered in Cherenkov detectors and data obtained through simulation. In this section we describe the simulations done in this context for the Ecuadorian project [6].

### 3.1 Corsika

We simulated the development of proton, photon and hadron initiated Extensive Air Showers for the different sites in Ecuador. The strategy used follows a general approach adopted by other LAGO sites for this kind of studies [7]. We considered the primary flux at the top of the atmosphere, taking into account the measured spectra for nuclei with $1 \leq A_p \leq 56$. Primary energies ($E_p$) were considered from 10 GeV to $10^6$ GeV with zenith angles between 0° to 60°. Showers were simulated using the Extensive Air Showers package CORSIKA v7.350 [8]. QGSJET II-04 model was used. No thinning was applied, thus all particles above a very small threshold were followed until absorbed in the atmosphere or reached the detector level.

We obtained the secondary particle flux as a function of the particle momentum for the EPN site (LAT 0° 12' S, LONG 78° 29' O and 2817 m.a.s.l.) The geomagnetic field at this location ($B_x = 27.09\,[\mu\mathrm{T}]$, $B_z = 11.05\,[\mu\mathrm{T}]$) was used and geomagnetic effects were taking into account by calculating the rigidity cut-off for the EPN site. We use MAGNETOCOSMICS, a GEANT4 based simulation package, to solve the equations of motion for single particles in the geomagnetic field at this latitude for different arrival directions [9].

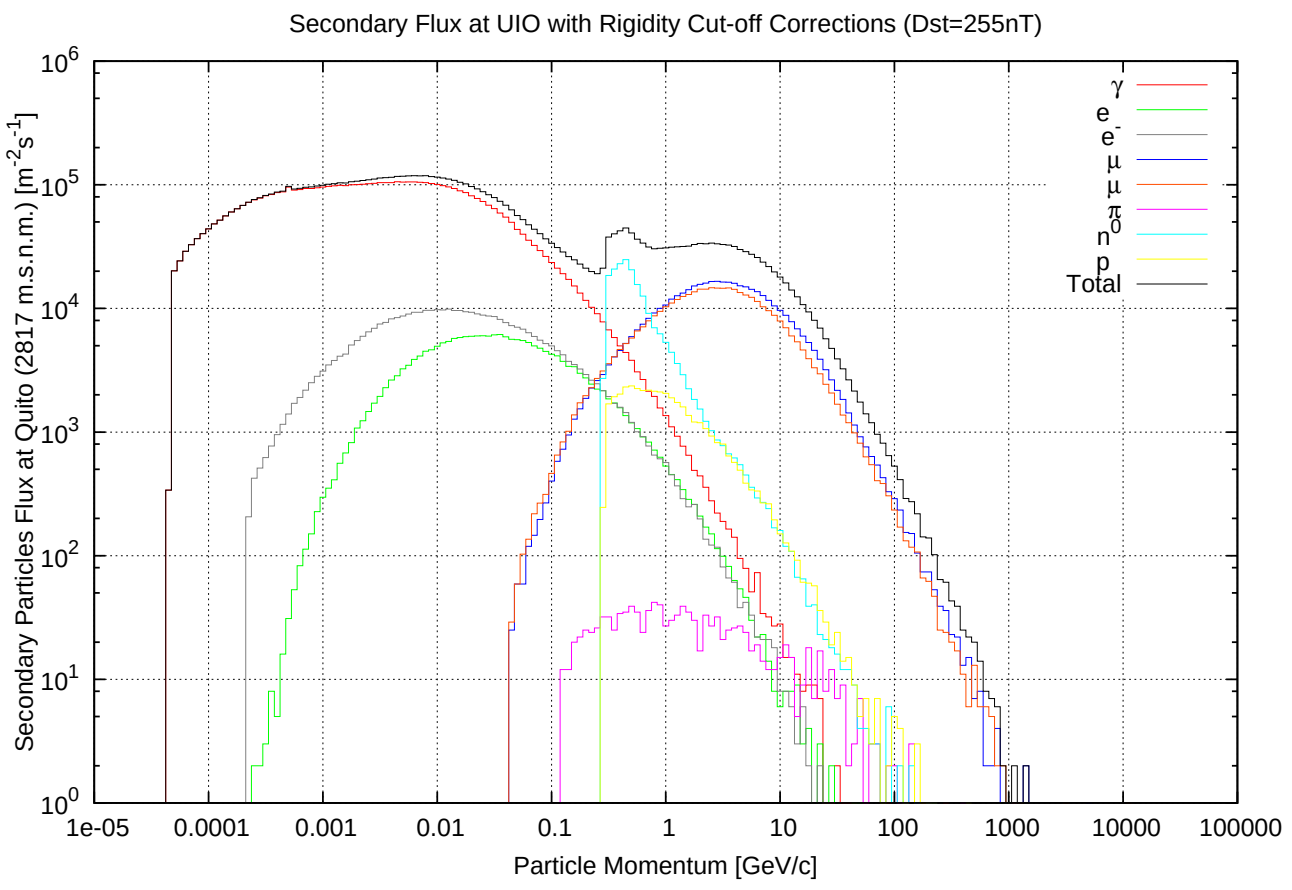

**Figure 6:** Simulated secondary flux at detector level for "Politanque" WCD at 2817 m.a.s.l. Rigidity Cut-Off corrections were taken into account.

### 3.2 Geant4 simulation

For the simulation we have used the Geant4 release 9.6.p02 [9]. We have modified the simulation included in the examples—optical, named LXe as the base code for the simulation, as well as the B1 example [10].

The world volume is declared as a box filled with "G4_Air", here the main volume is built using the polycone function, the resulting volume is shown in figure 8. The main volume is filled with "G4_Water" and covered with a layer of "G4_POLYETHYLENE". The surface between the polyethylene and the water is declared as an optical surface, here we can modify the properties of the material, for instance, we declared an efficiency between 90% and 95% for the reflectivity of the "Tyvek".

The simulation done so far shows the particles created from the interaction of the VEM with the water, the detector response is still under development.

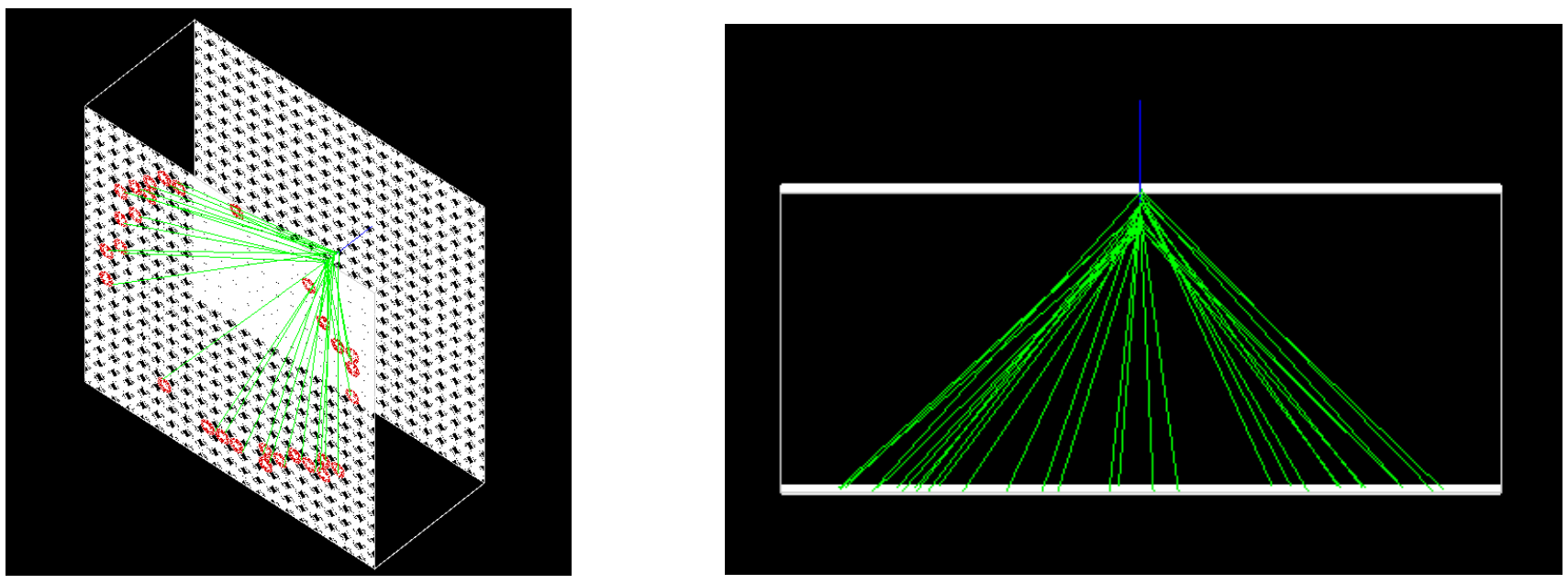

**Figure 7:** *Left:* Simulated cherenkov event in the LXe example. *Right:* Lateral view of the event.

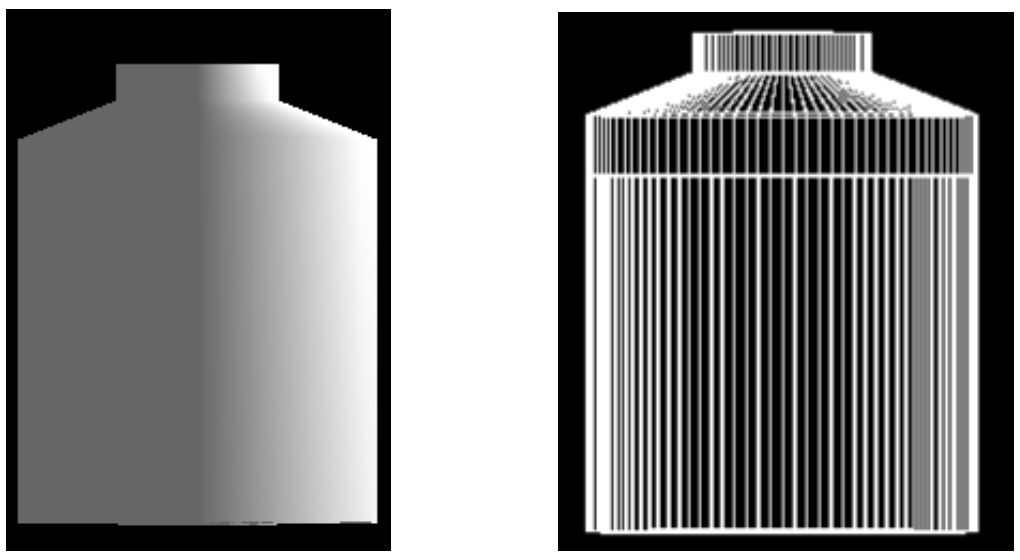

**Figure 8:** Construction of the main volume shown as solid(left) and wireframe(right).

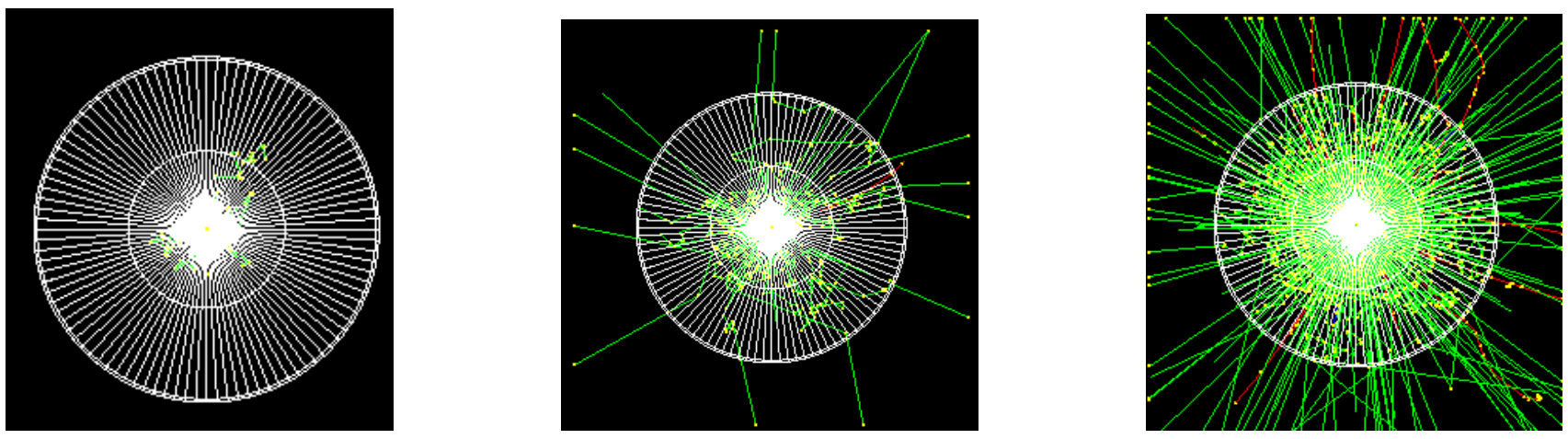

(a) (b) (c)

**Figure 9:** Development of the VEM in the WCD for an incident: (a) muon 50 [MeV], (b) muon of 1 [GeV], (c) 4000 [GeV] muon. The shown view is from above the WCD.

## 4. Conclusion

Data obtained with "politanque" shows the first calibration for the WCD, a longer period of acquisition is needed in order to get a better fit for the first VEM.

Also we found the relationship between the number of events registered and the thresholds for the same exposure times. This is expected because of the characteristic shape of the energy spectrum for Cosmic Rays. For a higher threshold, it's required a more energetic event, events of higher energies have the property that their frequency diminishes.

We have succesfully simulated the secondary flux at detector level for "Politanque" WCD at 2817 m.a.s.l. taking into account the Rigidity Cut-Off using Corsika [8]. We found that the principal contribution in energy of the particles that get to the detector are muons. The simulations in Geant4 [9]are still under development, but the results so far are promising.

The cosmic ray physics community in Ecuador is a growing collaboration. There is a long term future plan to install a national detector network in order to take advantage of this unique location. This work shows the first steps toward the realization of this plan.

**Acknowledgements:**

The LAGO Collaboration is very thankful to the Pierre Auger Collaboration for its continuous support. The authors of this paper gratefully acknowledge the financial support from Escuela Politécnica Nacional, Universidad San Francisco de Quito USFQ, Fundación CEDIA through CEPRA IX research grants and SENECYT for the prometeo program.

# A Project to Install Water-Cherenkov Detectors in the Antarctic Peninsula as part of the LAGO Detection Network

**Dasso S.**[a,b,c]**, Gulisano A.M.**[a,c,d]**, Masías-Meza J.J.**[c] **and Asorey H.**[e]**, for the LAGO Collaboration**[f]

[a] *Instituto de Astronomía y Física del Espacio (UBA-CONICET), Buenos Aires, Argentina.*
[b] *Departamento de Ciencias de la Atmósfera y los Océanos, Facultad de Ciencias Exactas y Naturales, Universidad de Buenos Aires (UBA), Buenos Aires, Argentina.*
[c] *Departamento de Física and IFIBA, Facultad de Ciencias Exactas y Naturales, Universidad de Buenos Aires (UBA), Buenos Aires, Argentina.*
[d] *Instituto Antártico Argentino, DNA, Buenos Aires, Argentina.*
[e] *Laboratorio de Detección de Partículas y Radiación, Instituto Balseiro y Centro Atómico Bariloche, Bariloche, Argentina. Escuela de Física - Universidad Industrial de Santander, Bucaramanga, Colombia.*
[f] lagoproject.org, *full list of members and institutions at* lagoproject.org/collab.html. *E-mail:* *lago-pi@lagoproject.org*

Due to the geomagnetic shielding, particle detectors located at high latitudes allow the observation of cosmic rays (CRs) with lower energies than those located at middle or low latitudes. Thus, Antarctica is a privileged place to study CRs having the lowest energies that can be observed from ground level. The Latin American Giant Observatory (LAGO) consists on a network of water-Cherenkov detectors (WCDs) located in nine countries of Latin America, to study with extreme detail the flux of CRs from ground level. The main scientific objectives are oriented to address several problems of astrophysics, space physics and atmospheric physics. In particular, LAGO has started to develop a Space Weather program. A project to install WCDs in the Argentinean Marambio station, located at the Antarctic Peninsula, is being developed as the first antarctic node of LAGO. In this work, we present several aspects of the project, including information about the site, the detector, and studies of geomagnetic and atmospheric properties, that permit us to characterize several useful properties of this location. Results from this new LAGO site will provide important insight for Space Weather, and will provide knowledge to better understand the Sun-Earth coupling.

## 1. Introduction

One of the aims of the Latin American Giant Observatory (LAGO, previsouly known as Large Aperture Gamma ray burst Observatory, see [1]) is the design, instalation, deployment and operation of an extended astroparticle observatory to a global scale.

The project is operated by the LAGO Collaboration, which is a collaborative, distributed, and not centralized network, composed of researchers and students from nine countries in Latin America: Argentina, Bolivia, Brazil, Colombia, Ecuador, Guatemala, Mexico, Peru and Venezuela. They have recently joined institutions in Spain. The LAGO detection network is formed by particle detectors, individual or in small arrays, located at ground level and installed at different sites. For a details of different LAGO detectors at each LAGO site, see [2]. The network covers a wide latitudinal distribution (currently from Mexico to Patagonia and soon in Antarctica, as will be seen in the development of this paper) and spans a huge range of altitudes (from sea level to over 5000 m asl), covering an extensive range of geomagnetic rigidities cutoff, and different levels of absorption and reaction at the atmosphere. In addition to the Antartic site that is presented in this paper, other LAGO sites are also starting or will start to operate in LAGO (for more details see [3, 4, 5]). The strategy of LAGO to be able to storage, catalogue and preserve its huge amount of data (from both, observations and simulations) can be found in [6].

One the the recent main scientific aims of LAGO is focussed on the study of astroparticles in the context of space weather and atmospheric radiation. Forbush events (see e.g., [7]) are associated with decreases of cosmic rays flux at ground level, due to the shielding in the solar wind, produced by huge interplanetary transient structures, that are recorded by neutron monitors as well as by other ground detectors, as for instance water Cherenkov detectors (WCDs) (e.g., [8]). Occasionally, solar emission of cosmic rays of sufficient energy and intensity as to raise radiation levels at ground level, can be detected by neutron monitors. These events are called Ground Level Enhancements (GLEs). GLEs are of major interest because an enhanced relativistic proton and electron flux can pose a hazard to astronauts, air crews, and also aircraft electronics, then its observation can provide the earliest direct indication of an impending space radiation storm (see e.g., [9]). They can be observed with proper surface detectors located at different appropriate locations of the terrestrial globe, mainly at high latitude regions.

The trajectory of cosmic rays in the far space ambient of the terrestrial environment is mainly affected by the electromagnetic fields that are there. Due to the fact that these particles do not achieve significant effects on the fields (i.e. they are not enough and they have not enough energy as to be significant sources of electric fields or electric currents), they can be treated as passive particles embeded in pre-set fields. In particular, interplanetary conditions or the presence of geomagnetic fields can produce very important effects on these particles, and frequently they produce particle modulation that can be observed using ground particle detectors. When these particles are getting closer to the terrestrial surface, primary cosmic rays interact with the background particles of the atmosphere, producing a huge amount of secondary particles, generated from the interaction between primaries and the atmospheric components. Thus, in order to be able to get a correct interpretation of the signals from the particle detectors, as the LAGO ones, it is necessary to quantify as accurately as possible these effects, in particular those produced by the geomagnetic field and by the atmosphere.

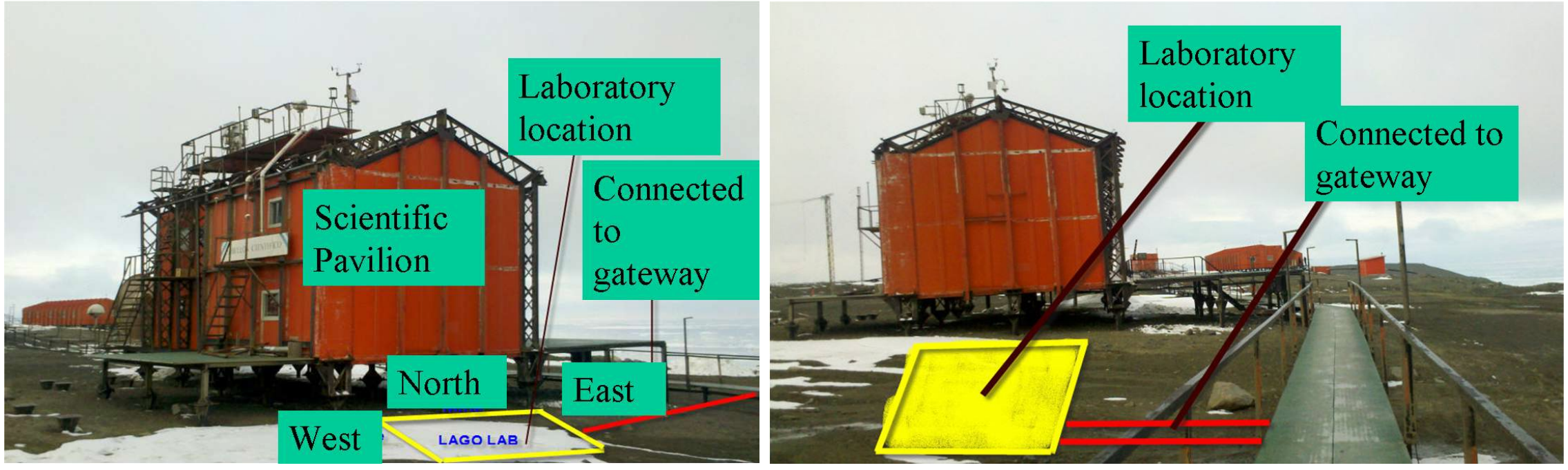

**Figure 1:** Two views of the LAGO Antarctic facility location.

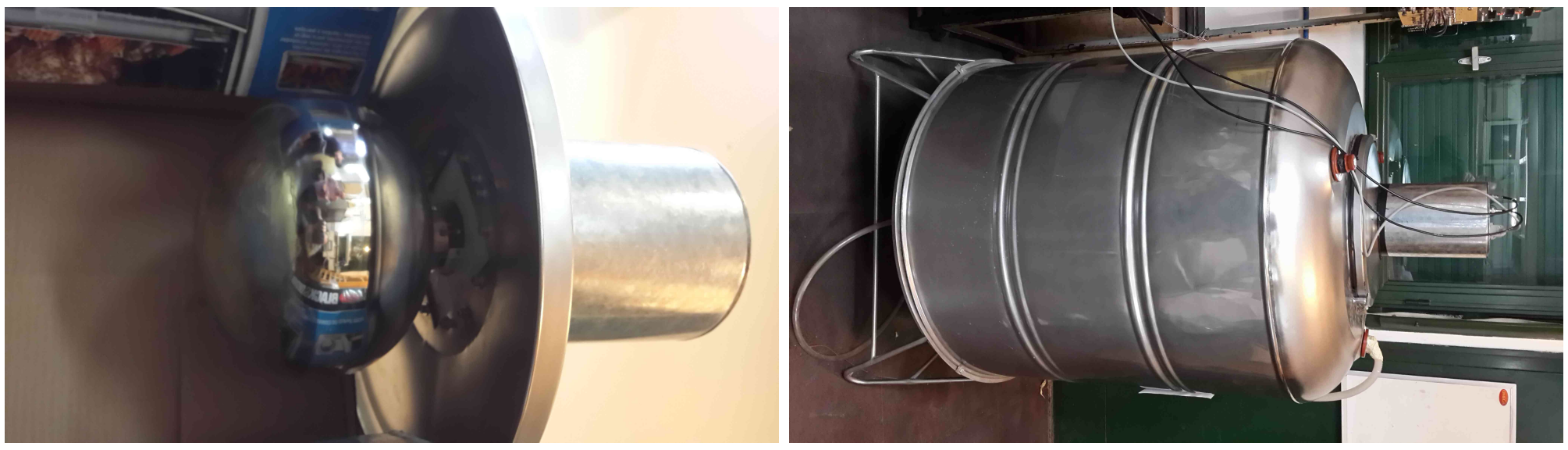

**Figure 2:** Protopype of the detector.

In Section 2 we describe the place where the detectors will be installed. We then present our studies to characterize the site. In particular, in Section 3 we quantify several effects of the geomagnetic field on primary particles, and in Section 4 we present a comparison of the pressure height profile observed in the local atmosphere with balloons with the typical ones available in cannonical shower simulators. Finally, in Section 5 we present a Summary and our conclusions of the present paper.

## 2. The site at Marambio

The Marambio station of the argentine Antarctic regions is located at Lat. 64°14'24.96"S and Long. 56°37'30.34"W, at 196 meters above sea level. Due to the low temperatures during winter, WCDs at Marambio should be thermally isolated to avoid the uncontrolled freezing of the water and the consequent lost in sensitivity. To do this, we designed facilities where the WCDs will be allocated. The laboratory design allows the installation of up to three WCDs and additional equipment, while in this first stage only two WCDs will be deployed, one detector provided by the Instituto de Astronomía y Física del Espacio (IAFE, UBA-CONICET) and the another one by the Centro Atómico Bariloche (CAB, CNEA). This building takes into account the gateway accesibility, it has permafrost characteristics, and optimal construction materials and assembly. The whole structure will be anchored to the permafrost with a depth of at least one meter. The steel base will be elevated 1 m from ground, to diminish the impact of snow accumulation. The new building will be built next to the current location of the central scientific pavillon of the Marambio base, and will have access by using an external corridor. Figure 1 shows the place where the LAGO building will be constructed and Figure 2 shows a prototype of the detector that was contructed at the Instituto de Astronomía y Física del Espacio, in Buenos Aires. In the left panel of the figure can be noticed that the base of the PMT is isolated above the water container, in order to decrease the level of humidity. Several pieces of the detector has been tested at temperatures well below zero Celsius, to ensure its quality after suffering any unforeseen situation with low temperatures regimes in the Antarctic soil.

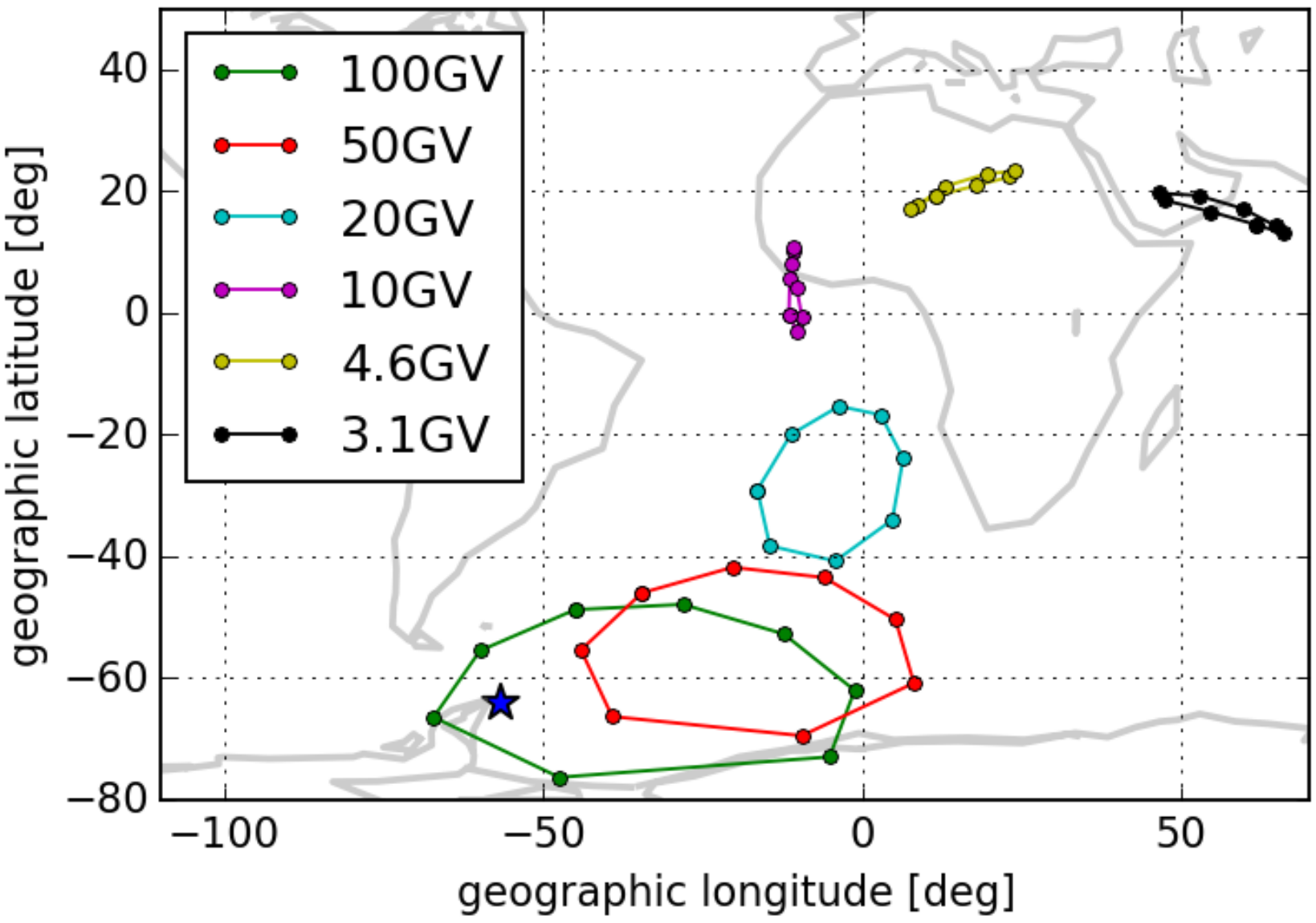

**Figure 3:** Asymptotic directions of primary protons before interacting with the geomagnetic field, for different energies and different incidence azimuth angles (MAGCOS numerical simulations).

The analysis of the data from this WCD at Antartic will allow us to make studies of the background cosmic rays rate, similar to the ones that LAGO is currently making in different sites (e.g., at Chacaltaya [10]), but with higher counting rates and with the impact from primary particles having lower energies.

## 3. Geomagnetic field effects: Numerical Simulations

In this section we present results of numerical simulations of the trajectory of charged particles arriving to Marambio site, using the MAGCOS code (http://cosray.unibe.ch/ laurent/magnetocosmics). For details of the simulations developed see [11]. Similar studies as the ones presented here for Marambio, but for another LAGO sites can be found in [12].

Figure 3 shows the asymptotic directions of protons arriving at the LAGO Marambio site (projected on the Earth's surface) for 15° zenith incidence and eight equispaced incidence azimuth values (45°, 90°, 135°, 180°, 225°, 270°, 315°, and 360°). For this calculation we use the International Geomagnetic Reference Field (IGRF10) for modeling the Earth magnetic field. Figure shows these asymptotic directions for rigidities 3.1, 4.6, 10, 20, 50 and 100 GV. We can see that as the particle rigidity decreases, the asymptotic trajectories get closer to the equatorial region. From a detailed analysis of the simulations done it is possible to conclude that these particles are mostly deflected at heights lower than $\sim 2$ RE, where the configuration of the geomagnetic field is strongly dominated by a dipolar component.

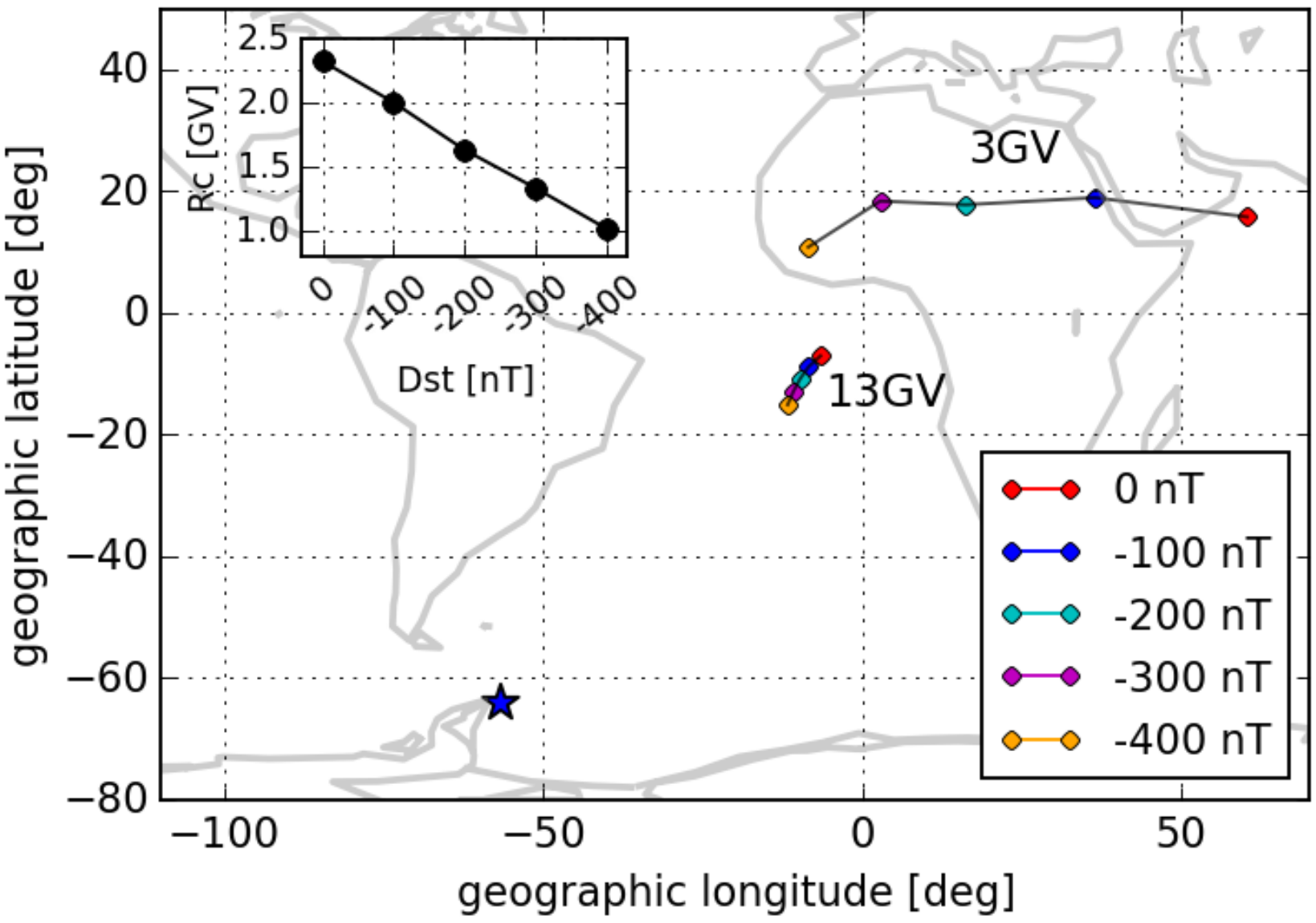

**Figure 4:** Asymptotic directions for vertical incidence, covering different levels of geomagnetic activity. The inset shows the rigidity cutoff for different geomagnetic conditions, according to the *Dst* index.

However, during periods of geomagnetic storms, the non-dipolar component of Bgeo (produced mainly by magnetospheric electric currents) can significantly affect the trajectory of these particles. We develope simulations to compute asymptotic directions for different geomagnetic conditions using the *Dst* index as the proxy and considering the Tsyganenko 2001 (TSY01) model [13]. In figure 4 we show the asymptotic directions of particles with vertical incidence and rigidities for different values of *Dst* (*Dst*=0, -100, -200, -300 and -400nT). We note the shift in longitude for these asymptotic directions, to eastward for higher geomagnetic activity. The inset of 4 also shows the rigidity cutoff for different *Dst* values, getting *Rc* as low as $\sim 1$ GV for *Dst* $\sim -400$ nT. We find that the rate of variation of *Rc* with *Dst* is $\sim$ -0.003GV/nT at Marambio.

## 4. Study of the Atmosphere at Marambio

Numerical simulations of cosmic particle cascades in the terrestrial atmosphere require a detailed knowledge of the pressure height profile in the site. Some numerical models (e.g. CORSIKA code) need specific information about the height profiles of the simulated site. Ten different MODTRAN models are commonly used in CORSIKA [14] simulations.

Presently, a typical CORSIKA shower simulation at the LAGO Marambio site can be done assuming one of the available atmospheric MODTRAN models. In order to accomplish a detailed depiction for the profile at this site, in this section we analyze balloon soundings data measured at 12 UTC by the argentine National Weather Service (Servicio Meteorológico Nacional, SMN) from 1998 until 2014, at Marambio. We computed the density height profile based on the height pressure and temperature profiles observed from the soundings as follows: we take into account the typical atmospheric composition, their molecular masses and the Dalton law for partial pressures for ideal gases. Thus, from the pressure and temperature data for each altitude, we compute the density height profile.

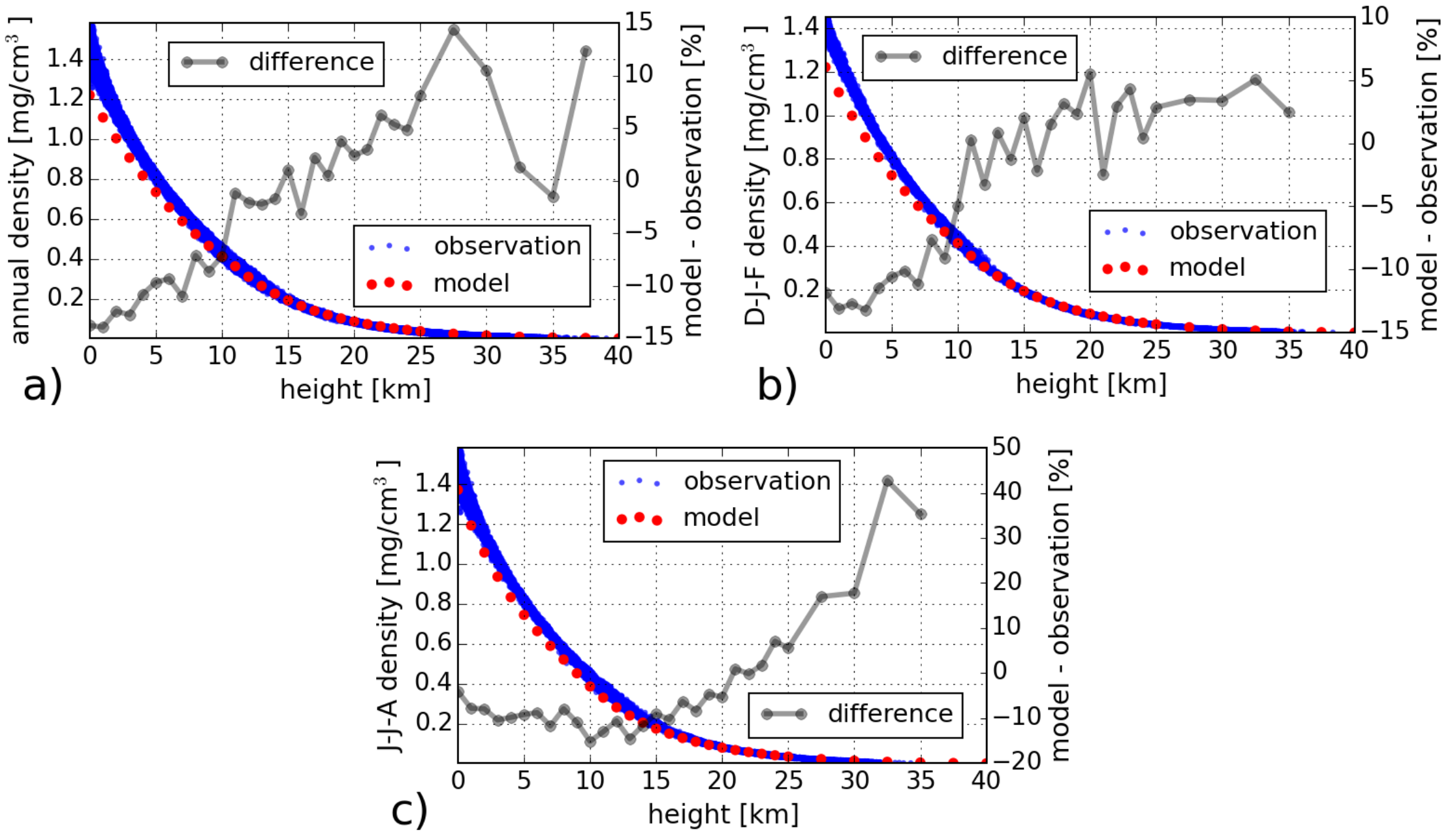

**Figure 5:** A comparison of atmospheric density models from MODTRAN with the observed with baloons at Marambio base. The atmospheric MODTRAN models compared here are: a) the standard one, b) the subartic summer, and c) the subartic winter.

A comparison between the characterization that we obtain and the ones frequently used in simulations performed with CORSIKA, is presented in Figure 5. In particular, we compare our results with three pre-defined atmospheric (MODTRAN) models of CORSIKA: a) 'Standard', b) 'Sub-Artic Summer', and c) 'Sub-Artic Winter'. We make the comparisons using appropriate different seasons. We find that the profile for the 'Sub-Artic Winter' model deviates significantly from the local profile observed (between 10% and 40% for altitudes larger than 25km). However, the deviation of the density profiles from the 'Standard' and 'Sub-Artic Summer' models, respect to the ones observed with the baloons of SMN at Marambio, is not so high. However, the found differences can affect a proper simulation. In particular it differs in less than $\sim$ 10% for the 'Standard' and even less (less than that 5%) for the 'Sub-Artic Summer' model. We notice that the larger differences correspond to larger altitudes. And in general the models lightly underestimate the observations at lower altitudes (lower than $\sim$ 10 km) and make significantly overestimations at altitudes higher than $\sim$ 15 km, where details of the shower simulations are crucial. The presented results can help to take the decision of what pre-set model to use, but also to construct a new one for Marambio, based on local observations. So that, they will improve the quality of numerical simulation for cosmic rays showers at the atmosphere at Marambio.

## 5. Summary and Conclusions

In this work we presented the project of the LAGO collaboration for installing WCDs in the argentine Antarctica. We also presented some studies for this site that will allow a better interpretation of the acquired data. Our results quantify asymptotic directions of primaries for different rigidities, and also rigidities cutoff for different geomagnetic conditions. We presented a study of the typical atmospheric conditions for this site using baloons observations, and compared them with the MODTRAN models commonly used in CORSIKA, quantifying relative differences. We conclude that the 'Sub-Artic Summer' model is the most appropriated one at Marambio for the months December, January and February. The results we found here will improve the development of showers numerical simulations we plan to do in the near future in the Antarctic node of the LAGO collaboration, getting a better simulation of these secondary showers. In particular, we plan to develope showers simulations using new models based on these balloon observations.

It is a unique moment for making combined studies of astroparticles flows, combining onboard spacecraft detectors (for nucleons with energies between a few tens of keV to 100 MeV per nucleon, e.g., instruments aboard the ACE, SOHO and SAMPEX probes) with ground particle detectors (covering energies from $\sim$ 1GeV up to $\sim$ 1 TeV, e.g., neutron monitors, muon telescopes and WCDs from the LAGO and Auger observatories).

While neutron monitors are proportional counters that measure the total number of particles (without energy discrimination), the modern WCDs that will be installed at Antarctic are able to measure the histogram of the energy deposited by the secondary particles observing only at a single geographic localization. These histograms can be re-interpreted in function of the energy of the primary cosmic rays from numerical simulations, and thus for instance will be possible to measure for the first time the effects of the passage of a magnetic cloud on the energy spectrum of galactic cosmic rays at the terrestrial environment.

The new site of LAGO in Antartic will be able to observe two particular classes of physical events (Forbush drecreases and GLEs), which are of major interest in space weather.

These combined studies are complementary and will make it possible to achieve a deeper understanding of the space environment. They will contribute to a better comprehension of different effects of the Sun-Earth coupling on the cosmic radiation levels at different levels of the terrestrial environment and also at the ground level of Earth.

## 6. Acknowledges

This work was partially supported by the Argentinean grants PICT-2013-1462 (FONCyT-ANPCyT), UBACyT 20020120100220 (UBA), and PIP-11220130100439CO (CONICET). We acknowledge SMN for providing baloons data for Marambio, and also to Viviana Lopez for helping us to understand its format. The LAGO Collaboration thanks the Pierre Auger Collaboration for its continuous support.

# 4

# High Energy Astrophysics

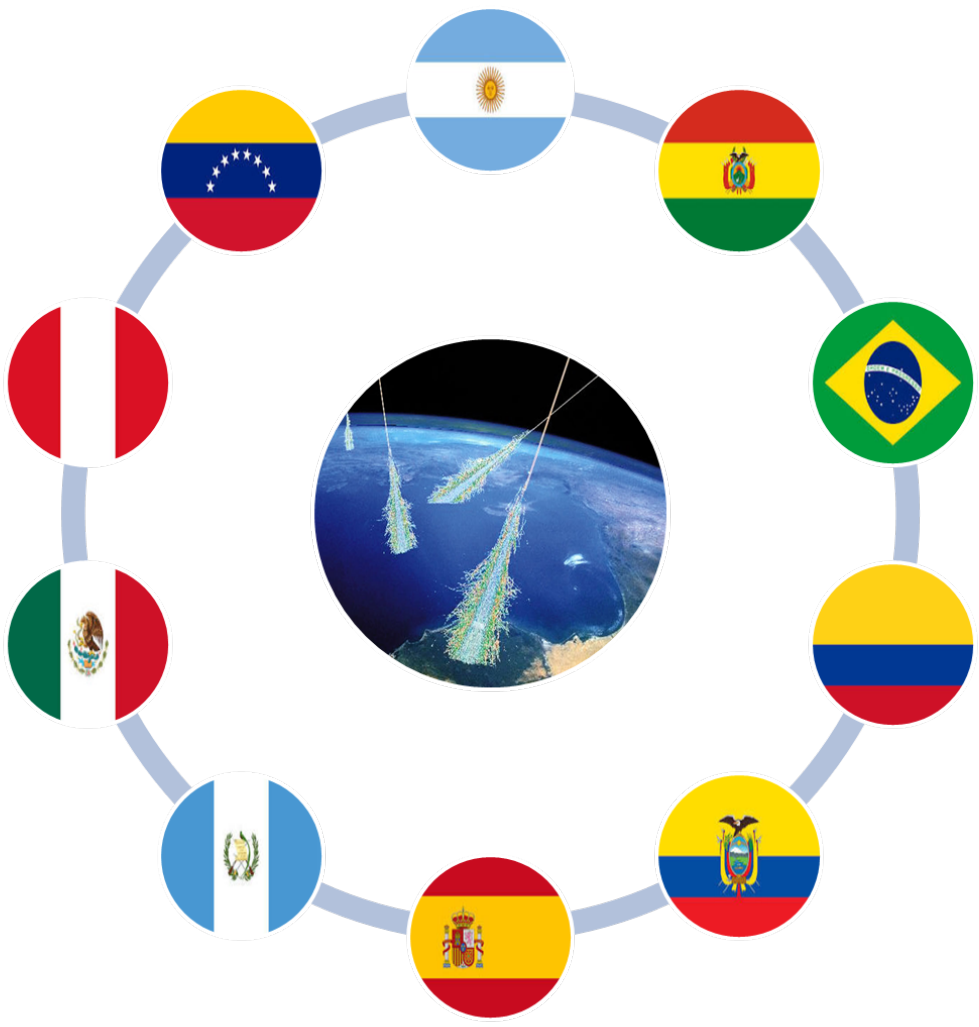

1

# Analysis of Background Cosmic Ray Rate in the 2010-2012 Period from the LAGO-Chacaltaya Detectors

**H. Asorey[1,2,3], P. Miranda[5], A. Núñez-Castiñeyra[4], L.A. Núñez[3,4], J. Salinas[5], C. Sarmiento-Cano*[3], R. Ticona[5], A. Velarde[5], for the LAGO Collaboration[6]**

[1] *Laboratorio Detección de Partículas y Radiación, Instituto Balseiro y Centro Atómico Bariloche, S.C. de Bariloche, Argentina.*
[2] *Sede Andina, Universidad Nacional de Río Negro, S.C. de Bariloche, Argentina.*
[3] *Escuela de Física, Universidad Industrial de Santander, Bucaramanga, Colombia.*
[4] *Centro de Física Fundamental, Dept. de Física, Universidad de Los Andes, Mérida Venezuela*
[5] *Instituto de Investigaciones Físicas, Universidad Mayor de San Andrés, La Paz, Bolivia*
[6] *lagoproject.org, see the full list of members and institutions at lagoproject.org/collab.html*
*e-mail:* `lago-pi@lagoproject.org`

The Latin American Giant Observatory (LAGO) is an extended Cosmic Rays observatory composed by a network of Cherenkov Detectors (WCDs) spread over Latin America. This work will report the analysis of three years of data from three LAGO WCD located in Cerro Chacaltaya, Bolivia, at 5200 m a.s.l. Background cosmic ray rate from these detectors is checked for DAQ issues and inconsistencies, and corrected for atmospheric effects. An analysis for short transients up to the minute timescale is performed, in search for coincidence with transients observed by satellites. Sidereal and solar long term epoch data analysis are also presented.

*Speaker.

## 1. Introduction

Due to the atmospheric absorption, it is very difficult to directly detect low energy cosmic rays or photons in 10 GeV $-$10 TeV energy range from ground level. Very large detection areas or very high altitude sites are currently used around the world to complement direct measurements made on board of satellites. Since 2006, the LAGO (Latin American Giant Observatory, formerly Large Aperture Gamma Ray Burst Observatory) project [1] has been developing and operating sites at high altitude montain in Latin America [2].

Nowadays, the LAGO Project is an extended Astroparticle Observatory at a continental scale [3], mainly oriented towards developing astroparticle physics at Latin America and doing basic research in three areas: search for the high energy component of GRBs at high altitude sites, Space Weather phenomena, and Background Radiation at ground level [4]. The LAGO detection network consists in particle detectors deployed at ground level, spanning over different sites located at significantly different latitudes (currently planned from Mexico down the Antarctic region) and different altitudes (from sea level up to more than 5000 meters over sea level), covering a large range of geomagnetic rigidity cut-offs and atmospheric absorption/reaction levels [5]. The current distribution and status of the LAGO detection network in the American continent is shown in figure 1. This network of detectors is operated by the LAGO Collaboration, a non-centralized and distributed collaborative network of more than 80 scientist from institutions of nine Latin American countries (currently Argentina, Bolivia, Brazil, Colombia, Ecuador, Guatemala, Mexico, Peru and Venezuela) and Spain. Due to its proved reliability, high detection efficiency to all components present in atmospheric extensive showers, and low cost, water Cherenkov detectors (WCD) are currently used at LAGO sites [3].

In this work we show a new analysis of the scalers rates (an implementation of the single particle technique [6]) of the LAGO WCDs installed at Mount Chacaltaya (section 2). This analysis is conducted for the search of short transient events ($\Delta t \lesssim 1$ minute) as we show in section 3). Finally, in section 4 we show our epoch analysis in solar and sidereal time for the search of periodic signals.

## 2. LAGO at Mount Chacaltaya

The WCD of the LAGO project in Bolivia are located at Mount Chacaltaya (16°21′00″S, 68°07′53″W), at 5270 m above sea level. In one of the main buildings of the Chacaltaya Astrophysical Observatory, three LAGO WCD (two of them with a detection area of 4 m$^2$ and the third of 2 m$^2$) register the flux of secondary particles as a function of time. Each WCD is buit by using a commercial water tank filled with purified water, and a single photomultiplier tube located in the center of the detector roof is submersed in the water volume. An internal coating made from Tyvek assure the diffusion of the Cherenkov radiation produced by the passage of ultra-relativistic charged particles trough the detector volume. This setup improve the detection efficiency by diminishing the dependence with the direction of propagation of the secondary particles inside the detector.

The spatial configuration of the array can be seen in the right panel of figure 1. The instrumented area and the high altitude of this site provide enough sensitivity to detect the high energy component of Gamma Ray Bursts (GRBs) [2].

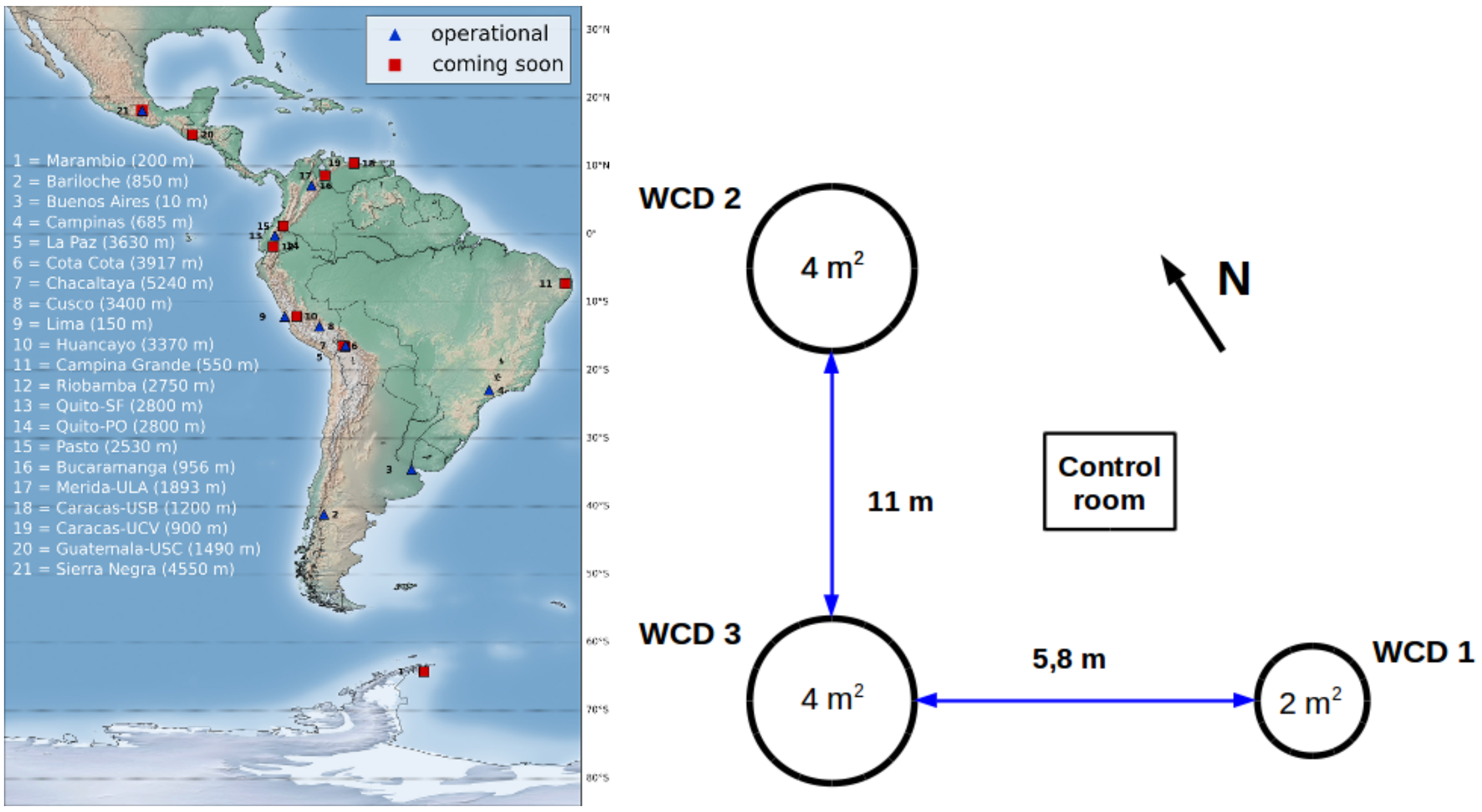

(a) (b)

**Figure 1:** (a): The Sites of the Latin American Giant Observatory, located at Latin America in eight countries. The first detectors are starting in operation (blue triangles), and some other detectors (red squares) are planned to start in operation in 2015-2016. (b): The spatial distribution of the LAGO WCD in the Chacaltaya Observatory. The WCD2 and WCD3 detectors have a detection area of $4\,\mathrm{m}^2$, while the effective area of the WCD1 is $2\,\mathrm{m}^2$.

Two generations of the data acquisition system (DAQ) have been developed and used in the LAGO WCDs. The first acquisition system used [2] was based on an adaptation of the electronics of the first engineering-array phase of the Pierre Auger Observatory [7]. Since Jun 2012 a new LAGO specific electronic system was installed at the Chacaltaya site. Since then this new data acquisition system has been being installed at the different sites of the LAGO project [5].

By using the first electronic system, in the period 2010-2012 the Chacaltaya station has collected data during $\sim 17000$ hours of detection for WCD1 and WCD2, and $\sim 15700$ hours for WCD3. Recent studies [8], based on CORSIKA [9] simulations, show that the angular aperture of the Chacaltaya site can be extended up to a zenith angle of $25^{\circ}$ in the energy range of interest for GRB and other Gamma originated signals at Chacaltaya. Combining these result with the uptime of each WCD at Chacaltaya in this period, the total exposure of this site accumulated during the 2010-2012 period was $2.7 \times 10^{8}\,\mathrm{m}^2\,\mathrm{s\,sr}$.

### 2.1 Data provenance of Chacaltaya LAGO data

The first data acquisition board can control up to three independent WCD and six independent signal channels. Usually, the last dynode and the anode of the PMT of each WCD are measured. Four different signals thresholds are set for each channel, three of them are for fixed values above the baseline and correspond to different limits in the deposited energy ($E_d$) inside the detector by the registered secondary particles. The first one of these sub-channels records particles with $E_d \lesssim 5\,\mathrm{MeV}$, the second sub-channel counts particles that deposited between 5 MeV and 10 MeV within the water volume, while the third one is for signals with $E_d \gtrsim 50\,\text{MeV}$ [10]. The fourth sub-channel is intended to register the occurrence of undershoot pulses at the PMT, which are associated with signal saturation, high frequency noise due to lightnings at the site and other possible noise electronic sources. The trigger condition is that the signal amplitude should be above (or below for sub-channel 4) the corresponding threshold.

The sampling frequency in the first LAGO DAQ is of 200 Hz, corresponding to time bins of 5 ms. The data is recorded in plain ascii compressed files containing one hour of data, with a single row per temporal bin with 25 columns: the first 24 columns account for the counting rates of each threshold for each channel, and an extra column is used to provide timing information from the GPS module on board the station. A sample of one of the 1-hour data files is shown in Figure 2. A very simple check of the quality of the acquisition procedure is to verify that every GPS should contain 200 lines, and so every 1-hour file must contain 720,000 lines and 3,600 different GPS labels.

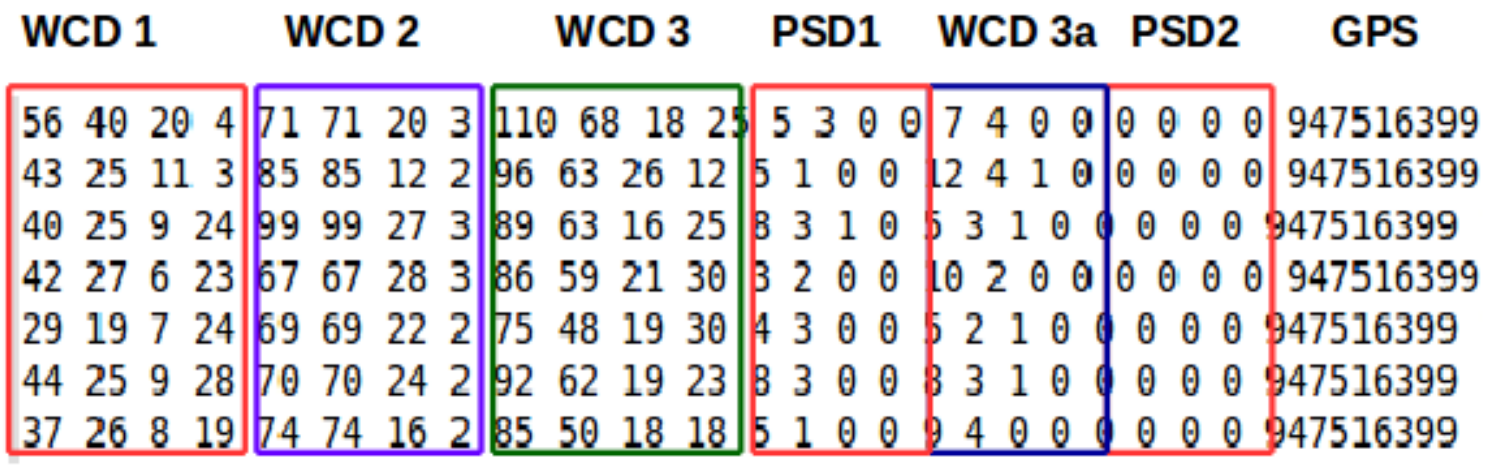

**Figure 2:** A 35 ms sample of a data file collected in the Chacaltaya station, where all six channels are operative. Each signal channel is represented by a four columns block, corresponding to the four sub-channels described in the text. Each line correspond to the count of signals that satisfied the trigger condition for each sub-channel within a 5 ms interval. Channels 1, 2, and 3 are connected to WCD1, WCD2 and WCD3 respectively. The remaining channels were used at that time to acquire data from detectors of the INCA project operating also at the Chacaltaya Astrophysical Observatory.

After introduced those simple quality cuts in the acquired data it was found that data collected in 2008 and part of 2009 have a sampling frequency that shifts between 128, 200 and 256 samples per second, this problem was identified and solved in 2010.

Another way to verify that data is being collected correctly is to examine the background behaviour, which variations have known sources, such as the atmospheric conditions at the site and can be corrected. Large deviations from the expected background are carefully analyzed to check for possible signals or inconsistencies. A possible source of this deviations could be the electronics system overload due to multiply programmed parallel tasks on the station board. To check for these issues, we stacked the data from several 1-hour files and found that 5 minutes after the beginning of each hour a sudden decrease in the counting rates occurs. This happens during the transfer of the previous data file from the station to the computer that controls the DAQ system. As the data acquired during this short period of time can't be used, then are labelled as bad data and are not used to perform physics analysis.

These validations and data quality cuts were introduced over all of the data collected in the Chacaltaya station [11] and after that, the surviving data was stored in the LAGO data repositories of the collaboration [12].

## 3. Search for Short Transient Events

Considering the typical variations on the flux at the time scales of interest, we base the first steps of our analysis in the moving window average (MWA) method [13]. Using this method, we look for $\geq |3\sigma|$ instantaneous deviations in the central 5 ms bin on a moving average window of 2 minutes (24,000 time bins) of duration. If such a deviation is observed, the data is tagged as a pontential transient candidate.

After that, we also required that such deviation must be observed simultaneously in at least two operating detectors in the site, and must be present in at least the low and the intermediate deposited energy sub-channels. Finally, an additional criterium is imposed: during the observed excesses in sub-channels 1 and 2 of the triggered detectors, no significant excess must be observed in the noise counting sub-channel. All these criteria are required to discard, for example, signals produced during lightning discharges that could be misinterpreted as potential candidates during the standard $\sigma - \delta$ analysis [2]. If all those criteria are fulfilled, the data is labeled as candidate and is separated for a detailed analysis.

After applying this method and imposed the quality cuts described, over 2 terabytes of data collected in the period 2010-2012 at Chacaltaya, we found a potential candidate, started on Wed Dic 07 15:45:49.675±0.005 UTC 2011 (unix time $(1323272749.675 \pm 0.005)$ s). At this time, the equatorial coordinates of the Chacaltaya zenith were RA/Dec (J2000) $16^{\mathrm{h}}17^{\mathrm{m}}31.3^{\mathrm{s}}/-16^\circ21'00''$, with an acceptance aperture of $\theta \lesssim 25^\circ$.

By integrating the signal shape, we determine a duration of 5.5 s ($T_{95}$, i.e, the time needed to reach 95% of the integrated signal), with a rise time ($T_{50} - T_{10}$) of 0.55 s and a fall time ($T_{90} - T_{50}$) of 3.5 s. This signal was observed on both WCD1 and WCD2 simultaneously in the sub-channels 1 ($E_d \leq 5$ MeV) and 2 ($5 \leq E_d \leq 10$ MeV) and a less significant excess on the sub-channel 3 of both detectors. After a careful examination of the signals shapes, calibration data and operation metadata of our detectors and the atmospheric database of this site, we discarded the possibility that this event was produced by detectors malfunctions, HF noises, electric lightnings or other phenomena of atmospheric origin.

After this blind analysis, we look for coincidences with events registered on both the SWIFT [14] and Fermi[15] satellites and also at the Gamma-ray Coordinates Network (GCN) database[1]. However, at this time the field of view of Fermi was outside of our acceptance cone in Chacaltaya.

## 4. Long Term Epoch Analysis

On the search for periodic signals we performed a series of analyses over large periods of time on the data collected at the Chacaltaya station. Data stacking or summation in two different time systems, solar and sidereal, were made over the data. The idea behind this process is based on the random, poissonian, nature of the majority of the radiation measured by our detectors. If any not random, periodic, signal were to exist, and if sufficient data were to be summed in a scale in which this signal happened at the same time every day or month or week. No matter how small this signal could be it will build an observable rise on the line that represent the summed data, as the random fluctuations will mutually cancel and the signal will grow [8].

[1] http://gcn.gsfc.nasa.gov/

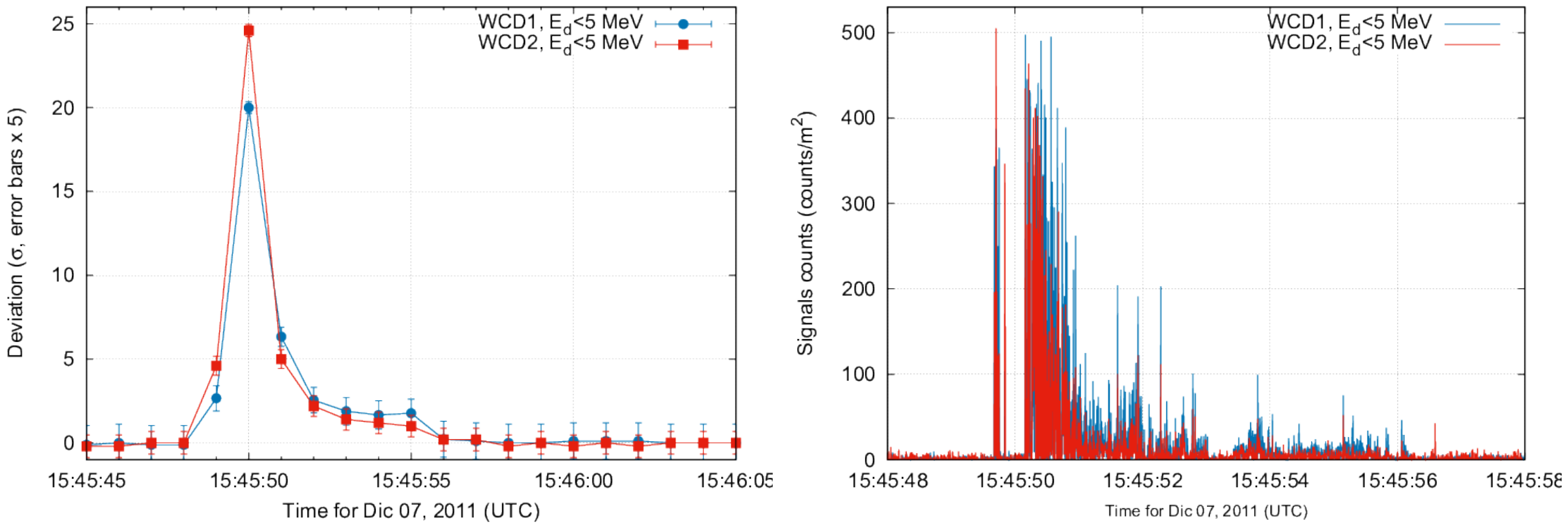

(a) (b)

**Figure 3:** Potential candidate signals for the event registered on Wed Dic 07 15:47:02.378 UTC 2011. In panel (a) we show our alert as described in the text, while in the panel (b) the registered signals in 5 ms temporal bins are shown

This summation of data were made on two different systems: sidereal time, for signals coming from outside of our Solar System and in solar time, for signals modulated by Solar activity. A minute average signals were summed with the average of the corresponding minute of the next day, solar or sidereal. The sudden drop in the counting rate described in the previous section produced a huge peak at every summed hour when it is not corrected. In figure 4 the results of the summation of the data of April 2011 are shown for the three signal related sub-channels of WCD1. A WMA analysis was performed over the stacked data on both time systems, looking for excesses on the summed data that could be associated with the passage of known point sources by the sky at Chacaltaya. In this way, as the single particle technique lacks of directional reconstruction, we are exploiting atmospheric absorption as a source selection tool. Additionally, a method based on CORSIKA simulations was developed to account for the sensitivity of the LAGO detectors at high altitude sites for different gamma sources in the tens of GeV-TeV energy range [8, 16].

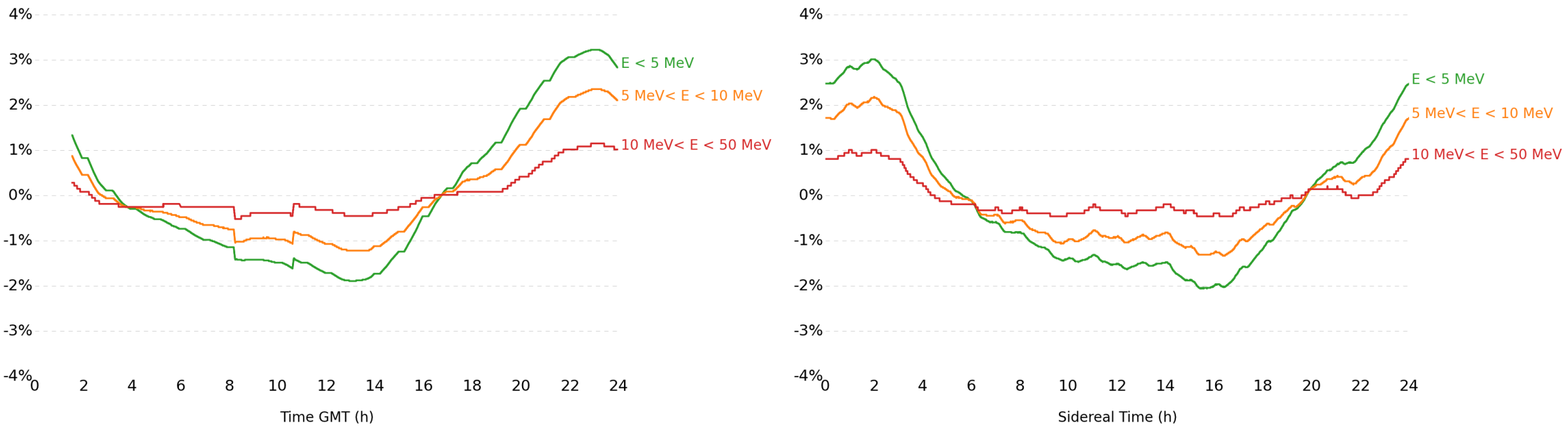

(a) (b)

**Figure 4:** Solar (a) and sidereal (b) summation over the data gathered in the month of April of 2011 for the WCD1 detector at Chacaltaya. The green lines represents the observed modulation in the low energy sub-channel. The orange lines represent the energy threshold corresponding to particles with energies between 5 MeV and 10 MeV and the red lines particles with energy higher than 10 MeV. The vertical axis represent the percentage of change form the middle value of each sub-channel in the solar or sidereal time analysis.

The results of the complete summation on the data collected in the Chacaltaya station in 2011 is show in figure 6 where both solar time and sidereal time are shown here, and have consistent results. In solar time it appears a rise on the counts that holds in the same stage of the solar day while in sidereal time the rise moves about two hour every month. These facts are clear indicators of the solar nature of this phenomenon. The amplitude and phase observed on the solar time variations are consistent with the well known daily modulation of solar origin in the flux of low energy cosmic rays.

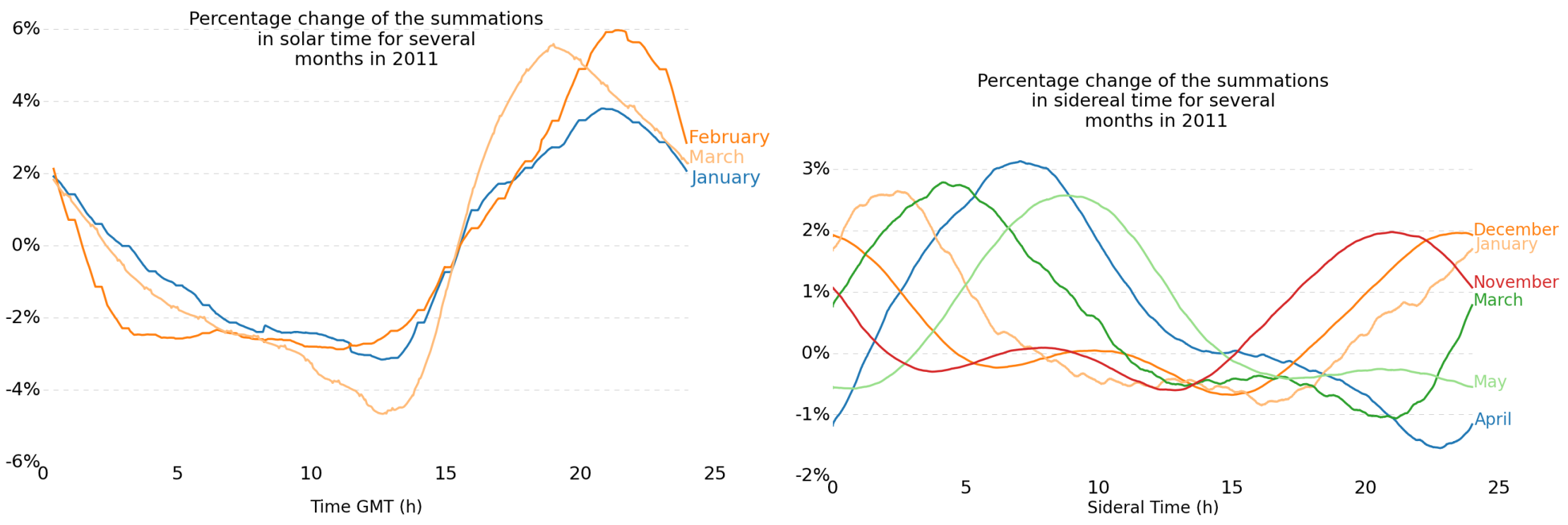

(a) (b)

**Figure 5:** Solar daily modulation in the flux of cosmic rays observed at the LAGO site in Chacaltaya. This phenomena was observed by stacking and summing the measured and corrected fluxes in solar (a) and sidereal (b) times.

This observations confirm that, by using an adapted analysis technique to the characteristics of our small detectors, it is possible to observe different type of signals of solar nature at different time scales from the ground level in the LAGO network of WCDs across Latin America.

**Acknowledgements:**

The LAGO Collaboration is very thankful to the Pierre Auger Collaboration for its continuous support. The authors of this paper gratefully acknowledge the financial support from CDCHT-ULA project C-1598-08-05-A and Vicerrectoría Investigación y Extensión Universidad Industrial de Santander.

# Development of a High Altitude LAGO Site in Peru

**L Otiniano [1], S Vargas*[2], F Quispe[1], W Guevara[1], For The LAGO Collaboration[3]**

[1] *Dirección de Astrofísica Comisión Nacional De Investigación y Desarrollo Aeroespacial, Lima, Perú*

[2] *Escuela Politécnica Nacional, Quito, Ecuador*

[3] *lagoproject.org, see the full list of members and institutions at lagoproject.org/collab.html*

*E-mail:* lotiniano@conida.gob.pe

The Latin American Giant Observatory (LAGO) Project is an extended Cosmic Ray Observatory mainly oriented to perform basic research in three branches: high energy phenomena, space weather and atmospheric radiation at ground level. To observe the high energy component (over 10 GeV) of Gamma Ray Bursts (GRBs), the LAGO Collaboration is installing Water Cherenkov Detectors ) in high altitude sites.

Extensive Air Showers produced in the atmosphere by Gamma Ray Bursts high energy photons could be detected by WCD arrays given their good sensitivity to secondary photons and other particles in the cascades, by looking for excesses over the secondary particle flux. In this work the current developments to build and characterize a high altitude (>4600 m a.s.l.) LAGO site at the central highlands of Peru are described.

*Speaker.

## 1. Introduction

One of the LAGO international collaboration goals is to observe high energy component (over 10 GeV) of Gamma Ray Bursts (GRBs) in arrays of Water Cherenkov Detectors (WCD) at high mountain sites [1], using the single particle technique [2]. When a GRB occurs, the atmosphere is impacted by a multitude of high energy gamma rays that produces decay showers. Together those showers could be detectable at ground level as an instant excess in the flux measured by a detector. Showers initiated by gamma rays have a composition of 90% photons, 9% electrons and $< 1\%$ muons, WCDs are sensitive to all those particles. These showers are absorbed high in the atmosphere, not being able to survive and reach the ground. In order to detect the greatest possible number of particles, the WCDs must be placed at high altitude.

In Peru a new array of the WCDs for the LAGO project is under development. In this article we describe the site chosen for this purpose and the detector prototype, then we describe a GEANT4[3] simulation of the detector and his validation using data of the prototype. Finally we discuss the perspectives for the installation at high altitude of the array at the top of Huaytapallana's mountain at 4800 m.a.s.l.

## 2. Huancayo Site and Experimental Setup

The Huayao Magnetic Observatory (OMH) is located in the Central Sierra of Peru (12"02.7′N 75"20.4′W) on a vast plain at 3300 m.a.s.l.. Since 1936 it has hosted secondary cosmic rays ground experiments (a Forbush Sphere and Neutron Monitor), currently makes geomagnetic and meteorological records, monitoring of solar activity and has a seismic station[4]. For monitoring detectors and storage of materials a mobile station has been built using steel containers.

The LAGO WCD is commonly made of a water commercial tank, filled with filtered and chemically purified water, a external coberture for light isolation and an internal liner of Tyvek, to reflect and diffuse Cherenkov photons produced in water by high energy secondary cosmic rays. The diffusion of Cherenkov photons reduces the signal dependence with the secondary particle trajectory within the detector. A large area photomultiplier PMT complete the detector[5]. Our previous experience in high altitude deploy of WCDs in Peru [6] has shown the need to minimize the size and weight of the detectors components in order to facilitate and optimize cost of deployment. Thus we developed a prototype made of a cylindrical surface (3 m diameter 1 m height) assembled using four stainless steel plates that support two cylindrical bags filled with treated water, the external bag, made of high density polyethylene, provides light isolation, the internal bag is made of Tyvek. All the detector is also covered with an external coating of plastic blanket for environmental and also light isolation. The detector is completed with an upper 9354KB Electron Tube (PMT), see Figure 3.

The typical signal produced by the detector being crossed by a particle is a negative pulse with a fast decay and an exponential recovery, the pulse amplitude is 500 mV and width of 150 ns. The LAGO project has development its own DAQ system with three channels at $40 MHz$ sampling rate 10 $bits$ digitalization (ADC unit $\approx 1\ mV$), time stamp via GPS signal, sampling of pressure and temperature each second and control for high voltage powers supply needed to operate and record the signal of PMTs[5].

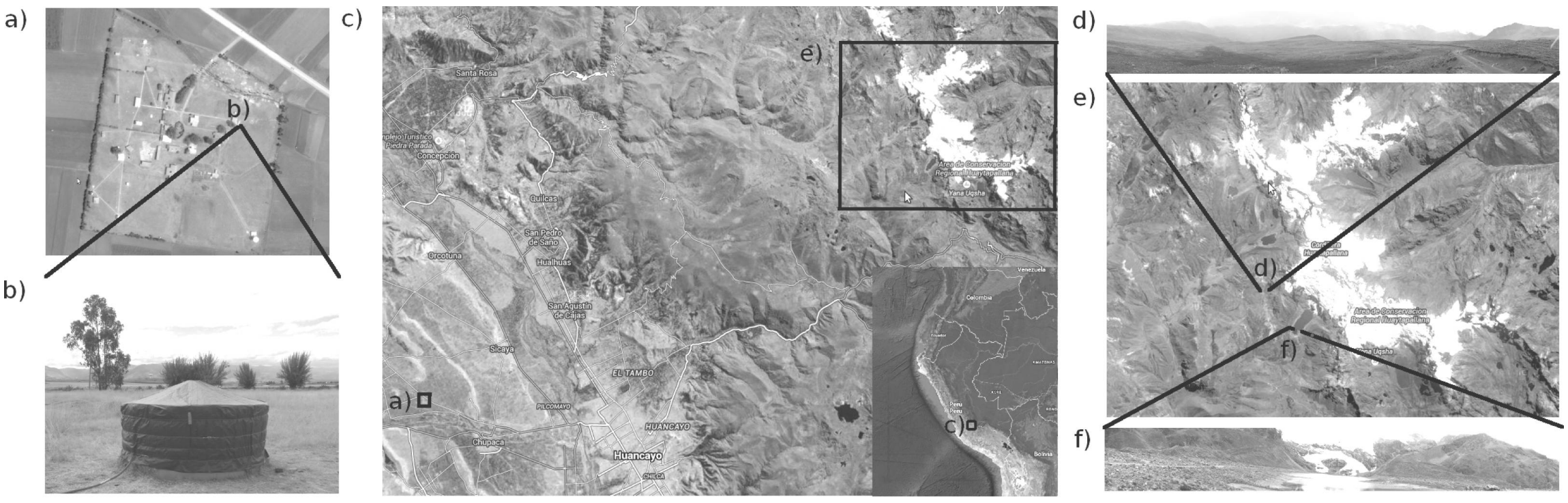

**Figure 1:** Huancayo site location: 30 km from the city of Huancayo (c) the Huayao Magnetic Observatory (OHM) (a) houses our WCD prototype (b). Nearby there are two sites competing in the Huaytapallana mountain to accommodate an array of detectors. One at. 4600 m.a.s.l. having a nearby lagoon (f), the other at 4800 m.a.s.l. with no nearby water sources, but with easy access by road (d).

## 3. Detector Background Simulation Validation

The LAGO collaboration has developed CORSIKA [7] based tools to simulate radiation environments of secondary cosmic rays[8]. Corrected for geomagnetic effects by cutting the spectrum of primaries reaching the atmosphere. A campaign for simulations of all sites of LAGO with WCDs is on development. Figure 2 shows the result of simulation spectrum of secondary cosmic rays for the OMH at 3300 m.a.s.l. and for the same coordinates but at 4800 m.a.s.l.

In order to calibrate the response of WCDs to this radiation environments, a GEANT4 detector simulation has been developed. This simulation must be versatile enough to include all models of WCDs used in the collaboration, which vary principally because of the local materials used in manufacturing on each site to reduce costs. So different geometries due to the use of water tanks acquired in each country, different types of PMTs and different water treatments should be taken into account [9].

The simulation uses a input table that allows to vary the geometry of the cylindrical water tank, the size and type (by varing the quantum efficiency model) of the PMT and the mean free path of photons in water (a percent of the mean free path in pure water, $0 < \alpha < 1$ parameter). Also a model of the tyvek reflection is implemented based on [10] with a total reflection parameter implemented ($0 < \beta < 1$).

In order to calibrate the WCD we tune up unknown parameters ($\alpha$ and $\beta$) to adjust the pulse shape signal response of the prototype WCD measured with the DAQ produced simulated vertical muons reaching the detector from top. Then we simulate the response of the detector to the secondary cosmic rays generated using CORSIKA simulation toolkit [8]. A comparison between these simulations and the spectrum measured with the detector of the OMH prototype is shown in Figure 3, we can notice that we are missing the low energy spectrum of gammas that reach the detector, this is probably due by the high level of trigger we are using for the pulses, 100 ADC units ($\approx 100mV$) to avoid noise.

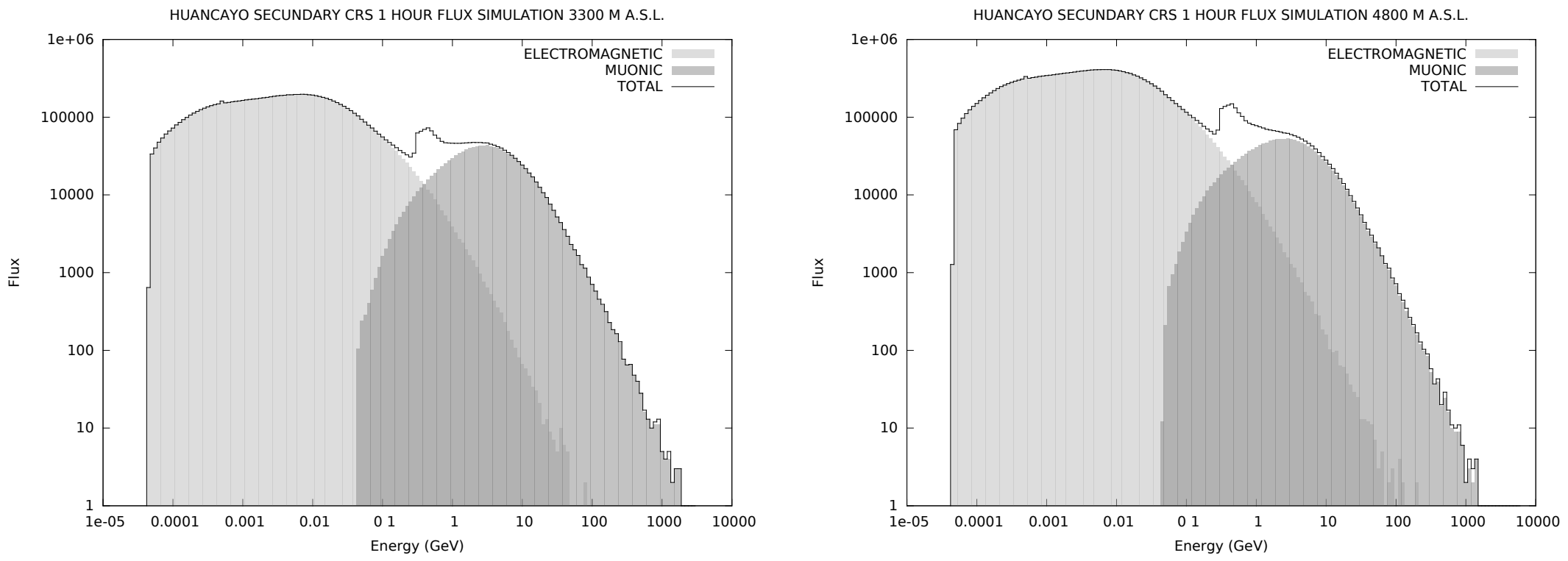

**Figure 2:** Simulated secondary cosmic ray spectrum in the location of the OMH at 3300 m.a.s.l. (left) and 4800 m.a.s.l. (right). In both we show the main components of the flux (electromagnetic and muon). Note also the expected increased electromagnetic component at higher altitude.

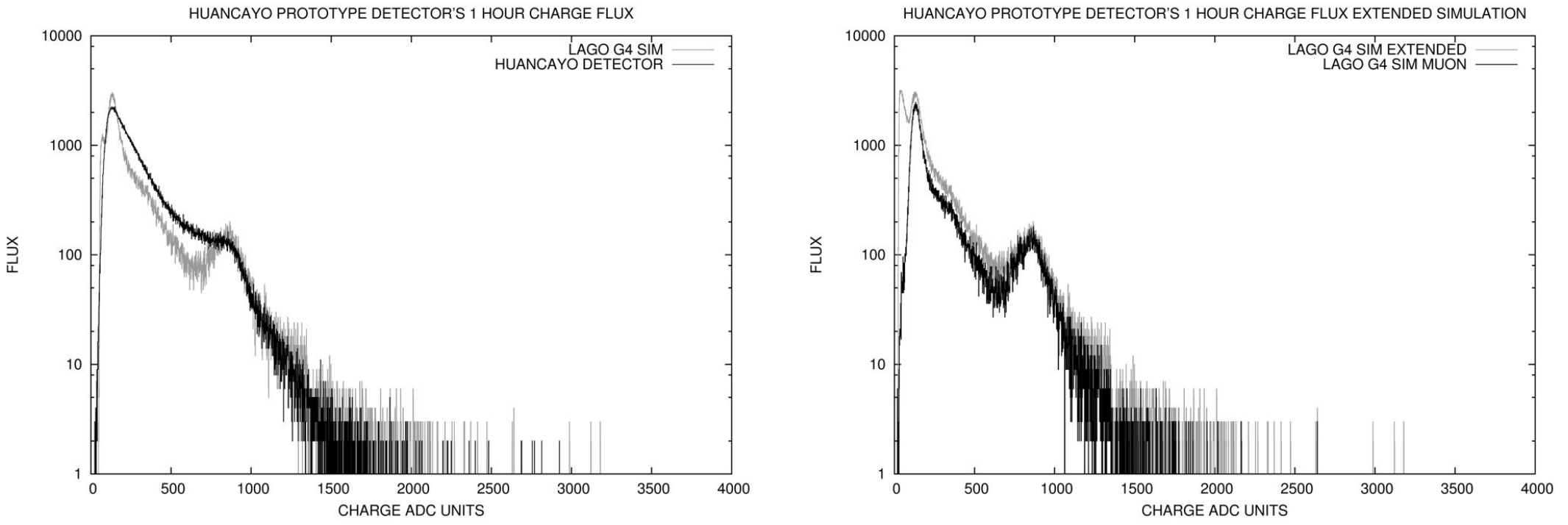

**Figure 3:** Simulated and measured charge spectra for the prototype WCD in the OMH. Left: Comparison of the spectra. Right: simulated muon spectrum component and the spectrum expected with no trigger level.

## 4. Discussion and Conclusions

We are planning the installation of an array of WCD at 4600 or 4800 m.a.s.l. in central Andes of Peru, for the LAGO project. In the coming months two more detectors will be added to form a triangular array in order to test coinciding acquisition mode.

The detector is in process of characterization using simulations and we have developed a protocol and a series of logistics process to easily install it. However, it is necessary to refine the simulation, in particular identifying the Vertical Equivalent Muon[1] (VEM ) on the measured load spectrum in the WCD pulses signal. Besides, it is necessary to amplify the signal to observe the electromagnetic spectrum at low energies, the simulation can be used to study two strategies to do that: adding a wavelength shifter in the water of the detector to increase the number of photons detected or including an electronic preamp. We also hope that the simulator allows us to estimate the response to the flux of high energy gamma generated at ground level by GRBs.

[1] Signal produced by vertical muons that cross the WCDs. The angular distribution of secondary cosmic rays arriving at ground has its maximum in vertical, so this is reflected in the measured flow of the detector, see[11].

## 5. Acknowledgments

The authors gratefully acknowledges the financial support from Fondo para la Innovación, Ciencia y Tecnología (FINCyT) trough the project PIBA-2-P-020-14. The LAGO Collaboration is very thankful to the Pierre Auger Collaboration for its continuous support.

# Calibration of large water-Cherenkov Detector at the Sierra Negra site of LAGO

**A. Galindo**[a], **E. Moreno**[b], **E. Carrasco**[a], **I. Torres**[a], **A. Carramiñana**[*a], **for the LAGO Collaboration**[c]
[a] *Instituto Nacional de Astrofísica, Óptica y Electrónica, México*
[b] *Benemérita Universidad Auntónoma de Puebla, México*
*E-mail:* alinegalindo@inaoep.mx
[c] *lagoproject.org, see the full list of members and institutions at lagoproject.org/collab.html*
*e-mail:* asoreyh@cab.cnea.gov.ar

The Latin American Giant Observatory (LAGO) is an international network of water-Cherenkov detectors (WCD) set in different sites across Latin America. On the top of the Sierra Negra volcano at 4530 m a.s.l., LAGO has completed its first instrumented detector of an array, consisting of a cylindrical WCD with 7.3 m in diameter and 1 m of height, with a total detection area of 40 $m^2$ and sectioned in four equal slices. Each one of these slices is instrumented with an 8" photo-multiplier tube installed at the top of the detector and looking downwards. The final setup will have three WCD as the one mentioned, distributed in triangular shape and one WCD with 7.3 m in diameter and 5 m of height located in the centre. The data acquisition of this first WCD started in June 2014. In this work the full calibration procedure of this detector will be discused and the preliminary measurements of stability in rate.

*Speaker.

## 1. Introduction

The Earth is constantly bombarded by particles, cosmic rays and photons, originated in outer space. These particles can be effectively detected in indirect mode on the Earth's surface by collecting the secondary particles generated from the interaction of primary particles at the top of the atmosphere. The collisions give rise to Extensive Air Showers (EAS), integrated by hadronic, electromagnetic, muon and neutrino components. If the primary is a photon, the electromagnetic component is dominant. The electromagnetic and muon components are identified with ground detectors, the hadronic component dies away soon after the primary collision transforming into other components and neutrinos do not interact enough to be detected.

Water Cherenkov detectors (WCD) are successfully used to detect secondary particles. They travel at relativistic velocities, leaving behind Cherenkov light easily detectable by photomultiplier tubes (PMT). The Latin American Giant Observatory uses WCD in order to detect gamma ray bursts (GRBs) using the single particle technique, which looks for increases in the background rate on all detectors in a short period of time due to the arrival of many photons during a burst [1]. LAGO detectors are also used to study the solar activity and space weather.

We describe the calibration of the detector, including the PMTs characterization. We present preliminary data of the stability of the rate over time.

## 2. Experimental setup

LAGO is located at the top of the Sierra Negra volcano at 4530 m a.s.l. It consists of three cylindrical aluminum WDCs situated each in a corner of a triangle array, with 1.2 m of height and 7.3 m in diameter. The detectors are divided in fourth by walls and a PMT of 8 " model 9354KBL is placed at 2 m of distance from the detector's center in the central axis of every fourth. The inner surfaces and the walls are covered by a high diffusive and reflective Tivek ® material, the detectors are filled up at 1 m with purified water. An aditional WDC with same diameter but 5.0 m in height will be installed at the barycenter of the array.

## 3. Calibration

### 3.1 Calibration of PMTs

Each PMT was tested against the specifications given by ET Enterprises, to ensure that only PMTs of desired quality are used in the detector. The measurement of the single photoelectron spectrum (SPE) was required in order to determinate the absolute gain of the PMT [2]. This is done by flashing a LED at an intensity such that there is a ratio of 1/10 of signal-noise relation present on the photocathode. A schematic of the SPE setup is shown in figure 1. The aquisition system was conformed by CAEN devices based on VME bus, such as crate VME mod. 8011, a digitizer card mod. V1751, HV supply mod. V6533 and a USB bridge mod. V1718.

To determine a suitable intensity for the light source to achieve the desired ratio, an offline calibration was performed. The light intensity was varied by changing the width of a pulsed signal with 800 mVpp, 1.2 V offset and frequency of 1 KHz. When it was observed one pulsed signal from the PMT per 10 noise signals, the intensity of the light source was fixed, this ensures that any signal observed above the noise is predominantly from SPEs [3]. The voltage supply for the PMT was set at the nominal value given by the manufacturer. The traces taken were processed to obtain charges distributions. A typical SPE distribution can be seen in figure 2.

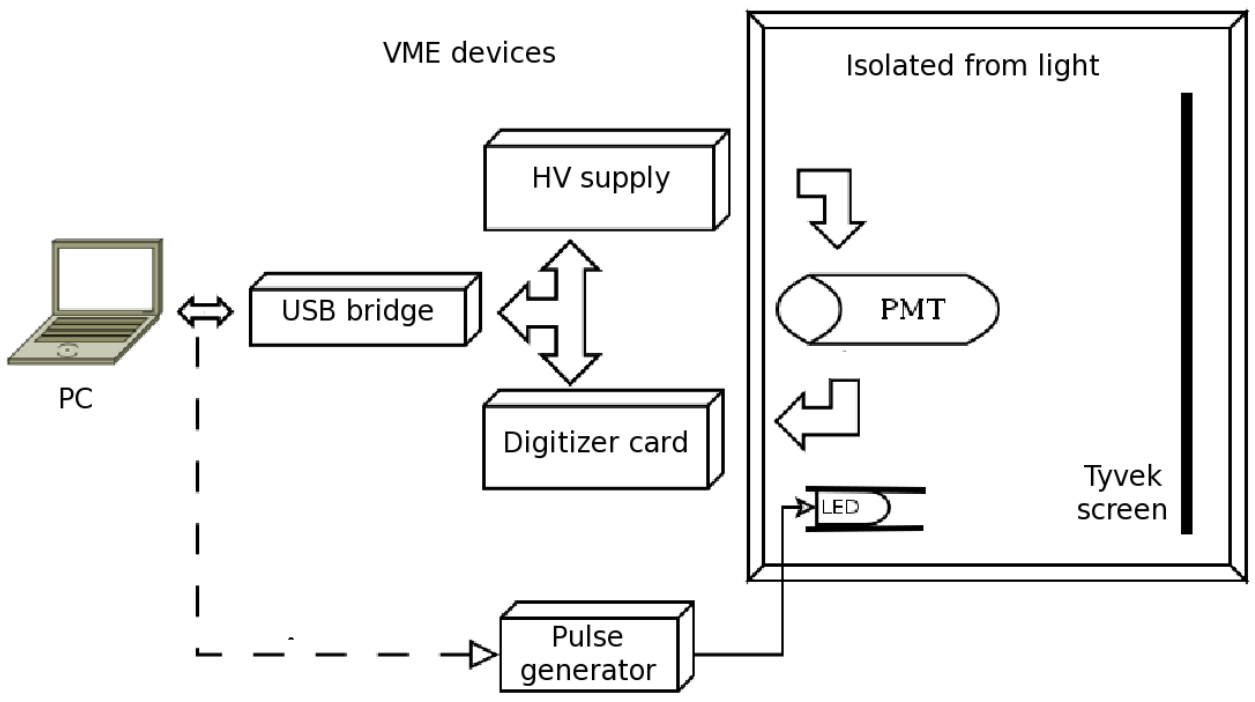

**Figure 1:** Schematic of the SPE setup.

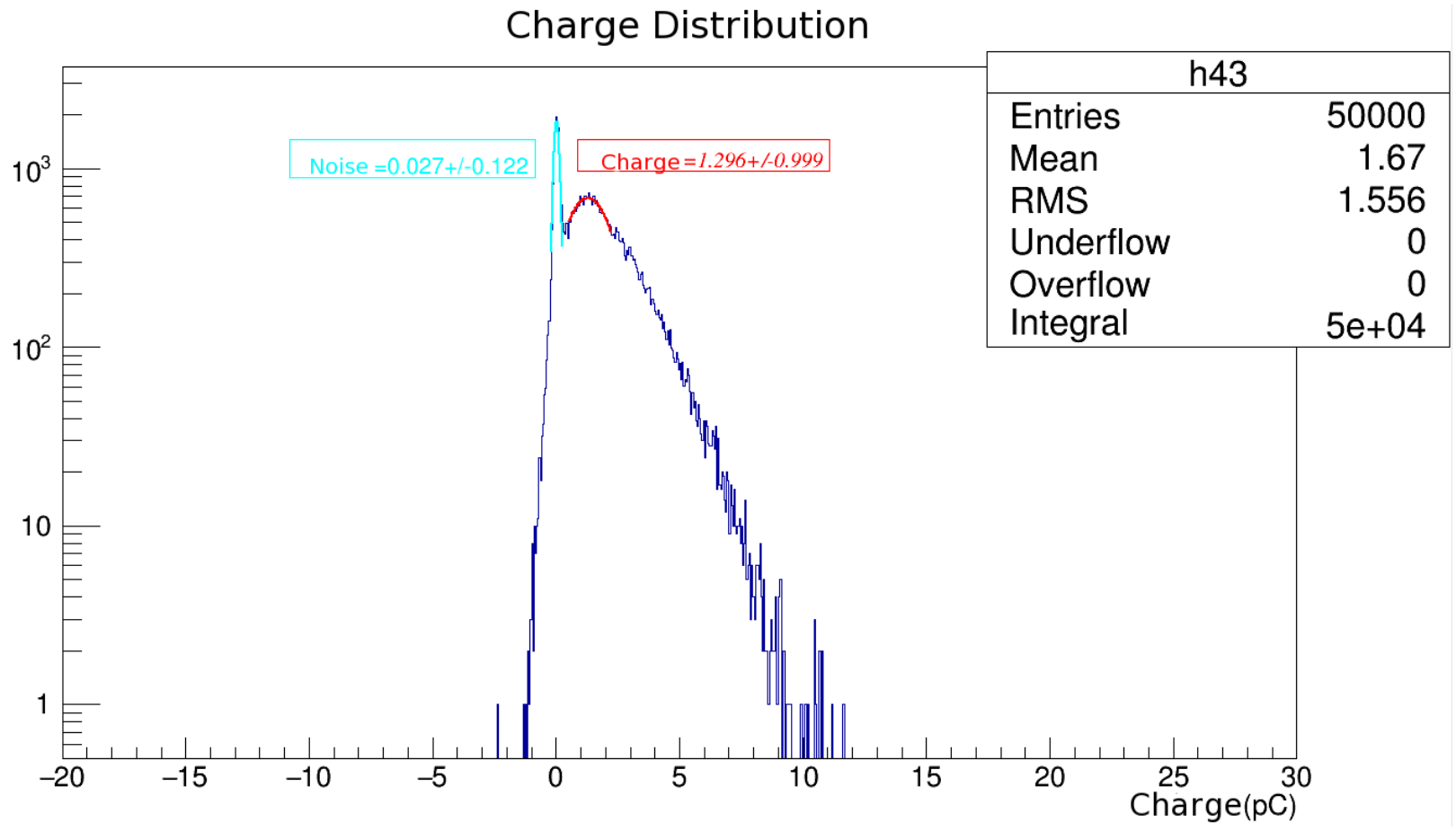

**Figure 2:** Typical single photoelectron distribution. The first peak is the noise and the second peak is due to 1 pe event, taken form the PMT with serial number 7089, feeded with 1500 Volts, 50000 traces.

To calculate the gain it was necessary to find the mean and the deviation of the photoelectron signal directly from the measured. The charge fit was divided by the known value of 1 pe charge, $1.602 \times 10^{-19}$ C [4]. Once the gain is calculated, the absolute gain at that voltage is known. However, it was necessary to have several points to determine the gain as a function of the input voltage, so we obtained distributions with different values of voltage supply. The results of this test for the PMTs working in the site are shown in the figure 3. Now, we have a reference to set a voltage supply for each one in order to have similar gain value in all the PMTs, or at least to have in account the proportionality of gain in the data process.

### 3.2 WCD calibration

A typical way to calibrate WCDs is done by measuring the natural flux of secondary particles through the detector. Nevertheless, calibrating and monitoring a WCD at high altitude is not easy.

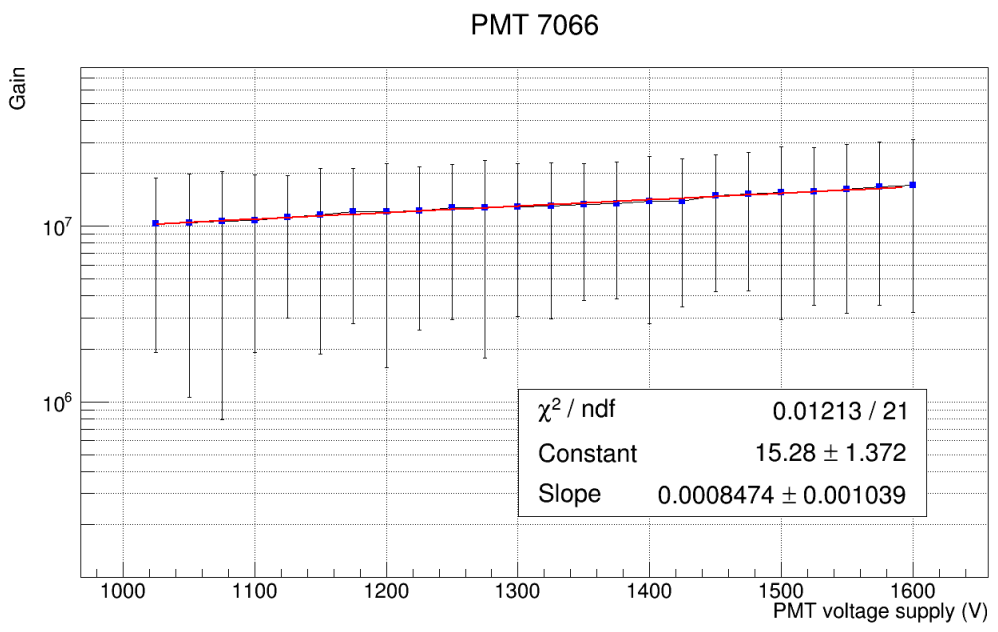

(a) PMT 7066.

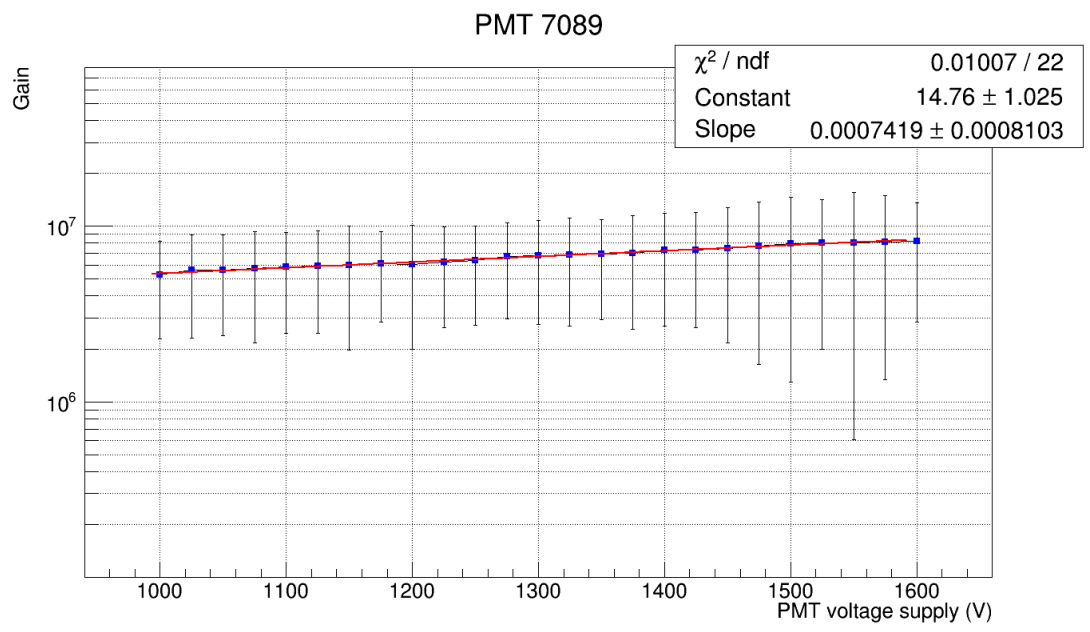

(b) PMT 7089.

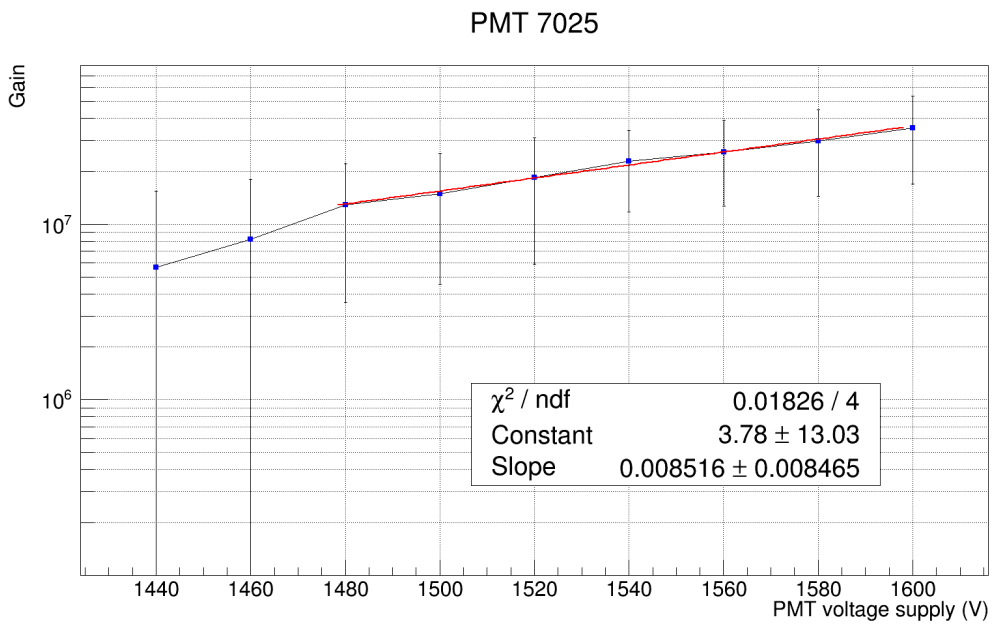

(c) PMT 7025.

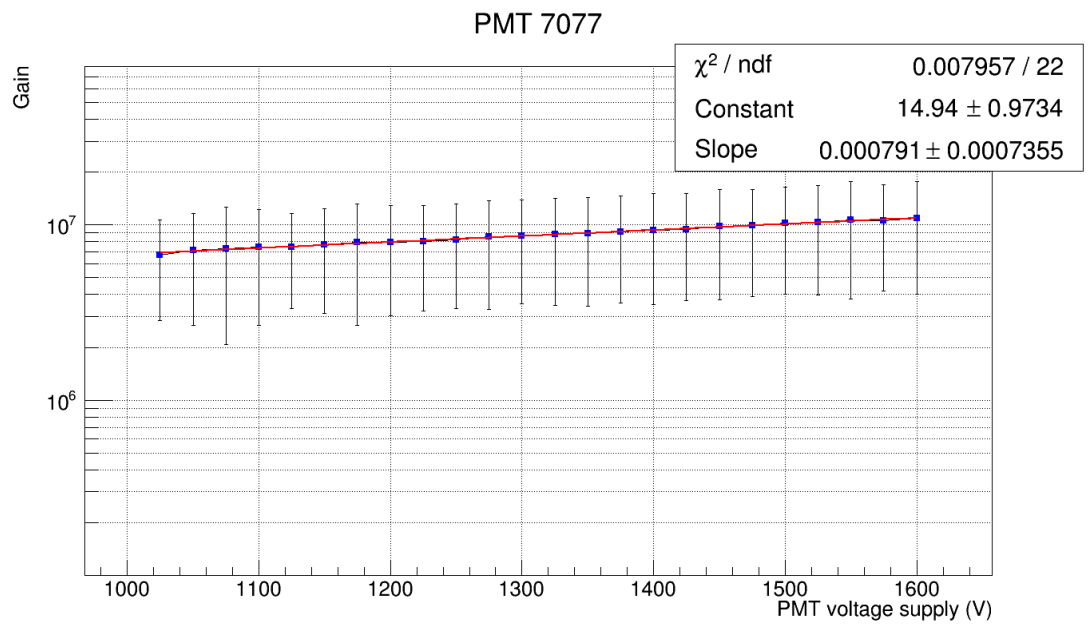

(d) PMT 7077.

**Figure 3:** Curves for the gain as a voltage function for the PMTs tested.

At low altitudes, background muons provides a natural mark easy to detect and calibrate [5], but at high altitude the electromagnetic component dominates over the mounic component.

At Sierra Negra, we need to use the muon peak as a calibration mode. We implemented a data acquisition system with modules NIM and VME in order to have a trigger capable to activate the acquisition only when the 4 PMTs have passed the threshold inside a window time. The setup of this system is shown in the figure 4.

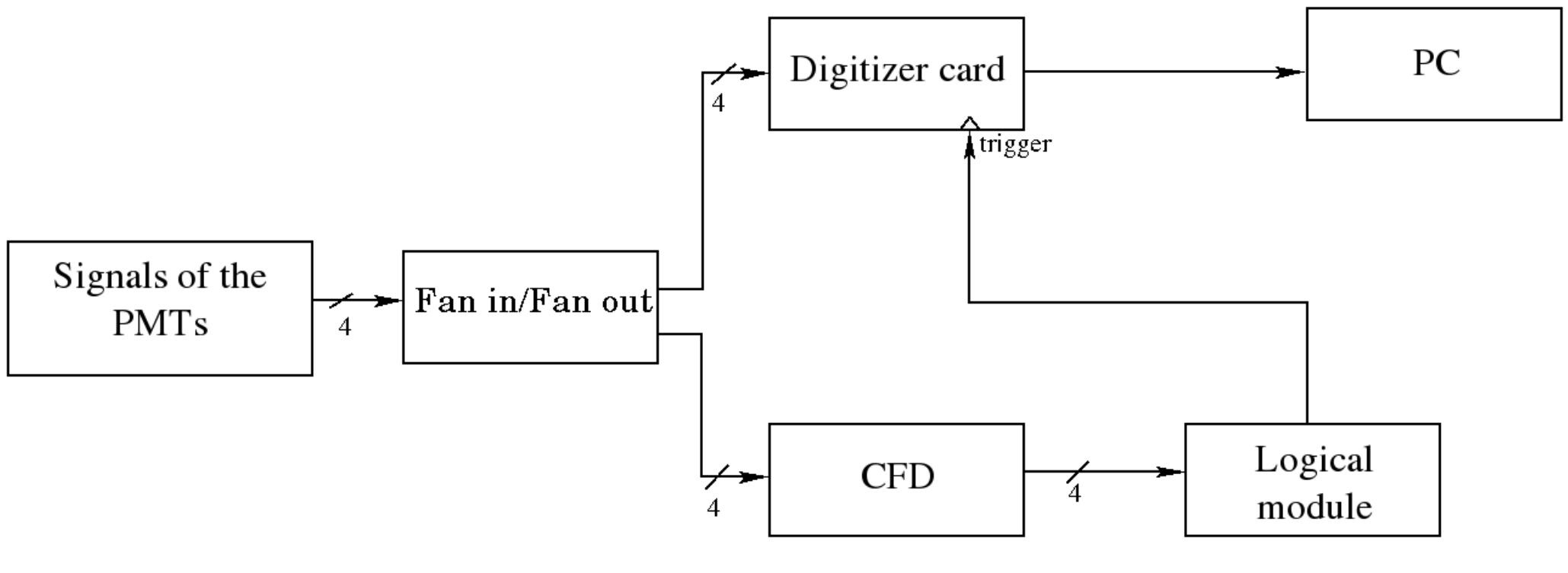

**Figure 4:** Block diagram for coincidence experiment.

The fan in/ fan out module mod.N625, divides the signal for digitizing and for generating the trigger signal at same time. The constant fraction discriminator module mod N842, generates a pulse with fixed width when the signal overtakes a threshold. A logical operation module mod.

V976, makes an AND operation for the four signals with pulses which assures the generation of the trigger for starting the acquisition. The resulting charge distribution of about 7 hours of data, is shown in figure 5.

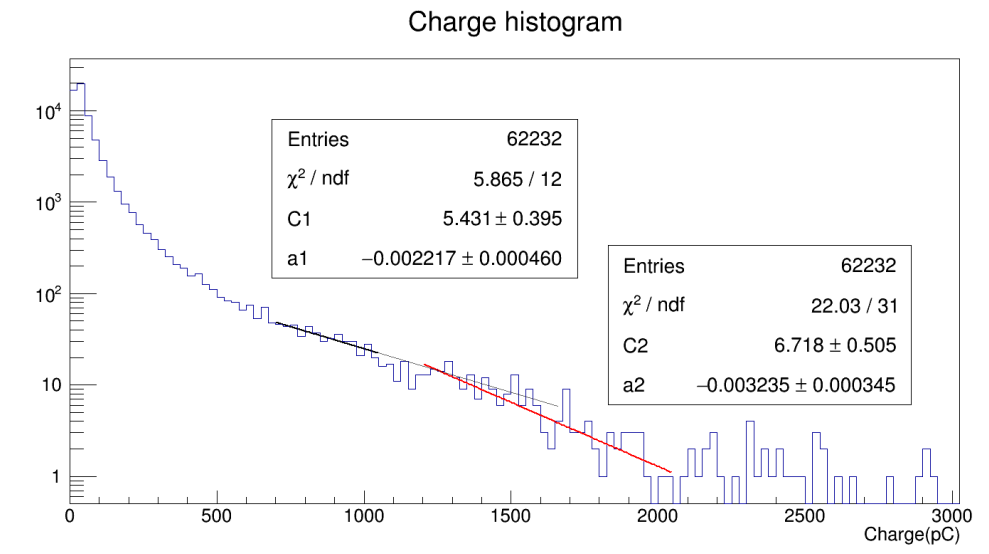

(a) Charge histogram.

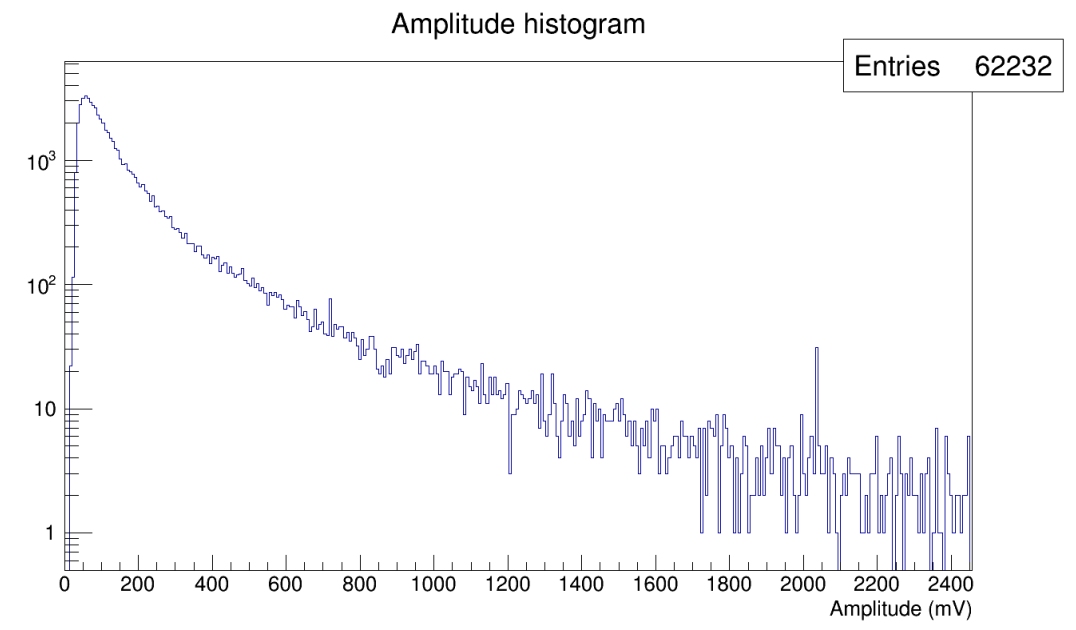

(b) Amplitude histogram.

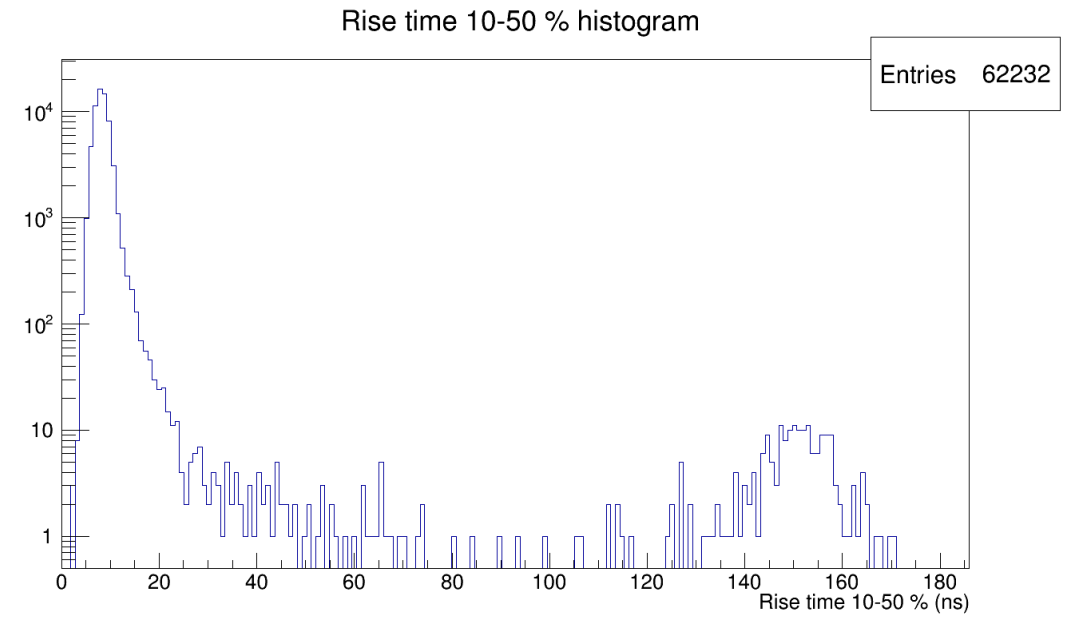

(c) Rise time 10-50 % histogram.

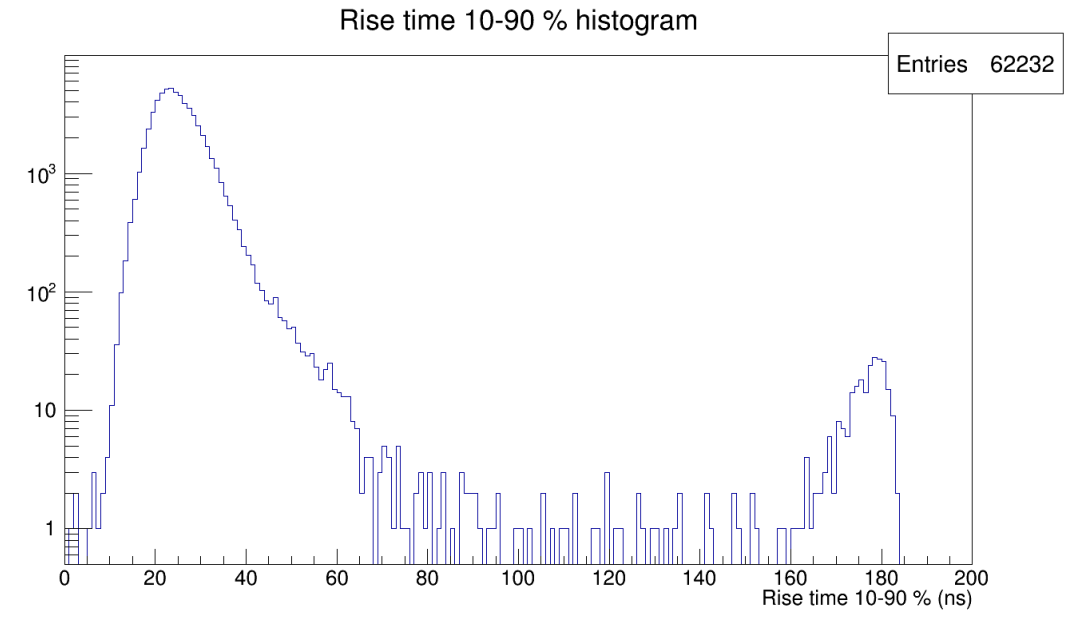

(d) Rise time 10-90 % histogram.

**Figure 5:** Calibration curves for the coincidence setup of the WCD.

A characteristic change in slope can be distinguished on the charge histogram, shown in figure 5 (a), this change can be used for calibration. The intersection of the two fitted lines of this detector can be used as a breakpoint intercalibration to the other two WCDs [6].

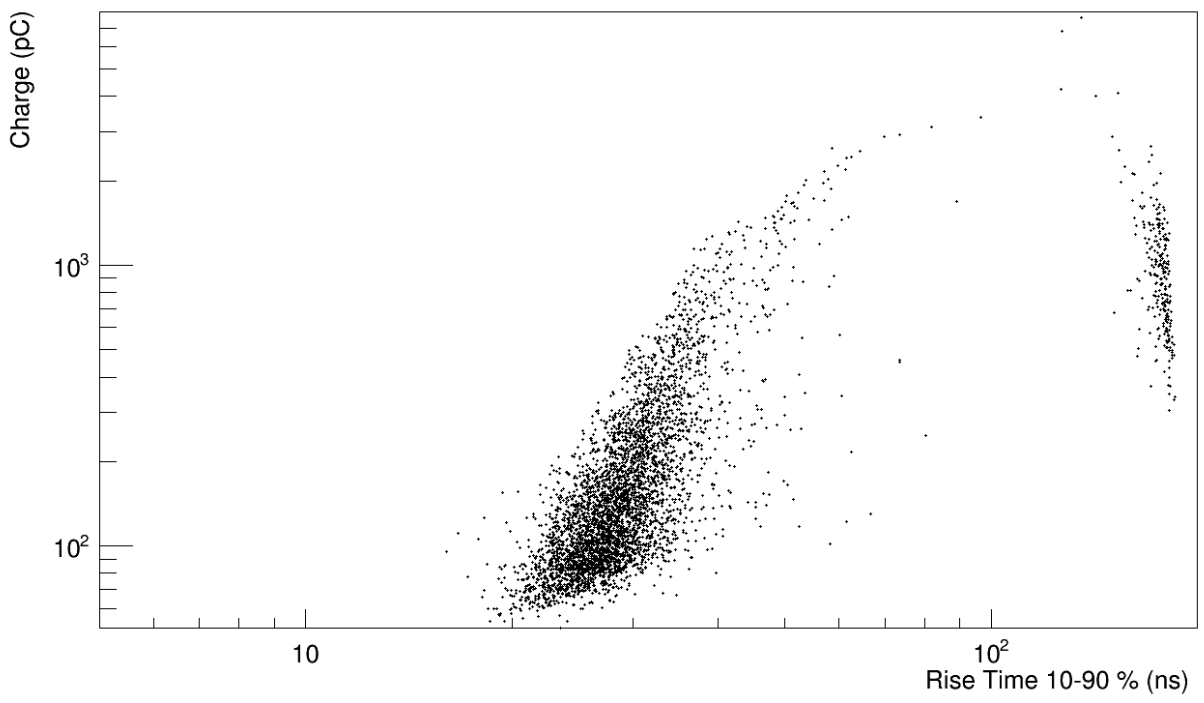

**Figure 6:** Correlation between charge and rise time 10-90 %, for events with amplitudes greater than 200 mV.

The bidimensional correlation plots, i.e., amplitude vs. rise time, charge vs. amplitude or charge vs. rise time as shown in figure 6, can be used to separate physical processes inside the

WCD [7] [8].

For the location of LAGO in Sierra Negra, the main contribution to the flux of secondary particles coming from cosmic rays are photons and electrons [1]. Events around 25 ns are due to electromagnetic particles. On the other hand, the measurement of the muon contents relative to the electromagnetic component give us an idea of the primary particle composition given that in heavier primary cosmic rays, the content of muons is higher [9].

## 4. Rate data

The acquisition rate of the first PMT started in June 2014. The installation and functionality of the entire WDC finished on January of this year. It is important to monitor the stability of the WCD. As an example we show the count rate at different threshold of one PMT. Figure 8 presents 9 day of data from February while figure 7 shows the entire month. The acquisition system is described in [10], it was implemented in rate mode and has 4 different thresholds for each PMT.

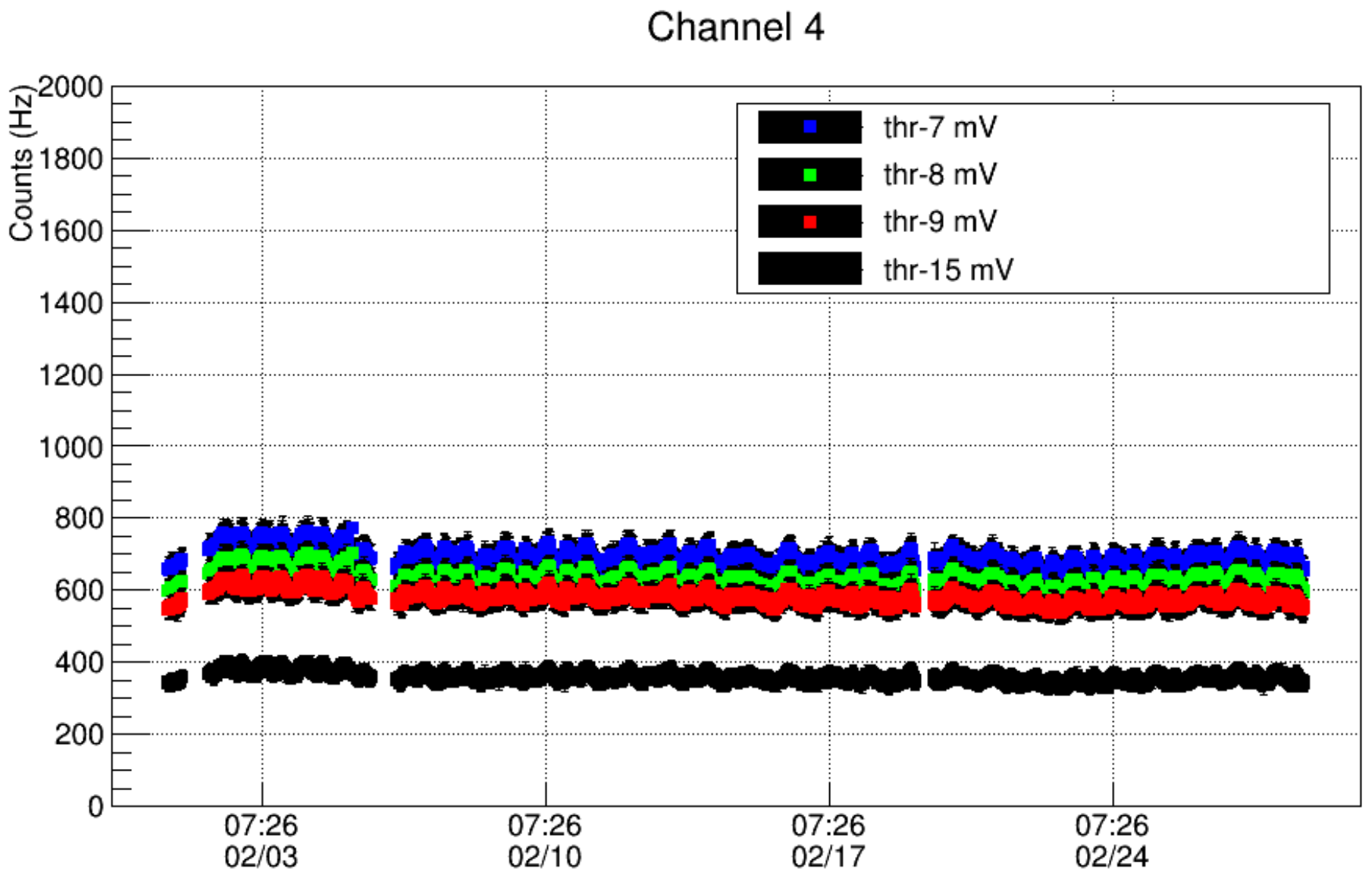

**Figure 7:** Data from February 2015 of the channel 4.

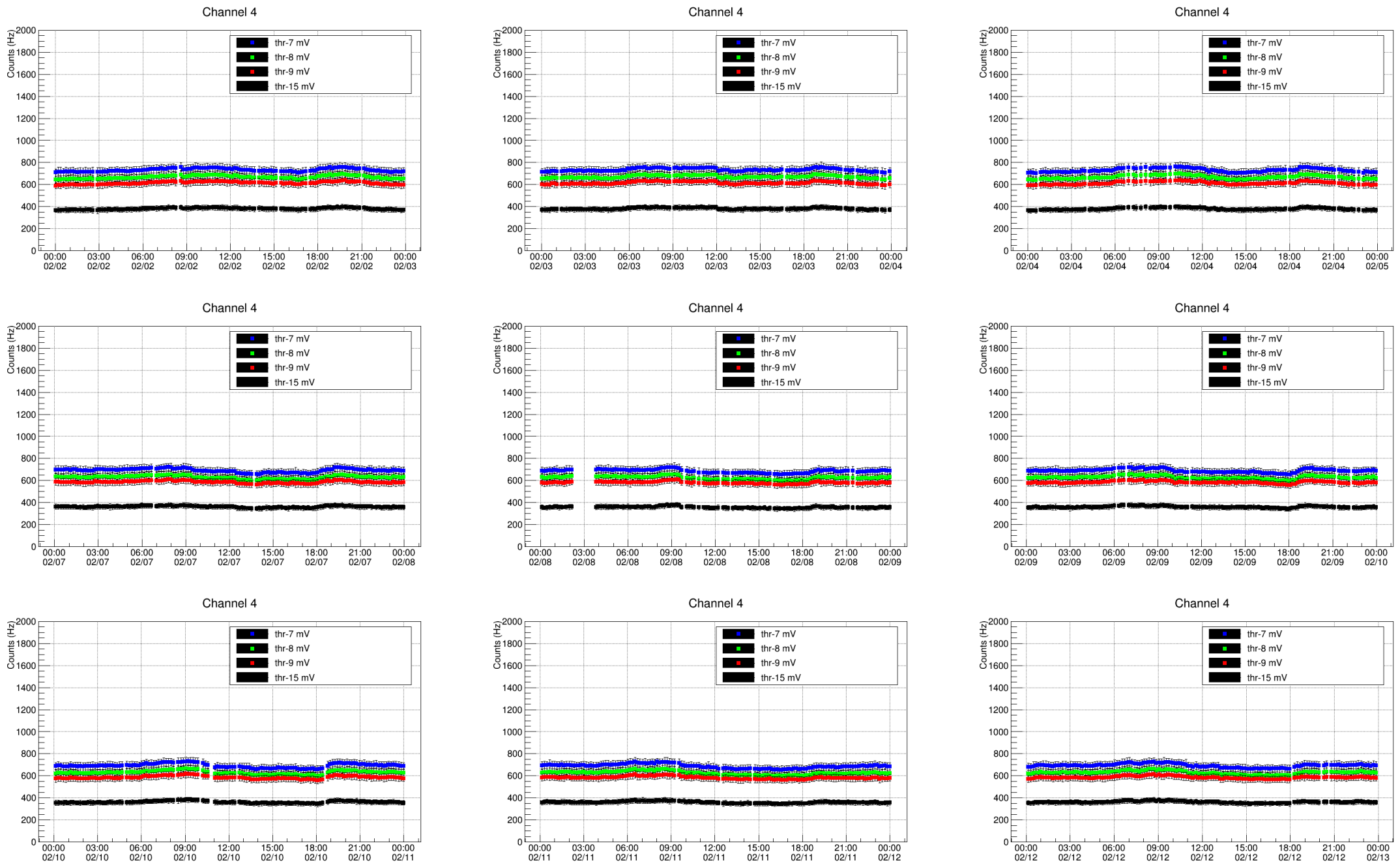

**Figure 8:** 9 days of data from Frebruary 2015, channel 4.

The rate mode based on single-particle rates was used as a simple estimator of the stability. The typical dispersion rms/mean was lower than 5% [7] for the analized data in the 4 PMTs.

## 5. Conclusions

An acquisition system to calibrate each PMT of the WCD has been implemented in order to obtain gain curves as a function of the voltage supply, using the SPE. After the installation of the calibrated PMTs, the characteristic curves of the WCD set in a coincidences mode were obtained and a breakpoint value of 1245 pC to fix the value for the other two WCDs. We also mentioned a method that migth be useful to separate electromagnetic from muon components from the installed WCD in coincidence setup. The stability from the rate counts of the detector with less than 5% in rms/mean value were discussed from preliminary data.